\documentclass[journal,10pt]{IEEEtran}
\usepackage{amssymb,amsmath,amsfonts,amsthm,bm,bbm}
\usepackage{mathrsfs}
\usepackage{graphicx}
\usepackage{cite}
\usepackage{color}
\usepackage{lipsum}
\usepackage[bookmarks]{hyperref}
\usepackage{booktabs}
\usepackage[mathscr]{euscript}
\usepackage[scr]{rsfso}
\usepackage{subcaption}
\usepackage{caption}
\usepackage{tablefootnote}
\usepackage{wrapfig}
\usepackage{listings}
\usepackage{array,threeparttable}
\usepackage{makecell}
\usepackage{relsize}
\usepackage{stackengine} 

\usepackage{enumitem} 

\usepackage{float}

\usepackage{multirow}
\usepackage[export]{adjustbox}
\makeatletter
\def\HyPsd@Warning#1{}%
\def\@font@warning#1{}%
\def\caption@Warning#1{}%
\def\caption@WarningNoLine#1{}%
\makeatother
\AtBeginDocument{\hfuzz=\maxdimen \vfuzz=\maxdimen \hbadness=10000 \vbadness=10000 \sloppy \raggedbottom \emergencystretch=3em}
\newcolumntype{M}[1]{>{\centering\arraybackslash}m{#1}}
\def\BibTeX{{\rm B\kern-.05em{\sc i\kern-.025em b}\kern-.08em
    T\kern-.1667em\lower.7ex\hbox{E}\kern-.125emX}}

\usepackage{float}
\usepackage{multirow}
\usepackage[export]{adjustbox}

\def\BibTeX{{\rm B\kern-.05em{\sc i\kern-.025em b}\kern-.08em
    T\kern-.1667em\lower.7ex\hbox{E}\kern-.125emX}}
\AtBeginDocument{\definecolor{ojcolor}{cmyk}{0.93,0.59,0.15,0.02}}

\newtheorem*{theorem}{Theorem}

\newtheorem{proposition}{Proposition}
\newtheorem{corollary}{Corollary}

\begin{document}

\title{A Unified Framework for Mixed  Fading Sums}

\author{Fernando Darío Almeida García \IEEEmembership{(Senior Member, IEEE)}, Michel Daoud Yacoub \IEEEmembership{(Member, IEEE)}, and José Cândido Silveira Santos Filho 
\IEEEmembership{(Member, IEEE)}%
\thanks{Fernando Darío Almeida García is with the Wireless and Artificial Intelligence (WAI) Laboratory, National Institute of Telecommunications (INATEL), Santa Rita do Sapucaí, MG, 37540-000, Brazil (e-mail: fernando.almeida@inatel.br).}%
\thanks{Michel Daoud Yacoub and José Cândido Silveira Santos Filho are with the Department of Communications, School of Electrical and Computer Engineering, State University of Campinas, Campinas, SP, 13083-852, Brazil.}%
\thanks{This work has been funded by the Brasil 6G Project with support from RNP/MCTI (Grant 01245.010604/2020-14), by the xGMobile Project (Code XGM-AFCCT-2025-8-1-1) with resources from EMBRAPII/MCTI (Grant 052/2023 PPI IoT/Manufatura 4.0), by CNPq (Grant No. 305038/2024-9), by FAPEMIG (Grant Nos. APQ-03162-24, APQ-04523-23, APQ-05305-23, PPE-00124-23, RED-00194-23), and by FINEP (nº 1060/2 contract 01.25.0883.00).}}

\maketitle

\begin{abstract}
This paper introduces a unified analytical framework that yields exact, compact, and tractable solutions for the probability density function (PDF) and cumulative distribution function (CDF) of the sum of a broad class of mixed independent positive random variables.
To demonstrate its breadth and practical value, the proposed framework is applied to the longstanding problem of deriving the exact statistics of mixed sums of fading variates.
Specifically, we derive novel expressions for the PDF and CDF of sums of Gaussian-class and non-Gaussian-class fading distributions, thereby covering a plethora of conventional, generalized, and recently introduced fading models.
The framework accommodates independent and identically distributed (i.i.d.) sums, independent but not necessarily identically distributed (i.n.i.d.) sums, and mixed-type sums. Furthermore, we conduct an exact performance analysis of equal-gain combining and maximal-ratio combining diversity receivers operating under mixed fading conditions. The proposed framework provides an exact and versatile analytical foundation for channel modeling and performance evaluation across a wide range of wireless fading scenarios. Finally, we introduce the novel \mbox{$\alpha$-$\mu$} \textit{Mixture distribution} that unifies all Gaussian-class fading models. 
\end{abstract}

\begin{IEEEkeywords}
Mixed sums of random variables, mixed sums of fading distributions, Gaussian-class and non-Gaussian-class fading, single fading, composite fading.
\end{IEEEkeywords}

\section{Introduction}
\label{sec:intro}
\IEEEPARstart{S}{ums} of positive random variables (RVs) have long been a fundamental topic in statistics and lie at the heart of a wide range of disciplines, including physics, engineering, and finance.
In the realm of wireless communications, the concept of the sum of RVs plays a pivotal role in understanding several wireless phenomena and crafting resilient communication systems capable of mitigating the negative impacts of fading.
Sums of random (fading) signals naturally arise in various wireless communication scenarios, including signal detection, intersymbol interference, phase jitter, diversity-combining receivers, linear equalization, false-alarm rate detection, and cooperative relaying.
More broadly, such sums are also fundamentally connected to signal-to-noise ratio (SNR)-oriented optimization in emerging wireless architectures, including satellite–aerial integrated networks and reconfigurable intelligent surface (RIS)- or active simultaneously transmitting and reflecting surface (ASTARS)-assisted systems~\cite{Lin2021SupportingIoT,Yue2026FederatedLearningSTARS,Lin2024SelfPoweredAbsorptiveRIS,Lin2022RefractingRIS,GarciaRIS26}.
In all these cases, the appropriately combined or processed signals aim to optimize the output SNR.
In wireless communications, the characterization of the statistics of the sum of fading signals---through quantities such as the probability density function (PDF), cumulative distribution function (CDF), and related metrics---has long been a central problem extensively investigated in the literature.
Because many key wireless operations rely on combining multiple random signal contributions, the PDF and CDF of fading sums remain fundamental for channel modeling and system design. Yet, despite decades of research, obtaining these statistics in an exact, unified, and computationally tractable form remains a major analytical challenge.

Sums of fading signals comprise two types: sums of single fading and sums of composite fading, the former usually referring to either short-term or long-term fading, and the latter to both combined. For the theory developed herein, it is useful to define two classes of fading models: the Gaussian class and the non-Gaussian class.

The Gaussian class of fading models comprises those constructed from the sum of independent and non-identically distributed (i.non-i.d.) squared complex Gaussian variates with either random or deterministic means and arbitrary variances.\footnote{The term i.non-i.d.\ refers to independent RVs drawn from the same distributional family whose PDF parameters may differ across variables, i.e., all PDF parameters may be different.} These models originate from the superposition of multiple Gaussian scattering components---each representing a distinct propagation path---and therefore naturally describe multipath fading environments containing diffuse and specular components.  
The Gaussian class encompasses single short-term (small-scale) and composite fading models. Representative examples include the one-sided Gaussian, Rayleigh, Nakagami-$m$, Rice, Weibull, $\bar{\alpha}$-$\bar{\mu}$~\cite{Yacoub07}, $\kappa$-$\bar{\mu}$, $\bar{\eta}$-$\bar{\mu}$~\cite{Yacoub07kappa}, $\bar{\alpha}$-$\kappa$-$\bar{\mu}$~\cite{Fraidenraich06}, Extended $\bar{\eta}$-$\bar{\mu}$~\cite{Gustavo20}, Extended $\bar{\alpha}$-$\bar{\eta}$-$\bar{\mu}$~\cite{Al_Hmood22}, $\bar{\alpha}$-$\bar{\eta}$-$\kappa$-$\bar{\mu}$~\cite{Silva20}, Beckmann ($\bar{\eta}$-$\kappa$)~\cite{Xie00}, two-wave with diffuse power (TWDP)~\cite{Ermolova16}, Shadowed-Rice~\cite{Alfano07}, $\kappa$-$\bar{\mu}$ Shadowed~\cite{Paris14}, $\bar{\alpha}$-$\kappa$-$\bar{\mu}$ Shadowed~\cite{Ramirez_Espinosa19}, fluctuating Beckmann~\cite{Ramirez18}, fluctuating two-ray (FTR)~\cite{Romero17}, independent FTR (IFTR)~\cite{Olyaee23}, multi-cluster FTR (MFTR)~\cite{david22}, and multi-cluster two-wave (MTW)~\cite{Olyaee25}.\footnote{Hereinafter, certain fading parameters and distribution names are overlined (e.g., $\bar{\alpha}$, $\bar{\mu}$, $\bar{\eta}$, $\bar{r}$) to prevent confusion with the PDF parameters in our derived expressions.}
In contrast, the non-Gaussian class of fading models includes all distributions that cannot be directly generated from Gaussian processes. These models typically emerge from nonlinear transformations---such as ratios, products, or inversions---of Gaussian-class variates~\cite{Garcia24} and are widely employed to describe composite or doubly shadowed propagation environments. Typical examples include ratios of $\bar{\alpha}$-$\bar{\mu}$ variates and all their particular cases, such as the Fisher-Snedecor $\mathcal{F}$~\cite{Yoo17} and $\bar{\alpha}$-$\mathcal{F}$~\cite{Badarneh20} models.

In this paper, we provide exact, tractable expressions for the PDFs and the CDFs of the sum of \emph{all} these cited models, including mixed combinations of them.

\subsection{Related Works}
\label{sec: Related Works}

In this paper, we aim to obtain the sum statistics of independent positive RVs in an \emph{exact} and \emph{unified} fashion, emphasizing Gaussian-class and non-Gaussian-class fading distributions.

\subsubsection{Sums of Gaussian-class Fading}
Sums of Gaussian-class fading distributions can be found at either power or envelope level. 
A considerable body of work has focused on the sum of fading variates at the power level---commonly referred to as power sums.
Such works include the sum of squared Nakagami-$m$ (gamma)~\cite{Moschopoulos85,Alouini01sumgamma,Al_Hussaini85,Efthymoglou95,Karagiannidis06,Ansari17}, squared Shadowed-Rice~\cite{Alfano07}, squared $\kappa$-$\bar{\mu}$~\cite{Bhatnagar15}, squared $\bar{\eta}$-$\bar{\mu}$~\cite{Peppas10,Ansari13}, squared two-wave with diffuse power (TWDP)~\cite{Ermolova16}, squared fluctuating two-ray (FTR)~\cite{Zheng19,Garcia23FTR}, squared multi-cluster fluctuating two-ray (MFTR)~\cite{Garzon24}.
Notably, in \cite{Garcia24}, it was shown that the power distributions of several Gaussian-class fading models belong to the same family of distributions: the mixture of gamma (MG) distribution. From that conclusion, the authors then derived the sum statistics  of MG variates, effectively covering the power sum statistics of a wide range of Gaussian-class fading distributions.
On the other hand,  sums at the envelope level---commonly referred to as envelope sums or simply as sums of fading distributions---are substantially more challenging because the Laplace transform (or moment-generating function (MGF)) of many Gaussian-class models is either unavailable in closed form or does not belong to analytically convenient families such as gamma or MG forms. Consequently, exact envelope-sum results remain fragmented and are largely confined to a limited set of conventional fading models.
Examples include sums of Rayleigh~\cite{Brennan59,Iskander02,Garcia22}, Nakagami-$m$~\cite{almeida21sbrt,Yacoub01,Dharmawansa07,Rahman11}, Rice~\cite{AlmeidaRice21}, Weibull~\cite{Yilmaz09conf,almeida2021}, $\bar{\alpha}$-$\bar{\mu}$~\cite{Rahama18,Almeida23_alpha_mu}, and \mbox{$\kappa$-$\bar{\mu}$}~\cite{Garcia23}.

\subsubsection{Sums of non-Gaussian-class Fading}
Sums of non-Gaussian-class fading distributions are also found at power and envelope levels. Non-Gaussian-class sums are gaining considerable attention due to their exceptional fit to experimental data in scenarios where multipath and shadowing coexist, particularly in the context of diversity-combining techniques. 
Notable studies  have examined power sums of non-Gaussian-class fading distributions, including the sum of squared Fisher-Snedecor $\mathcal{F}$~\cite{Du20,Badarneh18} and squared \mbox{$\bar{\alpha}$-$\mathcal{F}$}~\cite{Silva22}. On the other hand, research on envelope sums of non-Gaussian-class fading distributions includes sums of \mbox{$\bar{\alpha}$-$\mathcal{F}$}~\cite{Silva22} variates and sums of ratios of \mbox{$\bar{\alpha}$-$\bar{\mu}$}~\cite{Lenin25}.\footnote{For the sake of notation, we denote the ratio of two $\bar{\alpha}$-$\bar{\mu}$ distributions simply as \mbox{$\bar{\alpha}$-$\bar{\mu}$/$\bar{\alpha}$-$\bar{\mu}$}, following the convention established in \cite{Carlos19}.}

\subsection{Drawbacks and Limitations of Existing Solutions}

\subsubsection{Computational Burden}
Most existing exact solutions for the sum statistics (PDF and CDF) of fading distributions rely on Lauricella hypergeometric series, confluent multivariate hypergeometric functions, the multivariate Fox $H$-function, nested series, or multifold integrals. However, these approaches face significant limitations, as none of these functions are implemented in standard mathematical software and their evaluation requires multidimensional infinite-range integrals or nested infinite series~\cite{mathai09,Srivastavabook}. Although often described as ``closed form,'' the multivariate Fox $H$-function is, in practice, implemented numerically and is highly sensitive to solver settings (e.g., accuracy goals, initialization, integration paths, and numerical methods). 
Furthermore, as the number of RVs in the sum increases, the computational burden and numerical instability of the Lauricella series, the multivariate Fox $H$-function, nested series, and multifold integrals grow rapidly, rendering these approaches inefficient or even impractical beyond moderate dimensions (e.g., more than six RVs).

\subsubsection{Homogeneous and Heterogeneous Sums}
Throughout the literature, sums of fading distributions are largely restricted to homogeneous sums, i.e., sums of independent RVs following the \emph{same statistical distribution}. Homogeneous sums include the independent and identically distributed (i.i.d.) and the independent but not necessarily identically distributed (i.n.i.d.) cases.\footnote{The term i.n.i.d.\ refers to independent RVs drawn from the same distributional family whose PDF parameters are not all identical across variables, i.e., at least one parameter differs. The term i.i.d.\ refers to independent RVs drawn from the same distributional family whose PDF parameters are identical across all variables.}
In contrast, heterogeneous (mixed) sums refer to sums of independent RVs following \emph{different statistical distributions}, encompassing the i.i.d.\ and i.n.i.d.\ cases as special instances.
To date, the only broadly applicable routes for obtaining the exact statistics of heterogeneous sums of fading distributions have been the numerical evaluation of the multifold Brennan integral~\cite{Brennan59} or the numerical inversion of the product of MGFs. However, the Brennan integral offers limited analytical insight, scales poorly with the number of summed variables, and becomes increasingly numerically unstable and eventually impractical as the dimensionality grows. Similar limitations affect the MGF-based approach, since the MGFs of many fading distributions are either unavailable in closed form or lead to inverse Laplace transforms that do not admit analytical evaluation and must instead be computed numerically. In practice, these shortcomings render such methods unsuitable as general-purpose tools for modern mixed-fading scenarios, where one seeks not only numerical values but also compact formulas, efficient evaluation, and design-oriented interpretation.

\subsection{Open Problems}

Despite important progress, the literature still lacks a general mechanism for deriving the exact sum statistics across conventional, generalized, and recently introduced fading models. This limitation becomes particularly critical in modern wireless scenarios, where heterogeneous channel components and richer fading structures render case-by-case derivations increasingly inadequate. The difficulty stems primarily from the inherent analytical complexity of many fading models, which makes the derivation of exact PDF and CDF expressions for their sums particularly challenging.
For instance, the sum statistics of several generalized and recently introduced fading models---tailored to millimeter-wave (mmWave) and terahertz (THz) bands---remain unknown, even for the i.i.d. case, let alone for the more intricate mixed-sum scenario.
This gap includes the sum of \mbox{$\bar{\alpha}$-$\kappa$-$\bar{\mu}$} Shadowed~\cite{Ramirez_Espinosa19}, Extended \mbox{$\bar{\alpha}$-$\bar{\eta}$-$\bar{\mu}$}~\cite{Al_Hmood22}, \mbox{$\bar{\alpha}$-$\bar{\eta}$-$\kappa$-$\bar{\mu}$}~\cite{Silva20},  fluctuating Beckmann~\cite{Ramirez18}, IFTR~\cite{Olyaee23}, MFTR~\cite{david22}, and MTW~\cite{Olyaee25} fading distributions.
In practical wireless deployments, different propagation paths may follow distinct fading laws and parameter settings, making the analysis of mixed sums not only mathematically challenging but also essential for realistic channel modeling.

\subsection{Our Contributions}

In this work, we tackle the challenging problem of summing independent positive RVs.
The primary contributions of this paper are as follows:
\begin{enumerate}
    \item We introduce an encompassing fading model: the \mbox{$\alpha$-$\mu$} Mixture fading model. Specifically, by building on \cite{Garcia24}, we show that \emph{all} Gaussian-class fading distributions can be represented, in a unified and tractable fashion, as an $\alpha$-$\mu$ Mixture distribution.

    \item We develop a general framework that converts a broad class of mixed-sum problems into exact single-series PDF and CDF representations. This approach unifies heterogeneous sums within a single analytical framework, enabling the exact and compact characterization of the sum of a wide range of independent positive RVs.

    \item Leveraging our framework, we address the longstanding challenge of summing Gaussian-class (i.e., \mbox{$\alpha$-$\mu$} Mixture) and non-Gaussian-class (i.e., \mbox{$\bar{\alpha}$-$\bar{\mu}$/$\bar{\alpha}$-$\bar{\mu}$}) fading distributions.
    The analysis accommodates independent and identically distributed (i.i.d.) sums, independent but not necessarily identically distributed (i.n.i.d.) sums, and mixed-type sums.
    This unprecedented result enables a comprehensive analysis of sums of a broad spectrum of conventional, generalized, and recently introduced fading models.

    \item We present a unified performance analysis of wireless diversity receivers operating under mixed fading conditions. In particular, we derive exact expressions for the outage probability (OP) of equal-gain combining (EGC) and maximal-ratio combining (MRC) receivers subject to diverse fading channels.

\end{enumerate}

These contributions provide a unified analytical basis and practical tools by enabling accurate channel modeling and establishing a unified framework for evaluating key performance metrics across diverse fading conditions and receiver architectures. 
Therefore, beyond its theoretical generality, the proposed framework provides a tractable analytical tool for evaluating reliability in emerging systems such as RIS-assisted links, cooperative networks, and diversity receivers operating over mixed fading environments.

\subsection{Notation}

In the sequel, $\text{Pr}[\cdot]$ denotes probability; $\mathbb{E} \left[ \cdot \right]$, expectation; $(\cdot)_{(\cdot)}$, the Pochhammer symbol; $\Gamma (\cdot)$, the gamma function; $\Upsilon \left( \cdot, \cdot \right)$, the lower incomplete gamma function; $\, _2F_1 \left( \cdot \right)$, the Gauss hypergeometric function; $\mathrm B(\cdot)$, the multivariate beta function; $L_{(\cdot)}^{(\cdot)} (\cdot)$, the generalized Laguerre polynomials; $\mathbb{R}$, the set of real numbers; $\mathbb{R}_0^+$, the set of positive real numbers including zero; $\mathbb{N}$, the set of natural numbers; $\mathbb{N}_0$, the set of natural numbers including zero; $I_{\nu}(\cdot)$, the modified Bessel function of the first kind and $\nu$th order; $P_{\left( \cdot \right)}^{\left( \cdot \right)} \left( \cdot \right)$, the Legendre polynomial of the first kind; and $\sim$, ``asymptotically equal to around zero,'' i.e., ${h(x)\sim g(x) \iff \lim\limits_{x\to 0} \, \frac{h (x)}{g (x)}=1}$.

\section{Proposed Sum Framework}
\label{sec: Main Result}

In this section, we introduce a novel analytical framework for the exact statistical characterization of sums of mixed independent positive RVs. A simplified formulation for the sum of i.i.d.\ positive RVs is also presented.
The main results are formally presented in the following theorem and corollary.

\subsection{Main Results}

\begin{theorem}
\label{theo: Main Theorem}
Let $\{X_\ell\}_{\ell=1}^{L}$ be independent, continuous,
nonnegative RVs. Assume that, for each $\ell\in\{1,\ldots,L\}$,
there exist parameters $\theta,\Psi_\ell,\beta_\ell>0$ such that
the PDF of $X_\ell$ admits the representation
\begin{align}
\label{eq: PDF marginal theorem}
f_{X_\ell}(x)
&=
\Psi_\ell
\sum_{i=0}^{\infty}
\frac{
\eta_{i,\ell}x^{i\theta+\beta_\ell-1}
}{
\Gamma(i\theta+\beta_\ell)
}, \qquad x>0,
\end{align}
where $\eta_{i,\ell}\in\mathbb R$ for all $i\in\mathbb N_0$ and
$\eta_{0,\ell}>0$ for each $\ell$.
Assume further that, for each $\ell$, there exist constants
$C_\ell,A_\ell,c_\ell>0$ and $0 \leq q_\ell<1$, all independent of
$i$, such that
\begin{align}
|\eta_{i,\ell}|
&\leq
C_\ell A_\ell^i
\frac{
\Gamma(i\theta+\beta_\ell)\Gamma(q_\ell i+c_\ell)
}{
i!
},
\qquad i\in\mathbb N_0.
\label{eq: generalized coefficient growth}
\end{align}

Define
\begin{align}
\label{eq: Sum def X}
X
\triangleq
\sum_{\ell=1}^{L}X_\ell .
\end{align}
Then, the PDF and CDF of $X$ are given, respectively, by
\begin{align}
\label{eq: PDF X Final}
f_X(x)
&=
\left(\prod_{\ell=1}^{L}\Psi_\ell\right)
\sum_{i=0}^{\infty}
\frac{
\delta_i x^{\theta i+B-1}
}{
\Gamma(\theta i+B)
}, & x>0
\\
\label{eq: CDF X Final}
F_X(x)
&=
\left(\prod_{\ell=1}^{L}\Psi_\ell\right)
\sum_{i=0}^{\infty}
\frac{
\delta_i x^{\theta i+B}
}{
\Gamma(1+\theta i+B)
},
& x\geq0,
\end{align}
where 
\begin{align}
    \label{eq: B def}
    B \triangleq
\sum_{\ell=1}^{L}\beta_\ell,
\end{align}
and the coefficients $\delta_i$ are determined by
\begin{subequations}
\label{eq: delta coefficients}
\begin{align}
\label{eq: delta 0}
\delta_0
&=
\prod_{\ell=1}^{L}\eta_{0,\ell},
\\
\label{eq: delta i}
\delta_i
&=
\frac{1}{i}
\sum_{h=1}^{i}
\delta_{i-h}
\sum_{\ell=1}^{L}
\phi_{h-1,\ell},
\qquad i\geq1.
\end{align}
\end{subequations}
The auxiliary coefficients $\phi_{h,\ell}$ in
\eqref{eq: delta i} are given by
\begin{subequations}
\label{eq: phi coefficients}
\begin{align}
\label{eq: phi 0}
\phi_{0,\ell}
&=
\frac{\eta_{1,\ell}}{\eta_{0,\ell}},
\\
\label{eq: phi i}
\phi_{h,\ell}
&=
\frac{1}{\eta_{0,\ell}}
\left[
(h+1)\eta_{h+1,\ell}
-
\sum_{t=1}^{h}
\eta_{t,\ell}\phi_{h-t,\ell}
\right],
\qquad h\geq1.
\end{align}
\end{subequations}
\end{theorem}

\begin{proof}
See Appendix~\ref{app: Theorem}. 
\end{proof}

The series in \eqref{eq: PDF X Final} and
\eqref{eq: CDF X Final} converge absolutely and locally uniformly
on $(0,\infty)$ and $[0,\infty)$, respectively,
as shown in Appendix~\ref{app: Absolute Convergence X}. The coefficient-growth condition in
\eqref{eq: generalized coefficient growth} is used only to
establish these $x$-domain convergence properties.

The Theorem also leads to the following corollary.

\begin{corollary}
\label{corol: IID case}
If $\left\{X_{\ell }\right\}_{\ell=1}^{L}$ is a set of i.i.d. continuous RVs (i.e., $\Psi _{\ell }= \Psi$,  $\beta_\ell = \beta$, $\eta_{i,\ell}= \eta_i$) with support $\mathbb{R}_0^+$, then the PDF and the CDF of $X$, given in \eqref{eq: PDF X Final} and \eqref{eq: CDF X Final}, reduce, respectively, to
\begin{align}
    \label{eq: PDF X Final iid}
    f_X (x) = & \Psi^L \sum _{i=0}^{\infty } \frac{\delta _i \, x^{-1+\theta  i+ L \beta }}{\Gamma \left(i \theta +L \beta \right)}, & x>0\\ \label{eq: CDF X Final iid}
    F_X (x) = & \Psi^L \sum _{i=0}^{\infty } \frac{\delta _i \, x^{\theta  i+L \beta }}{\Gamma \left(1+i \theta + L \beta \right)}, & x\geq0
\end{align}
\normalsize
where 
\begin{subequations}
\label{eq: delta coefficients iid}
\begin{align}
    \label{eq: delta 0 iid}
    &\delta_0= \eta _{0}^L & \\ 
    \label{eq: delta i iid}
    & \delta_i=  \frac{1}{i \eta _0 } \sum _{h=1}^i \delta _{i-h} \, \eta _h  \left(h L+h-i \right), & i  \geq 1.
\end{align}
\end{subequations}
\normalsize

\end{corollary}

\begin{proof}
See Appendix~\ref{app: IID case}.
\end{proof}

\textit{Remark 1:}
The Theorem provides a unified framework for sums of heterogeneous independent
nonnegative RVs whose marginal PDFs admit the canonical series
representation in \eqref{eq: PDF marginal theorem}. It is important to emphasize that
\eqref{eq: generalized coefficient growth} is used only to establish
absolute and locally uniform convergence in the $x$-domain.
The final PDF and CDF expressions
are obtained directly from the convolution structure and
absolute-convergence arguments, rather than from the manipulation
or inversion of marginal Laplace series.
Accordingly, for each marginal family under consideration, both
the canonical PDF representation in \eqref{eq: PDF marginal theorem}
and the generalized coefficient-growth condition in
\eqref{eq: generalized coefficient growth} must be verified
independently. As shown in Sections~\ref{sec: Sum of Mixtures} and
\ref{sec: Sum of Ratios of alpha-mu Distributions}, these
requirements can be established for broad classes of
Gaussian-class and non-Gaussian-class fading models. This includes,
in particular, the fading models studied in
\cite{Yacoub07,Yacoub07kappa,Fraidenraich06,Gustavo20,Al_Hmood22,
Silva20,Xie00,Ermolova16,Alfano07,Paris14,Ramirez_Espinosa19,
Ramirez18,Romero17,Olyaee23,david22,Olyaee25,Yoo17,Badarneh20,
Moschopoulos85,Alouini01sumgamma,Al_Hussaini85,Efthymoglou95,
Karagiannidis06,Ansari17,Bhatnagar15,Peppas10,Ansari13,Zheng19,
Garcia23FTR,Garzon24,Garcia24,Brennan59,Iskander02,Garcia22,
almeida21sbrt,Yacoub01,Dharmawansa07,Rahman11,AlmeidaRice21,
Yilmaz09conf,almeida2021,Rahama18,Almeida23_alpha_mu,Garcia23,
Du20,Badarneh18,Silva22,Lenin25,Carlos19}.
The i.i.d.\ case follows as a special instance through
Corollary~\ref{corol: IID case}.

Some fundamental distinctions between our solutions and those in \cite[Theorem 2]{Garcia24} are worth highlighting.
The results in \cite[Theorem 2]{Garcia24} are limited to power sums of RVs whose PDFs admit an MG representation, thereby excluding (i) power sums of several other Gaussian-class RVs, (ii) envelope sums of \emph{all} Gaussian-class RVs, and (iii) power and envelope sums of \emph{all} non-Gaussian-class RVs. In contrast, the proposed solutions are considerably more general, as they accommodate random variables (RVs) with diverse PDF structures beyond the MG form, enabling the analysis of power and envelope sums across \emph{all} Gaussian-class RVs and a wide range of non-Gaussian-class RVs.
In the special case $\theta=1$ (i.e., power sums), \eqref{eq: PDF X Final} and \eqref{eq: CDF X Final} yield alternative and complementary solutions to those in~\cite{Garcia24} for sums of MG distributions.

\subsection{Complexity Analysis}
\label{sec: Complexity Analysis}

Here, we assess the computational cost for  the heterogeneous and i.i.d. cases in units of \emph{sum iterations}---the number of terms traversed by explicit summations in the coefficient recursions and in the outer series. For the analysis, we truncate the PDF and CDF of $X$ to $t_0$ terms.
The following complexity analysis assumes that the marginal coefficients ${\eta_{i,\ell}}$ are known, as is the case for all fading models considered herein, for which these coefficients are available in closed form.

\subsubsection{Heterogeneous Case [Eqs. \eqref{eq: PDF X Final} and \eqref{eq: CDF X Final}]}
\label{sec: Complexity Analysis Heterogeneous}

Using \eqref{eq: phi coefficients}, for a fixed $\ell$ and truncation to $t_0$ terms, constructing $\{\phi_{h,\ell}\}_{h=1}^{t_0-1}$ entails the inner sum of length $h$ in \eqref{eq: phi i}; hence the cost per $\ell$ is $\sum_{h=1}^{t_0-1} h = t_0(t_0-1)/2$ sum iterations. Summed over all $\ell$, this gives $\mathcal{O}(L\,t_0(t_0-1)/2 )= \mathcal{O}(L\,t_0^2)$.  
From \eqref{eq: delta i}, for each fixed $i$ the sum over $h=1,\dots,i$ contains an inner sum over $\ell=1,\dots,L$; thus, computing $\delta_i$ requires $\mathcal{O}(i\,L)$ sum iterations, and summing $i=1,\dots,t_0-1$ yields $\mathcal{O}(L\,t_0(t_0-1)/2 )= \mathcal{O}(L\,t_0^2)$.  
The remaining work---$\delta_0$ (one pass over $L$ terms, $\mathcal{O}(L)$) and the two outer series in \eqref{eq: PDF X Final} and \eqref{eq: CDF X Final} (each requires the evaluation of $t_0$ terms, totaling $\mathcal{O}(t_0)$)---is lower order. Therefore, the overall complexity for evaluating the PDF and CDF is dominated by the coefficient construction and scales as $\mathcal{O}(L\,t_0^2)$. In short, the heterogeneous case grows \emph{linearly} with $L$ and \emph{quadratically} with $t_0$.

\subsubsection{I.I.D. Case [Eqs. \eqref{eq: PDF X Final iid} and \eqref{eq: CDF X Final iid}]}
\label{sec: Complexity Analysis Homogeneous}

From \eqref{eq: delta i iid}, for each fixed $i$ the sum over $h=1,\dots,i$ has $i$ terms and the factor $(hL+h-i)$ is scalar (there is no inner summation over $L$). Hence computing $\delta_i$ requires $\mathcal{O}(i)$ sum iterations, and summing $i=1,\dots,t_0-1$ gives $\mathcal{O}(t_0(t_0-1)/2) = \mathcal{O}(t_0^2)$.  
Computing $\delta_0=\eta_0^{L}$ costs $\mathcal{O}(L)$, and evaluating the two $t_0$-term outer series contributes $\mathcal{O}(t_0)$, both lower order. Consequently, the overall complexity for \eqref{eq: PDF X Final iid} and \eqref{eq: CDF X Final iid} is dominated by the coefficient recursion and scales as $\mathcal{O}(t_0^2)$. In the i.i.d. case, the dominant cost is \emph{independent} of $L$ and remains \emph{quadratic} in $t_0$.
Thus, moving from the heterogeneous to the i.i.d. case removes the multiplicative $L$ factor in the coefficient stage,
reducing the leading-order work from $\mathcal{O}(L\,t_0^2)$ to $\mathcal{O}(t_0^2)$.

\subsubsection{Complexity of Competing Solutions}

As noted earlier, many sum solutions appearing in the literature are expressed in terms of an $L$-dimensional Lauricella hypergeometric series, confluent multivariate hypergeometric functions, or $L$-fold nested series~\cite{Efthymoglou06,Yilmaz09conf,Dharmawansa07,Rahman11,Zheng19,Badarneh21}.
When these series are truncated to $t_0$ terms and analyzed following the approach in Sections~\ref{sec: Complexity Analysis Heterogeneous} and~\ref{sec: Complexity Analysis Homogeneous}, their computational complexity is found to be at least $\mathcal{O}(t_0^L)$.
This indicates that their complexity grows \emph{combinatorially} with $L$, which in practice behaves similarly to exponential growth for moderate and large values of $L$.

Other exact solutions are expressed in terms of the $L$-variate Fox-H function~\cite{Badarneh21,Rahama18,Du20}.
A direct one-to-one complexity comparison is not
possible because the numerical cost depends on the contour
parametrization, truncation range, quadrature rule, and prescribed
accuracy.
Nevertheless, under standard tensor-product quadrature with
\(\epsilon_{\mathrm{points}}\) quadrature points per integration dimension, the number of quadrature nodes scales as $O\!\left(\epsilon_{\mathrm{points}}^L\right)$.
Thus, this direct implementation also exhibits exponential growth with
\(L\). More advanced techniques, including sparse-grid quadrature and
adaptive cubature, may reduce the practical cost, but the computational
burden generally still increases rapidly with dimensionality.
The highly reduced complexity of our derived expressions, compared with state-of-the-art alternatives, highlights their  analytical and computational advantages.

\subsection{Truncation Error Analysis}

The PDF and CDF expressions in
\eqref{eq: PDF X Final} and \eqref{eq: CDF X Final} are given as infinite-series
representations. For numerical evaluation, these series must
be truncated to a finite number of terms. As in
Section~II-\ref{sec: Complexity Analysis}, suppose that
$t_0\in\mathbb N$ terms are retained.
Thus, the corresponding truncation errors for the PDF and CDF of the sum can be defined, respectively, as
\begin{align}
    \label{eq: Truncation Error PDF}
    \Xi_{f}^{(t_0)}  (x)\triangleq  & \left( \prod _{\ell =1}^L \Psi _{\ell } \right) \sum _{i=t_0}^{\infty } \frac{\delta _i \, x^{-1+\theta  i+B }}{\Gamma \left(i \theta +B \right)} \\ \label{eq: Truncation Error CDF}
    \Xi_{F}^{(t_0)} (x)\triangleq  & \left( \prod _{\ell =1}^L \Psi _{\ell } \right) \sum _{i=t_0}^{\infty } \frac{\delta _i \, x^{\theta  i+ B }}{\Gamma \left(1+i \theta + B \right)},
\end{align}
\normalsize
or, in compact form, as 
\begin{align}
    \label{eq: Truncation Error PDF CDF}
    \Xi _{\vartheta }^{(t_0)} (x)  \triangleq  & \left( \prod _{\ell =1}^L \Psi _{\ell } \right) \sum _{i=t_0}^{\infty } \frac{\delta _i x^{-1+\varrho+ \theta  i+B }}{\Gamma \left(i \theta +\varrho +B \right)},
\end{align}
\normalsize
where $\vartheta \in \{f,F\}$ and 
\begin{align}
\varrho
\triangleq
\begin{cases}
0, & \vartheta=f,\\
1, & \vartheta=F.
\end{cases}
\label{eq: varrho truncation main}
\end{align}

Since the coefficients $\delta_i$ need not be nonnegative, the
truncation errors are bounded in absolute value. As shown in
Appendix~\ref{app: Truncation error Bounds}, under the generalized
coefficient-growth condition in
\eqref{eq: generalized coefficient growth}, the absolute truncation
error satisfies
\begin{align}
& \left|\Xi_{\vartheta}^{(t_0)}(x)\right|
\leq
C_{\boldsymbol\beta}
\left(\prod_{\ell=1}^{L}C_\ell\Psi_\ell\right)
(t_0\theta+B)^{-\varrho}
x^{B+\varrho-1}
\chi^{-t_0}
\nonumber\\
&\quad\times
\prod_{\ell=1}^{L}
\Gamma(c_\ell)(1-\varepsilon_\ell)^{-c_\ell}
\exp\!\left[
\sum_{\ell=1}^{L}
\omega_\ell\!\left(
\varepsilon_\ell,
\chi A_\ell x^\theta
\right)
\right],
\label{eq: closed truncation bound main}
\end{align}
where
$\chi>1$, and $\varepsilon_\ell\in(0,1)$ are arbitrary. Here,
$C_{\boldsymbol\beta}$ is defined in
\eqref{eq: gamma ratio bound generalized}, and
\begin{align}
\omega_\ell(\varepsilon_\ell,y)
&\triangleq
(1-q_\ell)
q_\ell^{\frac{q_\ell}{1-q_\ell}}
\varepsilon_\ell^{-\frac{q_\ell}{1-q_\ell}}
y^{\frac{1}{1-q_\ell}} .
\end{align}
For $q_\ell=0$, $\omega_\ell(\varepsilon_\ell,y)$ is understood by
continuity as $y$.

The bound in \eqref{eq: closed truncation bound main} is valid only
when the generalized coefficient-growth condition
\eqref{eq: generalized coefficient growth} has been independently
verified. Accordingly, valid model-specific constants
$C_\ell$, $A_\ell$, $q_\ell$, and $c_\ell$ must be established
before \eqref{eq: closed truncation bound main} can be interpreted
as a rigorous numerical error bound. If such constants are
unavailable, the truncation order must instead be determined using
a different analytical majorant or an a posteriori numerical
convergence criterion.

In the subsequent sections, we demonstrate the framework's versatility and broad applicability by obtaining novel, exact solutions for the PDF and CDF of \mbox{$\alpha$-$\mu$} Mixture and \mbox{$\bar{\alpha}$-$\bar{\mu}$/$\bar{\alpha}$-$\bar{\mu}$} fading distributions.
To ensure clarity and logical progression, the analysis is divided into three scenarios: 1) sums of i.n.i.d. and i.i.d. \mbox{$\alpha$-$\mu$} Mixture distributions;  2) sums of i.n.i.d. and i.i.d. \mbox{$\bar{\alpha}$-$\bar{\mu}$/$\bar{\alpha}$-$\bar{\mu}$} distributions; and 3) the mixed-sum scenario, involving the combined sum of \mbox{$\alpha$-$\mu$}  Mixture and \mbox{$\bar{\alpha}$-$\bar{\mu}$/$\bar{\alpha}$-$\bar{\mu}$} fading distributions.

\section{Scenario 1: Sum of $\alpha$-$\mu$ Mixtures}
\label{sec: Sum of Mixtures}

In this section, we demonstrate that all Gaussian-class fading models are governed by an \mbox{$\alpha$-$\mu$} Mixture distribution.
We then obtain the chief statistics for the sum of \mbox{$\alpha$-$\mu$} Mixture  distributions.

\subsection{The $\alpha$-$\mu$ Mixture Distribution}

The $\alpha$-$\mu$ Mixture distribution is formally defined and characterized in the next proposition.

\begin{proposition}
\label{prop: common structure}
Let $\left\{R_{n}\right\}_{n=1}^{N}$ be a set of i.non-i.d. RVs, where each RV $R_{n}$ is defined as
\begin{align}
    \label{eq: Rn def}
    R_n \triangleq \sum _{m=1}^{T_n} \left(Q_{m,n}+ P_{m,n}\right)^2,
\end{align}
\normalsize
where $\left\{Q_{m,n}\right\}_{m=1,n=1}^{T_n,N}$ is a set of independent Gaussian RVs with zero mean and variance $\sigma _n^2$, and $\left\{P_{m,n}\right\}_{m=1,n=1}^{T_n,N}$ is a set of continuous i.non-i.d. RVs with finite variance and support $\left(-\infty,+\infty \right)$.
The two sets $\left\{Q_{m,n}\right\}$ and $\left\{P_{m,n}\right\}$ are also mutually independent. 
The variables $P_{m,n}$ can also be
deterministic.
Define $P_n \triangleq \frac{1}{\sigma_{n}^2} \sum _{m=1}^{T_n} P_{m,n}^2$, and $R^{\alpha}$ as the weighted sum of all RVs $R_n$, namely,
\begin{align}
    \label{eq: R def}
    R^{\alpha} \triangleq \sum _{n=1}^N v_n R_n,
\end{align}
\normalsize
where $v_n \triangleq \tilde{r}^\alpha  /(\sigma_{n}^2 N \left( T_n +\bar{P}_n\right))$, with $\bar{P}_n \triangleq \mathbb{E} \left[ P_n\right]$, $\tilde{r} \triangleq \sqrt[\alpha]{\mathbb{E} \left[ R^\alpha\right]}$, and $a_n
\triangleq
\sigma_n^2v_n$.
The term $v_n$ simply  acts as a normalizing factor to ensure that the expectation of $R^\alpha$ equals a given constant $\tilde{r}^\alpha$.
Then, regardless of the random or deterministic nature of $P_{m,n}$, the PDF and CDF of $R$ can be expressed in the form of a mixture of $\alpha$-$\mu$ distributions, given, respectively, by 
\begin{align}
    \label{eq: PDF R general}
    f_{R}(r)=& \sum _{i=0}^{\infty } \varphi_i f_{R_i} \left(\alpha, \mu+i, \hat{r}, r\right) \\ \label{eq: CDF R general}
    F_{R}(r)=& \sum _{i=0}^{\infty }\varphi _i F_{R_i} \left(\alpha, \mu+i, \hat{r}, r\right),
\end{align}
\normalsize
where $\hat{r}>0$, and
\begin{align}
    \label{eq: PDF R marginal}
     f_{R_i} \left(\alpha, \mu+i, \hat{r},r\right)= & \frac{\alpha}{r}
\frac{
\left(r^\alpha/\hat r^\alpha\right)^{\mu+i}
}{
\Gamma(\mu+i)
}
\exp\left(
-\frac{r^\alpha}{\hat r^\alpha}
\right) &  \\ \label{eq: CDF R marginal}
    F_{R_i} \left(\alpha, \mu+i, \hat{r},r\right) =& \frac{
\Upsilon\left(
\mu+i,
r^\alpha/\hat r^\alpha
\right)
}{
\Gamma(\mu+i)
}
\end{align}
\normalsize
denote the $i$th $\alpha$-$\mu$ PDF and CDF of the mixture, respectively, $\mu \triangleq \frac{1}{2} \sum _{n=1}^N T_n$, and $\varphi _i$ are the mixture weights, given by 
\begin{align}
    \label{eq: varphi k def}
    \varphi _i = \frac{\left(-\mu \right)^i}{i! } \sum _{j=i}^{\infty } \frac{j! h_j}{\Gamma (j-i+1)}.
\end{align}
\normalsize
Furthermore, $\sum _{i=0}^{\infty } \varphi _i=1$ with $\varphi _0> 0$, and $h_j$ in \eqref{eq: varphi k def} is obtained as
\begin{align}
    \label{eq: dj def}
    h_j = \int _0^{\infty }\cdots \int _0^{\infty }c_j \, \prod_{n=1}^{N}\mathit{f}_{P_n}\left(\rho_n\right) \text{d}\rho_1 \cdots \text{d}\rho_N,
\end{align}
\normalsize
in which $f_{P_n} (\rho_n)$ is the PDF of $P_n$, and $c_j$ is given by
\begin{subequations}
\label{eq: Coef cj}
\begin{align}
    \label{eq: c0}
    \nonumber c_0=& \mu ^{\mu } \exp \left(-\frac{1}{2} \sum _{n=1}^N \frac{(\mu -1) \rho _n a_n}{(\mu -1) a_n+\frac{\hat{r}^{\alpha }}{2 }}\right) & \\
    & \times \prod _{n=1}^N \left(\frac{2 \mu   a_n}{\hat{r}^{\alpha }}-\frac{2 a_n}{\hat{r}^{\alpha }}+1\right)^{-\frac{T_n}{2}} & \\ \label{eq: cj}
    c_j=& \frac{1}{j}\sum _{l=0}^{j-1} c_l \, d_{j-l}, & j \geq 1.
\end{align}
\end{subequations}
\normalsize

Finally, $d_l$ in \eqref{eq: cj} is given by
\begin{align}
    \label{eq: di coef}
    \nonumber  d_l & = -\frac{l \mu  \hat{r}^{\alpha }}{4} \sum _{n=1}^N \rho _n a_n  \left(\frac{1}{a_n(\mu -1)+\frac{\hat{r}^{\alpha }}{2}}\right)^{l+1} \\
     \nonumber & \times \left(\frac{\hat{r}^{\alpha }}{2}-a_n\right)^{l-1} +  \frac{1}{2}\sum _{n=1}^N T_n \left(\frac{2  a_n}{\hat{r}^{\alpha }}(\mu -1)+1\right)^{-l} \\
     & \times \left(1-\frac{2 a_n}{\hat{r}^{\alpha }}\right)^l, \ \ l\geq 1.
\end{align}
\normalsize

\end{proposition}

\begin{proof}
Following the derivation in \cite[App. A]{Garcia24} and after replacing $p_0=1$ and $p_1=\hat{r}^{\alpha} /2$ into \cite[Eqs. (44)--(46)]{Garcia24}, we are able to find the PDF and CDF of $W \triangleq R^{\alpha} $. Then, after a straightforward transformation of variables $ R = W^{1/\alpha} $, \eqref{eq: PDF R general} and \eqref{eq: CDF R general} are attained. Moreover, as the PDF and CDF of $W$ converge absolutely and uniformly, as outlined in~\cite{Garcia24}, so do the series in \eqref{eq: PDF R general} and \eqref{eq: CDF R general}.
    This concludes the proof.
\end{proof}

\begin{table*}[t!]
\caption{PDF Mapping}\label{tab: PDF-CDF parameters}
  \centering\begin{threeparttable}
    \resizebox{\textwidth}{!} {\begin{tabular}{M{20mm} M{110mm} M{5mm} M{5mm} M{10mm}} \toprule
{\textbf{Distribution}}
&\shortstack{$\varphi _i$}& \shortstack{$\alpha$}&{$\mu$} &{$\hat{r}^\alpha$} \\ \toprule \toprule

    Extended $\bar{\eta}$-$\bar{\mu}$~\cite{Gustavo20}
    & $\frac{1}{i!} \left(\frac{p}{\bar{\eta} }\right)^{\frac{\bar{\mu}  p}{p+1}} \left(\frac{\bar{\eta} -p}{\bar{\eta} }\right)^i \left(\frac{\bar{\mu}  p}{p+1}\right)_i$
    & $2$
    &  $\bar{\mu}$
    &  $\frac{\bar{\gamma} (1+p)}{\mu (1+\eta)}$ \\ \midrule 

    TWDP~\cite{Ermolova16}
    & $\frac{\exp (-K) K^i}{i!} \sum _{k=0}^i \left(\frac{\Delta }{2}\right)^k \binom{i}{k} \sum _{l=0}^k \binom{k}{l} I_{2 l-k}(-K \Delta )$
    & $2$
    &  $1$
    &  $2 \bar{\sigma}^2$ \\ \midrule

    FTR~\cite{Romero17}
    & $\begin {array} {c} \frac{m^m K^i}{i! \Gamma(m)}  \sum _{k=0}^i \left(\frac{\Delta }{2}\right)^k \binom{i}{k} \sum _{\nu =0}^k \binom{k}{\nu } \Gamma (i-k+m+2 \nu ) \exp \left(\frac{1}{2} \pi  i (2 \nu -k)\right)  \\ \times \, \left((K+m)^2-(\Delta  K)^2\right)^{-\frac{1}{2} (i+m)}  P_{i+m-1}^{k-2 \nu }\left(\frac{K+m}{\sqrt{(K+m)^2-(\Delta  K)^2}}\right)  \\ \end {array}$
    & $2$
    & $1$ 
    & $2 \bar{\sigma}^2$ \\ \midrule

    $\bar{\alpha}$-$\kappa$-$\bar{\mu}$ Shadowed~\cite{Ramirez_Espinosa19}
    & $\frac{m^m (m)_i}{i! (\kappa  \bar{\mu} +m)^m} \left(\frac{\kappa  \bar{\mu} }{\kappa  \bar{\mu} +m}\right)^i$
    & $\bar{\alpha}$
    &  $\bar{\mu}$ 
    &  $\frac{\bar{\gamma }}{(\kappa+1) \bar{\mu} }$ \\ \midrule 
    
    MFTR~\cite{david22}
    & $\begin {array} {c} \frac{m^m (1-\Delta )^i (K \bar{\mu} )^i \Gamma (i+m)}{\sqrt{\pi } \Gamma (i+1) \Gamma (m) ((1-\Delta ) K \mu+m)^{i+m}}\sum _{\bar{q}=0}^i \binom{i}{\bar{q}} \frac{ \Gamma \left( \bar{q}+\frac{1}{2}\right) }{\Gamma (q+1)} \left(\frac{2 \Delta }{1-\Delta }\right)^{\bar{q}} \\ \times \, _2F_1\left(i+m,\bar{q}+\frac{1}{2};\bar{q}+1;-\frac{2 \Delta  K \bar{\mu} }{(1-\Delta ) K \bar{\mu} +m}\right) \\\end {array}$
    & $2$
    &  $\bar{\mu}$ 
    &  $\frac{\bar{\gamma }}{(K+1) \bar{\mu} }$ \\ \midrule  

	$\bar{\alpha}$-$\bar{\eta}$-$\kappa$-$\bar{\mu}$~\cite{Silva20}
    & $\exp \left[- \frac{\kappa  \bar{\mu}  (p \bar{q}+1)}{\delta }\right] \frac{\bar{\eta}^{\bar{\mu}} \left(\frac{p}{\bar{\eta} }\right)^{\frac{\bar{\mu}  p}{p+1}}}{ p^{\bar{\mu} +i}} \sum _{k=0}^i \frac{(p-\bar{\eta} )^k \left(\frac{\kappa  \bar{\mu} p^2 \bar{q}}{\delta }\right)^{i-k} L_k^{\frac{\bar{\mu}}{p+1}-1}\left(\frac{\bar{\eta}  \kappa  \bar{\mu} }{\delta  (\bar{\eta} -p)}\right)}{\Gamma (-k+i+1)}$
	& $\bar{\alpha}$
    & $\bar{\mu}$ 
    & $\frac{\bar{\eta} \bar{r}^{\bar{\alpha }}}{\bar{\mu}  \bar{\xi}  p}$ 
 \\\bottomrule
    \end{tabular}}
\end{threeparttable}
\end{table*}

\textit{Remark} 2:
Proposition~1 reveals that the newly introduced \mbox{$\alpha$-$\mu$} Mixture, formally defined in \eqref{eq: PDF R general} and \eqref{eq: CDF R general}, is not merely another fading model but a unifying representation for the entire Gaussian class. In this sense, it serves as the analytical backbone of the proposed framework, extending the MG viewpoint in~\cite{Garcia24} from a partial power-domain characterization to a broader exact treatment that also supports envelope-sum analysis.
According to Proposition~\ref{prop: common structure}, any Gaussian-class fading distribution can be constructed from \eqref{eq: R def}. Consequently, the \mbox{$\alpha$-$\mu$} Mixture emerges as a compact and comprehensive model that unifies \emph{all} Gaussian-class fading distributions within a single analytical form. 
This model therefore constitutes the analytical cornerstone of the proposed framework, generalizing the MG representation in~\cite{Garcia24} from a subset of Gaussian-class models to \emph{all} members of this class.
Examples encompassed by this representation include the one-sided Gaussian, Rayleigh, Rice, \mbox{Nakagami-$m$}, Weibull, Nakagami-$q$ (Hoyt), \mbox{$\bar{\alpha}$-$\bar{\mu}$}, \mbox{$\kappa$-$\bar{\mu}$}, \mbox{$\bar{\eta}$-$\bar{\mu}$}, Beckmann, \mbox{$\bar{\alpha}$-$\bar{\eta}$-$\bar{\mu}$}, fluctuating Beckmann, \mbox{$\bar{\alpha}$-$\kappa$-$\bar{\mu}$}, Shadowed-Rice, $\kappa$-$\bar{\mu}$ Shadowed, Extended \mbox{$\bar{\eta}$-$\bar{\mu}$}, FTR, TWDP, $\bar{\alpha}$-$\kappa$-$\bar{\mu}$ Shadowed, Extended \mbox{$\bar{\alpha}$-$\bar{\eta}$-$\bar{\mu}$}, \mbox{$\bar{\alpha}$-$\bar{\eta}$-$\kappa$-$\bar{\mu}$}, MFTR, IFTR, and MTW distributions.

For illustration, Table~\ref{tab: PDF-CDF parameters} presents the mapping between the PDF parameters of six general Gaussian-class parent distributions and the corresponding parameters of the \mbox{$\alpha$-$\mu$} Mixture in \eqref{eq: PDF R general}. These examples are representative rather than exhaustive: many conventional and generalized fading laws arise from these parent distributions as special or limiting cases through standard parameter reductions. Accordingly, the mappings of the other Gaussian-class models listed in the manuscript can be obtained either by specializing the mappings in Table~\ref{tab: PDF-CDF parameters} or, equivalently, by applying the constructive formulation of Proposition~1 to their underlying Gaussian signal representation.

The \mbox{$\alpha$-$\mu$} Mixture should not be interpreted as a new scattering mechanism, but rather as a unified physical-statistical representation of Gaussian-class fading models. Fading models are mathematical abstractions that describe the random effects imposed by the propagation medium on the radio signal. Each model, from classical Rayleigh and Rice fading to more general formulations such as \mbox{$\bar{\alpha}$-$\bar{\eta}$-$\kappa$-$\bar{\mu}$} and composite fading distributions, captures a particular combination of propagation mechanisms, such as diffuse multipath, dominant components, clustering, nonlinearity, shadowing, and fluctuation. Since no finite-parameter model can represent all possible propagation environments, increasing physical generality often comes at the cost of substantial mathematical complexity.

The \mbox{$\alpha$-$\mu$} Mixture provides a tractable alternative. Instead of constructing an increasingly complex fading law that attempts to incorporate all phenomena simultaneously, it represents Gaussian-class fading distributions as weighted combinations of analytically manageable \mbox{$\alpha$-$\mu$} components. In this representation, the components provide the common mathematical basis, while the mixture weights carry the information inherited from the corresponding parent fading model. Hence, the connection between the parameters of the \mbox{$\alpha$-$\mu$} Mixture and the propagation medium can be traced through the original fading distributions. This preserves physical interpretability while enabling a unified and computationally tractable treatment of heterogeneous scenarios in which different links or branches may experience distinct fading mechanisms.

Proposition~\ref{prop: common structure} theoretically justifies why the $\bar{\alpha}$-$\bar{\mu}$ distribution closely approximates various fading models~\cite{Costa08aplha,Costa08,Filho06}. It shows that any Gaussian-class fading PDF can be represented as a linear combination of \mbox{$\bar{\alpha}$-$\bar{\mu}$} distributions, making it a fundamental basis for such models.

\subsection{Sum of $\alpha$-$\mu$ Mixtures}
\label{sec: Sum a-u Mixture Distributions}

Capitalizing on the Theorem and Proposition~\ref{prop: common structure}, the next proposition derives novel, exact expressions for the sum statistics of $\alpha$-$\mu$ Mixture distributions.  

\begin{proposition}
\label{lemma: Marginal LT}
Let $\left\{ R_\ell \right\}_{\ell=1}^{L}$ be a set of i.n.i.d. $\alpha$-$\mu$ Mixture RVs with marginal PDFs
\begin{align}
    \label{eq: PDF marginal def}
    \mathit{f}_{R_{\ell }} (r) = \alpha  \exp \left[-\frac{r^{\alpha }}{\hat{r}_{\ell }^{\alpha }}\right] \sum _{i=0}^{\infty } \frac{\varphi _{i,\ell } \, r^{\alpha  \left(i+\mu_{\ell }\right)-1}}{\hat{r}_{\ell }^{\alpha  \left(i+\mu_{\ell }\right)} \Gamma \left(i+\mu_{\ell }\right) } , \qquad r>0,
\end{align} 
\normalsize
where $\alpha,\mu_\ell>0$,
$\varphi_{j,\ell}\in\mathbb R$, $\hat{r}_\ell >0$, and
$\varphi_{0,\ell}>0$ for each $\ell$.
Let $R$ denote the sum of all RVs $R_{\ell }$, i.e.,
\begin{align}
    \label{eq: sum alpha-like}
    R \triangleq \sum_{\ell=1}^{L} R_\ell.
\end{align}
\normalsize
Then, the  PDF and CDF of $R$ are given, respectively, by
\begin{align}
    \label{eq: sum PDF alpha-like}
    \nonumber f_R (r) = &  \alpha^L \left( \prod _{\ell =1}^L \frac{1}{\hat{r}_{\ell }^{\alpha  \mu_{\ell }}}   \right)\\
    & \times \sum _{i=0}^{\infty } \frac{\delta _i \, r^{-1+\alpha  i+\alpha \sum _{\ell =1}^L   \mu_{\ell}}}{\Gamma \left(i \alpha + \alpha \sum _{\ell =1}^L  \mu_{\ell}\right)}, & r >0 \\ \label{eq: sum CDF alpha-like}
    \nonumber F_R (r) = & \alpha^L \left( \prod _{\ell =1}^L \frac{1}{\hat{r}_{\ell }^{\alpha  \mu_{\ell }}} \right) \\
    & \times \sum _{i=0}^{\infty } \frac{\delta _i \, r^{\alpha  i+ \alpha \sum _{\ell =1}^L  \mu_{\ell}}}{\Gamma \left(1+i \alpha + \alpha \sum _{\ell =1}^L  \mu_{\ell}\right)}, & r \geq 0,
\end{align}
\normalsize
where 
\begin{subequations}
\label{eq: delta coefficients alpha-like}
\begin{align}
    \label{eq: lambda 0}
    &\delta_0=\prod_{\ell=1}^L \lambda_{0,\ell} & \\ 
    \label{eq: lambda i}
    & \delta_i=  \frac{1}{i}\sum _{h=1}^i \delta _{i-h} \sum _{\ell =1}^L \phi_{h-1,\ell }, & i  \geq 1.
\end{align}
\end{subequations}
\normalsize
Additionally, 
\begin{subequations}
\begin{align}
    \phi_{0,\ell } =& \frac{\lambda _{1,\ell }}{\lambda _{0,\ell }} & \\ 
     \phi_{h,\ell } = & \frac{1}{\lambda _{0,\ell }} \left[(h+1) \, \lambda _{h+1,\ell }-\sum _{t=1}^h \lambda _{t,\ell } \, \phi_{h-t,\ell }\right], & h\geq1,
\end{align}
\end{subequations}
\normalsize
in which
\begin{align}
    \label{eq: lambda coefficient}
    \lambda _{i,\ell } = &  \frac{\Gamma \left(\alpha  \left(i+\mu_{\ell }\right)\right)}{\hat{r}_{\ell }^{\alpha  i}} \sum _{j=0}^i \frac{(-1)^{i-j} \varphi _{j,\ell }}{(i-j)! \Gamma \left(j+\mu_{\ell }\right)}, & i\in\mathbb N_0.
\end{align}
\normalsize

\end{proposition}

\begin{proof} 
See Appendix~\ref{app: table theorem verification}.
\end{proof}

\begin{corollary}
\label{corol: sum iid alpha-like}
If $\left\{R_{\ell }\right\}_{\ell=1}^{L}$ is a set of i.i.d. $\alpha$-$\mu$ Mixture distributions (i.e., $\hat{r}_{\ell }= \hat{r}$,  $\mu_{\ell} =\mu$, $\lambda_{i,\ell}= \lambda_{i}$), then the PDF and CDF of $R$, given in \eqref{eq: sum PDF alpha-like} and \eqref{eq: sum CDF alpha-like}, reduce to
\begin{align}
    \label{eq: sum PDF alpha-like iid}
    f_R (r) = &    \frac{\alpha^L}{\hat{r}^{\alpha  \mu L}}  \sum _{i=0}^{\infty } \frac{\delta _i \, r^{-1+\alpha  i+\alpha L   \mu}}{\Gamma \left(i \alpha + \alpha L \mu\right)}, & r >0 \\ \label{eq: sum CDF alpha-like iid}
    F_R (r) = &  \frac{\alpha^L}{\hat{r}^{\alpha  \mu L}} \sum _{i=0}^{\infty } \frac{\delta _i \, r^{\alpha  i+ \alpha L \mu}}{\Gamma \left(1+i \alpha + \alpha L  \mu\right)}, & r \geq 0
\end{align}
\normalsize
where 
\begin{subequations}
\label{eq: delta coefficients alpha-like iid}
\begin{align}
    \label{eq: delta 0 alpha-like iid}
    &\delta_0= \lambda _{0}^L & \\ 
    \label{eq: delta i alpha-like iid}
    & \delta_i=  \frac{1}{i \lambda _0 } \sum _{h=1}^i \delta _{i-h} \, \lambda _h  \left(h L+h-i \right), & i  \geq 1.
\end{align}
\end{subequations}
\normalsize
\end{corollary}

\begin{proof}
    After a direct application of Corollary~\ref{corol: IID case}, \eqref{eq: sum PDF alpha-like iid} and \eqref{eq: sum CDF alpha-like iid} are derived.
\end{proof}

\textit{Remark} 3: Proposition \ref{lemma: Marginal LT} and Corollary~\ref{corol: sum iid alpha-like} provide novel, tractable expressions for the sum statistics---at either the power or envelope level---for any Gaussian-class fading distributions, thereby covering conventional, generalized, and recently introduced fading models. 
In particular, they offer the first reported exact solutions for the PDF and CDF of the envelope sums of i.i.d.\ and i.n.i.d.\ \mbox{$\bar{\alpha}$-$\bar{\eta}$-$\kappa$-$\bar{\mu}$}, \mbox{$\bar{\alpha}$-$\kappa$-$\bar{\mu}$} Shadowed, and Extended \mbox{$\bar{\alpha}$-$\bar{\eta}$-$\bar{\mu}$} fading models, as well as for the envelope sums of i.i.d.\ and i.non-i.d.\ $\kappa$-$\bar{\mu}$ Shadowed, fluctuating Beckmann, FTR, MFTR, IFTR, and MTW fading distributions.

\section{Scenario 2: Sum of $\bar{\alpha}$-$\bar{\mu}$/$\bar{\alpha}$-$\bar{\mu}$}
\label{sec: Sum of Ratios of alpha-mu Distributions}

In this section, we derive new, exact expressions for the sum statistics of \mbox{$\bar{\alpha}$-$\bar{\mu}$/$\bar{\alpha}$-$\bar{\mu}$} distributions.

\begin{proposition}
\label{prop: sum ratio}
Let $\left\{ Z_\ell \right\}_{\ell=1}^{L}$ be a set of i.n.i.d. RVs, where each $Z_\ell$ is formed by the ratio of two i.non-i.d. $\bar{\alpha}$-$\bar{\mu}$ variates $M_\ell$ and $Q_\ell$, namely, $Z_\ell =M_\ell/ Q_\ell$, where $M_\ell$ and $Q_\ell$ have a PDF given, respectively, by~\cite{Yacoub07}
\begin{align}
    \label{eq: PDF marginal M}
    f_{M_\ell} (m) =  & \frac{\alpha_M \mu_{M,\ell}^{\mu_{M,\ell}} m^{\alpha_M \mu_{M,\ell} -1}}{ \Omega_{M,\ell}^{\alpha_M \mu_{M,\ell}} \Gamma \left(\mu_{M,\ell} \right)} \exp \left( - \mu_{M,\ell} \frac{m^{\alpha_M}}{\Omega_{M,\ell}^{\alpha_M}}\right)  \\ \label{eq: PDF marginal S}
    f_{Q_\ell} (q) =&  \frac{\alpha_Q \mu_{Q,\ell}^{\mu_{Q,\ell}} q^{\alpha_Q \mu_{Q,\ell} -1}}{ \Omega_{Q,\ell}^{\alpha_Q \mu_{Q,\ell}} \Gamma \left(\mu_{Q,\ell} \right)} \exp \left( - \mu_{Q,\ell} \frac{q^{\alpha_Q}}{\Omega_{Q,\ell}^{\alpha_Q}}\right),
\end{align}
\normalsize
where $\alpha_a,\mu_{a,\ell},\Omega_{a,\ell}>0$ represent the nonlinearity of the medium,  the number of multipath clusters, and the $\alpha_a$-root mean value, respectively, with $a\in \{M,Q\}$. The PDF of $Z_\ell$, provided $\alpha_M<\alpha_Q$, is given by~\cite{Carlos19}
\begin{align}
    \label{eq: pdf_ratio}
    f_{Z_\ell}(z)&=\frac{\alpha_M }{z \, \Gamma \left(\mu _{M,\ell}\right) \Gamma \left(\mu_{Q,\ell}\right)}\sum _{i=0}^{\infty} \frac{(-1)^i}{i!} \left(\tilde{\varrho}_\ell \, z \right)^{\alpha_M \left(i+\mu_{M,\ell}\right)}\nonumber\\
    &\times \Gamma \left(\frac{\alpha_M \left(i+\mu_{M,\ell}\right)}{\alpha_Q}+\mu _{Q,\ell}\right), \qquad z>0
\end{align}
\normalsize
where $\tilde{\varrho}_\ell \triangleq ( \Omega _{Q,\ell} \, \mu_{M,\ell}^{1/\alpha_M})/( \Omega_{M,\ell} \, \mu_{Q,\ell}^{1/\alpha_Q})$.
Let $Z$ denote the sum of all RVs $Z_{\ell }$, namely,
\begin{align}
    \label{eq: sum z}
    Z \triangleq \sum_{\ell=1}^{L} Z_\ell.
\end{align}
\normalsize
Then, the  PDF and CDF of $Z$ are given, respectively, by
\begin{align}
    \label{eq: sum PDF z}
    \nonumber f_Z (z) & =   \alpha _M^L\left( \prod _{\ell =1}^L \frac{1}{\Gamma \left(\mu _{M,\ell }\right) \Gamma \left(\mu _{Q,\ell }\right)}  \right) \\
    & \times \sum _{i=0}^{\infty } \frac{\delta _i \, z^{-1+i \alpha _M+\alpha _M \sum _{\ell =1}^L \mu _{M,\ell }}}{\Gamma \left(i \alpha _M+\alpha _M \sum _{\ell =1}^L \mu _{M,\ell }\right)}& z >0  \\ \label{eq: sum CDF z}
    \nonumber F_Z (z) = & \alpha _M^L\left( \prod _{\ell =1}^L \frac{1}{\Gamma \left(\mu _{M,\ell }\right) \Gamma \left(\mu _{Q,\ell }\right)}  \right) \\
    & \times \sum _{i=0}^{\infty } \frac{\delta _i \, z^{i \alpha _M+\alpha _M \sum _{\ell =1}^L \mu _{M,\ell }}}{\Gamma \left(1+i \alpha _M+\alpha _M \sum _{\ell =1}^L \mu _{M,\ell }\right)}, & z  \geq 0
\end{align}
\normalsize
where 
\begin{subequations}
\label{eq: delta coefficients Z}
\begin{align}
    \label{eq: delta 0 Z}
    &\delta_0=\prod_{\ell=1}^L u_{0,\ell} & \\ 
    \label{eq: delta i Z}
    & \delta_i=  \frac{1}{i}\sum _{h=1}^i \delta _{i-h} \sum _{\ell =1}^L \phi_{h-1,\ell }, & i  \geq 1.
\end{align}
\end{subequations}
\normalsize
Moreover, 
\begin{subequations}
\begin{align}
    \phi_{0,\ell } =& \frac{u _{1,\ell }}{u_{0,\ell }} & \\ 
     \phi_{h,\ell } = & \frac{1}{u_{0,\ell }} \left[(h+1) \, u_{h+1,\ell }-\sum _{t=1}^h u_{t,\ell } \, \phi_{h-t,\ell }\right], & h \geq 1,
\end{align}
\end{subequations}
\normalsize
in which
\begin{align}
    \label{eq: u coefficient}
    \nonumber u_{i,\ell } = & \frac{(-1)^i  }{i! }\Gamma \left(\alpha _M \left(i+\mu _{M,\ell }\right)\right) \tilde{\varrho }^{\alpha _M \left(\mu _{M,\ell }+i\right)}_\ell &\\
    & \times \Gamma \left(\frac{\alpha _M \left(i+\mu _{M,\ell }\right)}{\alpha _Q}+\mu _{Q,\ell }\right), & i\in\mathbb N_0.
\end{align}
\normalsize

\end{proposition}

\begin{proof}
See Appendix~\ref{app: sum ratio}.  
\end{proof}

Eqs.~\eqref{eq: sum PDF z} and \eqref{eq: sum CDF z}
constitute novel contributions, encompassing the sum of various non-Gaussian-class fading models under the condition
$\alpha_M<\alpha_Q$. The equal-exponent case $\alpha_M=\alpha_Q$ does not lie directly within the admissible parameter domain of Proposition~\ref{prop: sum ratio}. Rather, it arises as the one-sided boundary limit $\alpha_Q\to\alpha_M^{+}$. Specifically, by setting $\alpha_Q=\alpha_M+\epsilon'$, with $\epsilon'>0$, Proposition~\ref{prop: sum ratio} remains exact for every fixed $\epsilon'$, while the equal-$\alpha$ case is approached as $\epsilon'\to0^{+}$, provided that the marginal coefficients and the resulting PDF and CDF series admit this limit and that the limiting operation can be interchanged with the corresponding infinite sums.
Under this limiting interpretation, the \mbox{$\bar{\alpha}$-$\mathcal{F}$} family is recovered as $\alpha_Q\to\alpha_M^{+}$, whereas the Fisher--Snedecor $\mathcal{F}$ distribution follows from the additional specialization $\alpha_M=\alpha_Q=2$. Equivalently, the Fisher--Snedecor $\mathcal{F}$ case can be approached by setting $\alpha_M=2$ and $\alpha_Q=2+\epsilon'$, or $\alpha_Q=2$ and $\alpha_M=2-\epsilon'$, and then taking $\epsilon'\to0^{+}$. Therefore, the \mbox{$\bar{\alpha}$-$\mathcal{F}$} and Fisher--Snedecor $\mathcal{F}$ models should be interpreted as limiting boundary cases of Proposition~\ref{prop: sum ratio}, rather than as direct specializations for finite $\epsilon'$.
Finally, if $\{Z_\ell\}_{\ell=1}^{L}$ is a set of i.i.d. ratios of \mbox{$\bar{\alpha}$-$\bar{\mu}$} variates, i.e., $\mu_{M,\ell}=\mu_M$, $\mu_{Q,\ell}=\mu_Q$, $\Omega_{M,\ell}=\Omega_M$, and $\Omega_{Q,\ell}=\Omega_Q$, then the PDF and CDF in \eqref{eq: sum PDF z} and \eqref{eq: sum CDF z} reduce, respectively, to the expressions reported in~\cite[Eqs.~(5) and~(7)]{Lenin25}.

\section{Scenario 3: Mixed Sums}
\label{sec: Mixed Sums}

In this section, by leveraging the Theorem, we obtain the mixed-sum statistics of Gaussian-class and non-Gaussian-class fading distributions.

Let $\left\{X_{\ell } \right\}_{\ell=1}^{L}$ be a set of independent RVs, where each RV $X_{\ell }$ may follow either an \mbox{$\alpha$-$\mu$} Mixture distribution or an \mbox{$\bar{\alpha}$-$\bar{\mu}$/$\bar{\alpha}$-$\bar{\mu}$} distribution. 
As established in Appendices~\ref{app: table theorem verification} and~\ref{app: sum ratio}, the marginal PDF representations satisfy all canonical-form assumptions of the Theorem, including the positivity of the leading coefficients and the generalized coefficient-growth condition. The mixed-sum framework is therefore applicable whenever the exponent increments of the marginal series are commensurate and can be embedded in a common lattice.
Thus, the sum $X = \sum_{\ell = 1}^{L} X_{\ell}$ can be formed by an arbitrary combination of \mbox{$\alpha$-$\mu$} Mixture and \mbox{$\bar{\alpha}$-$\bar{\mu}$/$\bar{\alpha}$-$\bar{\mu}$} fading models by simply adjusting the PDF parameters $\Psi_{\ell}$, $\beta_{\ell}$, $\theta$, and $\eta_{i,\ell}$ to match the desired marginal distributions.
Consequently, expressions \eqref{eq: PDF X Final} and \eqref{eq: CDF X Final} offer the mixed-sum statistics for a plethora of fading models.

\textit{Remark} 4: 
To the best of our knowledge, the Theorem provides the first exact unified treatment of sums formed by arbitrary combinations of Gaussian-class and non-Gaussian-class fading distributions, provided that their exponent increments share a common lattice. This significantly expands the analytical scope of the literature, moving beyond isolated i.i.d. and i.n.i.d. (homogeneous) results to enable the exact analysis of mixed (heterogeneous) fading scenarios.

\section{Outage Analysis of EGC and MRC Receivers over Mixed Fading Channels}
\label{sec: Sample Applications}

In this section, we present a unified performance analysis of diversity-combining receivers. Specifically, we derive exact expressions for the OP of EGC and MRC receivers under mixed fading conditions.

Consider EGC and MRC diversity receivers, each equipped with $L$ branches.
The instantaneous output SNR for both combining schemes can be expressed, in compact form, as
\begin{align}
\label{eq: instantaneous SNR def}
\Phi = \bar{\rho} \, g_{\zeta} \, \tilde{X}^{2/\vartheta_{\zeta}},
\end{align}
where $\tilde{X} \triangleq \sum_{\ell=1}^{L} X_\ell^{\vartheta_{\zeta}}$, and $X_\ell$ denotes the Gaussian-class or non-Gaussian-class fading envelope at the $\ell$th diversity branch.
Here, $\bar{\rho}=E_s/N_0$ denotes the average transmit SNR per symbol, where $E_s$ is the transmit symbol energy and $N_0$ is the one-sided noise power spectral density.
The index $\zeta \in \{\text{EGC}, \text{MRC}\}$ specifies the combining scheme, where $\vartheta_{\text{EGC}} = 1$, $\vartheta_{\text{MRC}} = 2$, $g_{\text{EGC}} = 1/L$, and $g_{\text{MRC}} = 1$.

The OP is defined as the probability that the instantaneous SNR at the combiner output falls below a prescribed SNR threshold $\gamma_{\text{th}}$, i.e., $P_{\mathrm{out}} \triangleq  \text{Pr} \left[ \Phi < \gamma_{\text{th}} \right]= \text{Pr} \left[ \tilde{X} < \left( \frac{\gamma_{\text{th}}}{\bar{\rho} \, g_{\zeta}} \right)^{\vartheta_{\zeta}/2} \right]$.

For the EGC scheme, since $\vartheta_{\text{EGC}} = 1$, the term $\tilde{X}$ in~\eqref{eq: instantaneous SNR def} reduces to $X$ as defined in~\eqref{eq: Sum def X}.
Hence, the OP can be expressed as
\begin{align}
\label{eq: OP EGC}
P_{\mathrm{out}}^{\mathrm{EGC}} = F_X \left( \sqrt{\frac{\gamma_{\text{th}} L}{\bar{\rho}}} \right),
\end{align}
\normalsize
where $F_X(\cdot)$ denotes the CDF of $X$ given in~\eqref{eq: CDF X Final}.

For the MRC scheme, define 
\begin{align}
    \label{eq: Y ell}
    Y_\ell &\triangleq X_\ell^2\\ \label{eq: X bar def}
    \widetilde{X} & \triangleq\sum_{\ell=1}^{L}Y_\ell
=\sum_{\ell=1}^{L}X_\ell^2.
\end{align}
Since $X_\ell\geq0$, the standard
change of variables gives
\begin{align}
f_{Y_\ell}(y)
&=
\frac{1}{2\sqrt{y}}f_{X_\ell}(\sqrt{y}),
\qquad y>0.
\end{align}
Using the marginal representation in
\eqref{eq: PDF marginal theorem} and relying on its absolute
convergence, we obtain
\begin{align}
f_{Y_\ell}(y)
&=
\frac{\Psi_\ell}{2}
\sum_{i=0}^{\infty}
\frac{
\eta_{i,\ell}
y^{(\beta_\ell+i\theta)/2-1}
}{
\Gamma(\beta_\ell+i\theta)
},
\qquad y>0.
\label{eq: MRC transformed marginal}
\end{align}
Therefore, \eqref{eq: MRC transformed marginal} can be rewritten in the canonical form
required by the Theorem as
\begin{align}
f_{Y_\ell}(y)
&=
\widetilde{\Psi}_\ell
\sum_{i=0}^{\infty}
\frac{
\widetilde{\eta}_{i,\ell}
y^{\widetilde{\beta}_\ell+i\widetilde{\theta}-1}
}{
\Gamma(\widetilde{\beta}_\ell+i\widetilde{\theta})
},
\qquad y>0,
\label{eq: MRC canonical marginal}
\end{align}
where $\widetilde{\Psi}_\ell=\Psi_\ell$,
$\widetilde{\theta}=\theta/2$,
$\widetilde{\beta}_\ell=\beta_\ell/2$, and
\begin{align}
\widetilde{\eta}_{i,\ell}
&=
\frac{\eta_{i,\ell}}{2}
\frac{
\Gamma\!\left(\frac{\beta_\ell+i\theta}{2}\right)
}{
\Gamma(\beta_\ell+i\theta)
}.
\label{eq: MRC transformed eta}
\end{align}
Moreover, since $\eta_{0,\ell}>0$, we also have
$\widetilde{\eta}_{0,\ell}>0$. In addition, the generalized
coefficient-growth condition is inherited by the transformed
coefficients. Indeed, using \eqref{eq: generalized coefficient growth}, we obtain
\begin{align}
|\widetilde{\eta}_{i,\ell}|
&\leq
\frac{C_\ell}{2}
A_\ell^i
\frac{
\Gamma\!\left(\frac{\beta_\ell+i\theta}{2}\right)
\Gamma(q_\ell i+c_\ell)
}{
i!
}
\nonumber\\
&=
\widetilde{C}_\ell A_\ell^i
\frac{
\Gamma(i\widetilde{\theta}+\widetilde{\beta}_\ell)
\Gamma(q_\ell i+c_\ell)
}{
i!
},
\end{align}
where $\widetilde{C}_\ell=C_\ell/2$. Hence, the transformed
marginals satisfy the assumptions of the Theorem.
The coefficients of the sum
in \eqref{eq: X bar def} must therefore be computed
from the transformed marginal coefficients. Specifically,
\begin{subequations}
\label{eq: MRC delta coefficients}
\begin{align}
\widetilde{\delta}_0
&=
\prod_{\ell=1}^{L}\widetilde{\eta}_{0,\ell},
\\
\widetilde{\delta}_i
&=
\frac{1}{i}
\sum_{h=1}^{i}
\widetilde{\delta}_{i-h}
\sum_{\ell=1}^{L}
\widetilde{\phi}_{h-1,\ell},
\qquad i\geq1,
\end{align}
\end{subequations}
where
\begin{subequations}
\label{eq: MRC phi coefficients}
\begin{align}
\widetilde{\phi}_{0,\ell}
&=
\frac{\widetilde{\eta}_{1,\ell}}
{\widetilde{\eta}_{0,\ell}},
\\
\widetilde{\phi}_{h,\ell}
&=
\frac{1}{\widetilde{\eta}_{0,\ell}}
\left[
(h+1)\widetilde{\eta}_{h+1,\ell}
-
\sum_{t=1}^{h}
\widetilde{\eta}_{t,\ell}
\widetilde{\phi}_{h-t,\ell}
\right],
\qquad h\geq1.
\end{align}
\end{subequations}
Applying the Theorem to
\eqref{eq: X bar def}, we get
\begin{align}
F_{\widetilde{X}}(x)
&=
\left(
\prod_{\ell=1}^{L}\widetilde{\Psi}_{\ell}
\right)
\sum_{i=0}^{\infty}
\frac{
\widetilde{\delta}_i
x^{i\widetilde{\theta}+\widetilde{B}}
}{
\Gamma\!\left(1+i\widetilde{\theta}+\widetilde{B}\right)
},
\qquad x\geq0,
\label{eq: CDF X tilde}
\end{align}
where $\widetilde{B}\triangleq
\sum_{\ell=1}^{L}\widetilde{\beta}_\ell
=\frac{1}{2}\sum_{\ell=1}^{L}\beta_\ell$. Therefore, the MRC
outage probability is
\begin{align}
P_{\mathrm{out}}^{\mathrm{MRC}}
&=
F_{\widetilde{X}}
\left(
\frac{\gamma_{\mathrm{th}}}{\bar{\rho}}
\right).
\label{eq: OP MRC}
\end{align}

Eqs.~\eqref{eq: OP EGC} and \eqref{eq: OP MRC} also
constitute original contributions, providing exact, unified, and
compact outage-probability expressions for diversity-combining
receivers over mixed fading channels. Beyond simplifying OP
evaluation, the underlying PDF and CDF enable the effects of the
combining scheme, number of branches, fading severity, channel
heterogeneity, and average SNR to be quantified without repeated
model-specific derivations or extensive Monte Carlo simulations.
Accordingly, the proposed framework supports reliability
assessment, comparison of diversity architectures, threshold selection,
and link-budget design in diversity-combining, cooperative,
RIS-assisted, and other multi-link wireless systems.

\subsection{High-SNR Analysis and Design Insights}

We now derive the high-SNR asymptotic OP, which corresponds to
the small-argument regime $x\to0$ or, equivalently,
$\bar{\rho}\to\infty$. We begin with the CDF in
\eqref{eq: CDF X Final}. Since $\theta>0$, the exponent of $x$
increases monotonically with $i$. Hence, the term with $i=0$
determines the leading-order behavior as $x\to0$. Therefore,
\begin{align}
F_X(x)
&\sim
\frac{
\left(\prod_{\ell=1}^{L}\Psi_\ell\right)\delta_0
}{
\Gamma(1+B)
}
x^{B},
\qquad x\to0.
\label{eq: high SNR CDF X asymptotic}
\end{align}
Defining
$\mathcal{G}\triangleq
\left(\prod_{\ell=1}^{L}\Psi_\ell\right)\delta_0/\Gamma(1+B)$,
the asymptotic OP of EGC follows from \eqref{eq: OP EGC} as
\begin{align}
P_{\mathrm{out}}^{\mathrm{EGC}}
&\sim
\mathcal{G}
(\gamma_{\mathrm{th}}L)^{B/2}
\bar{\rho}^{-B/2}.
\label{eq: OP EGC asymptotic}
\end{align}
Thus, the EGC diversity order is
\begin{align}
d_{\mathrm{EGC}}
&\triangleq
-\lim_{\bar{\rho}\to\infty}
\frac{\log P_{\mathrm{out}}^{\mathrm{EGC}}}
{\log\bar{\rho}}
=
\frac{B}{2}.
\label{eq: diversity EGC}
\end{align}
Equivalently, we can rewrite \eqref{eq: OP EGC asymptotic} as
\begin{align}
    \label{eq: OP EGC asymptotic final }
    P_{\mathrm{out}}^{\mathrm{EGC}}
\sim (G_{\mathrm c}^{\mathrm{EGC}}\bar{\rho})^{-d_{\mathrm{EGC}}},
\end{align}
where
\begin{align}
G_{\mathrm c}^{\mathrm{EGC}}
&=
\left(
\mathcal{G}
(\gamma_{\mathrm{th}}L)^{B/2}
\right)^{-2/B}
\label{eq: coding gain EGC}
\end{align}
is the corresponding coding gain.

For MRC, we use the CDF in \eqref{eq: CDF X tilde}. Since
$\widetilde{\theta}>0$, the exponent of $x$ also increases
monotonically with $i$, and the term with $i=0$ determines the
leading-order behavior as $x\to0$. Hence,
\begin{align}
F_{\widetilde X}(x)
&\sim
\frac{
\left(\prod_{\ell=1}^{L}\widetilde{\Psi}_\ell\right)
\widetilde{\delta}_0
}{
\Gamma(1+\widetilde B)
}
x^{\widetilde B},
\qquad x\to0,
\label{eq: high SNR CDF X tilde asymptotic}
\end{align}
where $\widetilde B\triangleq
\sum_{\ell=1}^{L}\widetilde{\beta}_\ell$. Defining
$\widetilde{\mathcal G}\triangleq
\left(\prod_{\ell=1}^{L}\widetilde{\Psi}_\ell\right)
\widetilde{\delta}_0/\Gamma(1+\widetilde B)$, the asymptotic OP of
MRC follows from \eqref{eq: OP MRC} as
\begin{align}
P_{\mathrm{out}}^{\mathrm{MRC}}
&\sim
\widetilde{\mathcal G}
\gamma_{\mathrm{th}}^{\widetilde B}
\bar{\rho}^{-\widetilde B}.
\label{eq: OP MRC asymptotic}
\end{align}

The MRC diversity order is
\begin{align}
d_{\mathrm{MRC}}
&\triangleq
-\lim_{\bar{\rho}\to\infty}
\frac{\log P_{\mathrm{out}}^{\mathrm{MRC}}}
{\log\bar{\rho}}
=
\widetilde B.
\label{eq: diversity MRC}
\end{align}
Thus, \eqref{eq: OP MRC asymptotic} can be equivalently written as
\begin{align}
    \label{eq: OP MRC asymptotic finl}
    P_{\mathrm{out}}^{\mathrm{MRC}}
\sim (G_{\mathrm c}^{\mathrm{MRC}}\bar{\rho})^{-d_{\mathrm{MRC}}},
\end{align}
where
\begin{align}
G_{\mathrm c}^{\mathrm{MRC}}
&=
\left(
\widetilde{\mathcal G}
\gamma_{\mathrm{th}}^{\widetilde B}
\right)^{-1/\widetilde B}
\label{eq: coding gain MRC}
\end{align}
is the corresponding coding gain.

For the squared-envelope transformation used in the MRC case, we have
$\widetilde{\beta}_\ell=\beta_\ell/2$. Therefore,
$\widetilde B=B/2$, and
\begin{align}
d_{\mathrm{MRC}}
&=
d_{\mathrm{EGC}}
=
\frac{1}{2}
\sum_{\ell=1}^{L}\beta_\ell .
\label{eq: diversity EGC MRC equality}
\end{align}
Thus, EGC and MRC attain the same asymptotic diversity order,
although their leading asymptotic coefficients, and hence their
coding gains, generally differ.
These expressions show that the diversity contributions of the
branches are additive. Specifically, the high-SNR outage slope is
determined by the sum of the leading small-argument exponents of
the heterogeneous marginals. Consequently, mixed fading
configurations with the same aggregate exponent achieve the same
diversity order, while differences in the leading coefficients
produce distinct coding gains and horizontal shifts in the outage
curves. Thus, any parameter variation that changes one or more
$\beta_\ell$ can change the diversity order. By contrast,
parameter variations that modify only the leading coefficients
$\Psi_\ell$, $\eta_{0,\ell}$, $\delta_0$, or their transformed
counterparts, while leaving the exponents $\beta_\ell$ unchanged,
affect the coding gain but not the asymptotic outage slope.

The proposed framework assumes independent diversity branches. This
assumption is relevant not only because independent or approximately
independent branches arise in several practical scenarios, but also because the independent case typically serves as a performance upper bound when correlation is present. Nevertheless, real channels may
exhibit spatial, frequency, or temporal correlation. Extending the
framework to correlated mixed fading sums would require an appropriate joint dependence model and is left for future research.

\section{Numerical Results}
\label{sec: Numerical Results}

\begin{table*}[t!]
\centering
\caption{Distribution Parameters}
\label{tab: PDF Parameters}
\begin{tabular}{cccccccccccccc} 
\toprule
\textbf{Fading Model} & \textbf{$\ell$} & $1$  & $2$  & $3$ & $4$ & $5$ & $6$ & $7$ & $8$ & $9$ & $10$ & $11$ & $12$  \\ \hline \hline  
\multirow{4}{*}{FTR} & $K_\ell$  &  $0.47$  & $0.32$  & $5.25$ & $10.25$  & $0.14$  & $0.76$  & $0.25$  & $2$  & $1.5$  & $1.31$  & $0.32$  & $2.06$  \\
&  $\Delta_\ell$  &  $0.7$  & $0.38$  & $0.76$ & $0.29$  & $0.88$  & $0.73$  & $0.88$  & $0.44$  & $0.66$  & $0.76$  & $0.76$  & $0.96$  \\
&  $m_\ell$  &  $1$  & $2$  & $0.5$ & $0.75$  & $2$  & $1$  & $0.25$  & $0.75$  & $2$  & $2$  & $0.5$  & $0.25$  \\
&  $\bar{\gamma}_\ell$  &  $3.31$  & $5.31$  & $6.25$ & $2.81$  & $18.25$  & $7.06$  & $11.25$  & $6.75$  & $2.5$  & $9.25$  & $21.25$  & $12.25$  \\ \hline
\multirow{6}{*}{$\bar{\alpha}$-$\bar{\eta}$-$\kappa$-$\bar{\mu}$} & $\bar{\alpha}$  &  $2$  & $2$  & $2$ & $2$  & $2$  & $2$  & $2$  & $2$  & $2$  & $2$  & $2$  & $2$  \\
&  $\bar{\eta}_\ell$  &  $0.07$  & $2$  & $3.55$ & $0.18$  & $0.72$  & $0.5$  & $0.06$  & $27$  & $0.22$  & $1.33$  & $0.25$  & $9$   \\
&  $\kappa_\ell$  &  $1.20$ & $2$  & $1.78$  & $0.25$ & $0.09$   & $0.05$  & $0.47$  & $0.14$  & $0.12$  & $1$  & $0.85$  & $0.42$   \\
&  $\bar{\mu}_\ell$  &  $2.5$  & $1.5$  & $1.5$ & $3.5$  & $1.5$  & $1.5$  & $4$  & $2$  & $1.5$  & $2$  & $2$  & $2.5$  \\
&  $p_\ell$  &  $0.66$  & $2$  & $0.5$ & $0.75$  & $2$  & $0.5$  & $1$  & $3$  & $2$  & $0.33$  & $1$  & $4$  \\
&  $q_\ell$  &  $144$  & $0.25$  & $20.25$ & $1$  & $0.04$  & $9$  & $16$  & $0.11$  & $20.25$  & $1$  & $0.25$  & $1.77$  \\
&  $\bar{r}_\ell$  &  $3.03$  & $2.40$  & $2.89$ & $3.55$  & $4.67$  & $2.76$  & $6.31$  & $4$  & $2.73$  & $2.87$  & $3.21$  & $5.03$ \\ \hline
\multirow{6}{*}{$\bar{\alpha}$-$\mathcal{F}$} & $\bar{\alpha}_{\mathcal{F}}$  &  $2$  & $2$  & $2$ & $2$  & $2$  & $2$  & $2$  & $2$  & $2$  & $2$  & $2$  & $2$  \\
&  $\bar{\mu}_{\mathcal{F},\ell}$  &  $2$  & $4$  & $5$ & $2$  & $3$  & $4$  & $2$  & $3$  & $2$  & $2$  & $3$  & $4$  \\
&  $m_{\mathcal{F},\ell}$  &  $2$  & $3$  & $2$ & $2$  & $3$  & $4$  & $2$  & $4$  & $5$  & $2$  & $3$  & $4$  \\
&  $\bar{\gamma}_{\mathcal{F},\ell}$  &  $2.5$  & $1$  & $2$ & $0.5$  & $0.25$  & $1.5$  & $2$  & $1$  & $2$  & $0.25$  & $1.5$  & $1$  \\
\bottomrule
\end{tabular}
\end{table*}
\begin{figure}[t!]
  \centering
  \begin{tabular}[c]{cc}
    \begin{minipage}[c]{0.22\textwidth}\centering
    \includegraphics[trim={0cm 0cm 0cm 0cm},clip,scale=0.30]{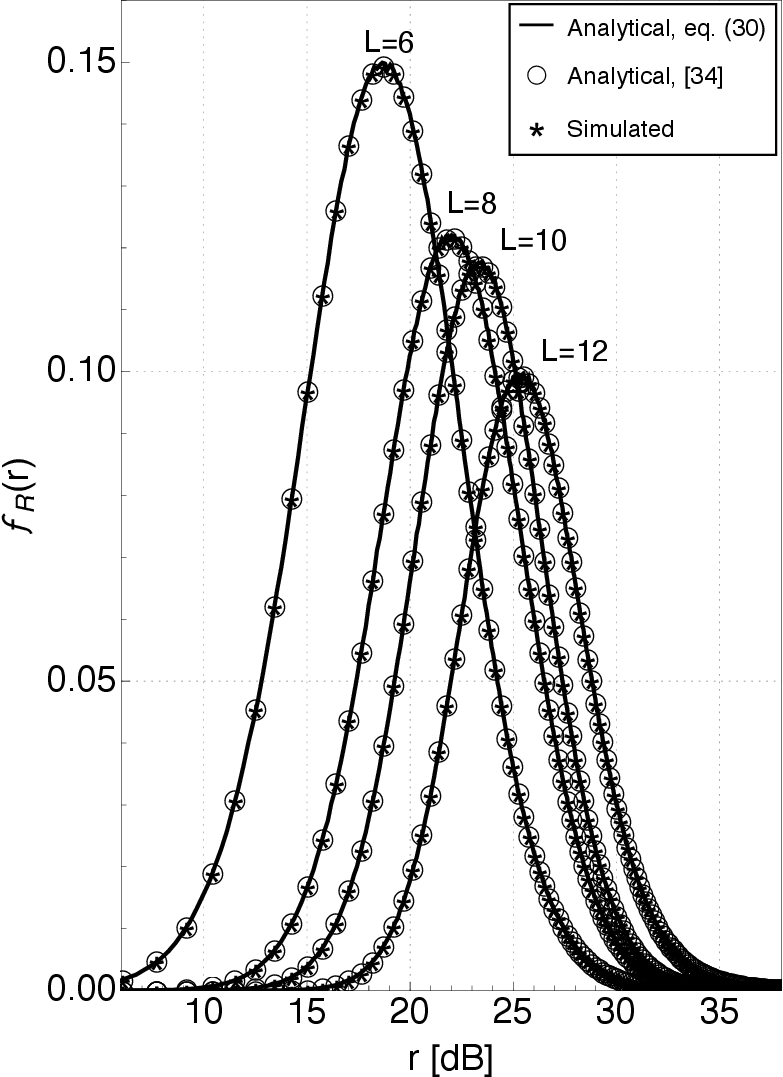}
      \\[-1mm]{\footnotesize (a) Squared FTR}
    \end{minipage}
    \hspace{-0.05cm}
    \begin{minipage}[c]{0.22\textwidth}\centering
    \includegraphics[trim={0cm 0cm 0cm 0cm},clip,scale=0.30]{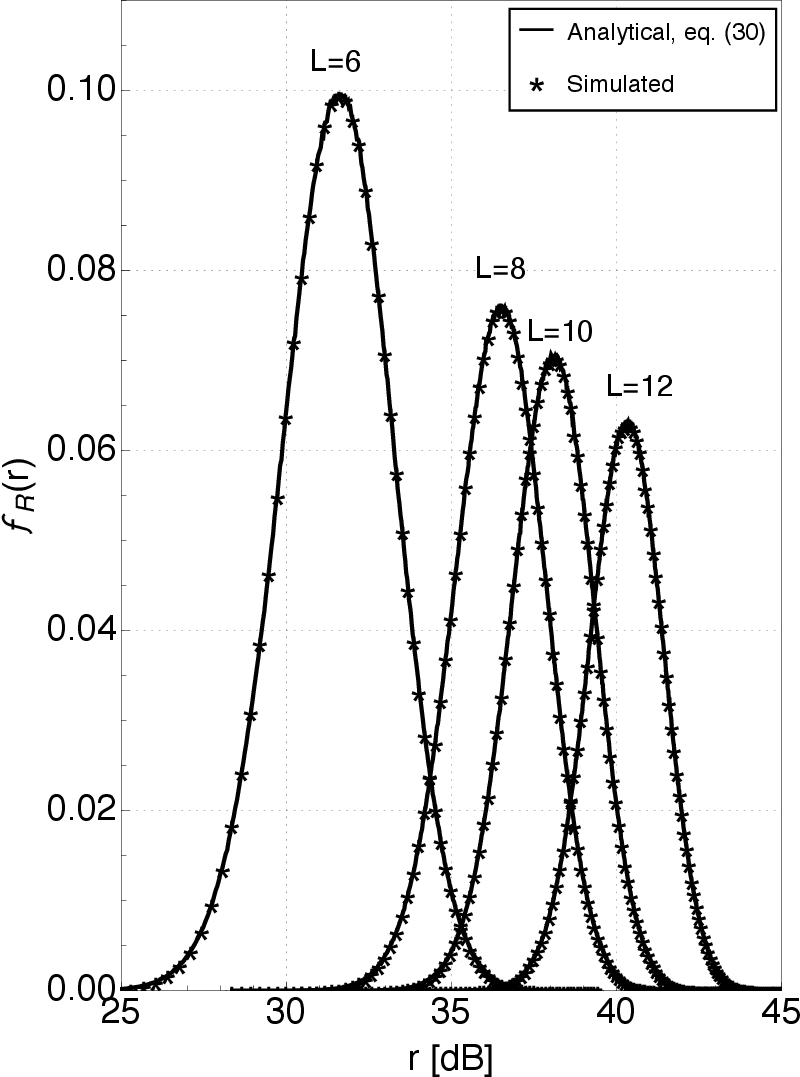}
      \\[-1mm]{\footnotesize (b) $\bar{\alpha}$-$\bar{\eta}$-$\kappa$-$\bar{\mu}$}
    \end{minipage}
    \vspace{0.3cm}
    \\
    \hspace{-0.4cm}
    \begin{minipage}[c]{0.22\textwidth}\centering
     \includegraphics[trim={0cm 0cm 0cm 0cm},clip,scale=0.30]{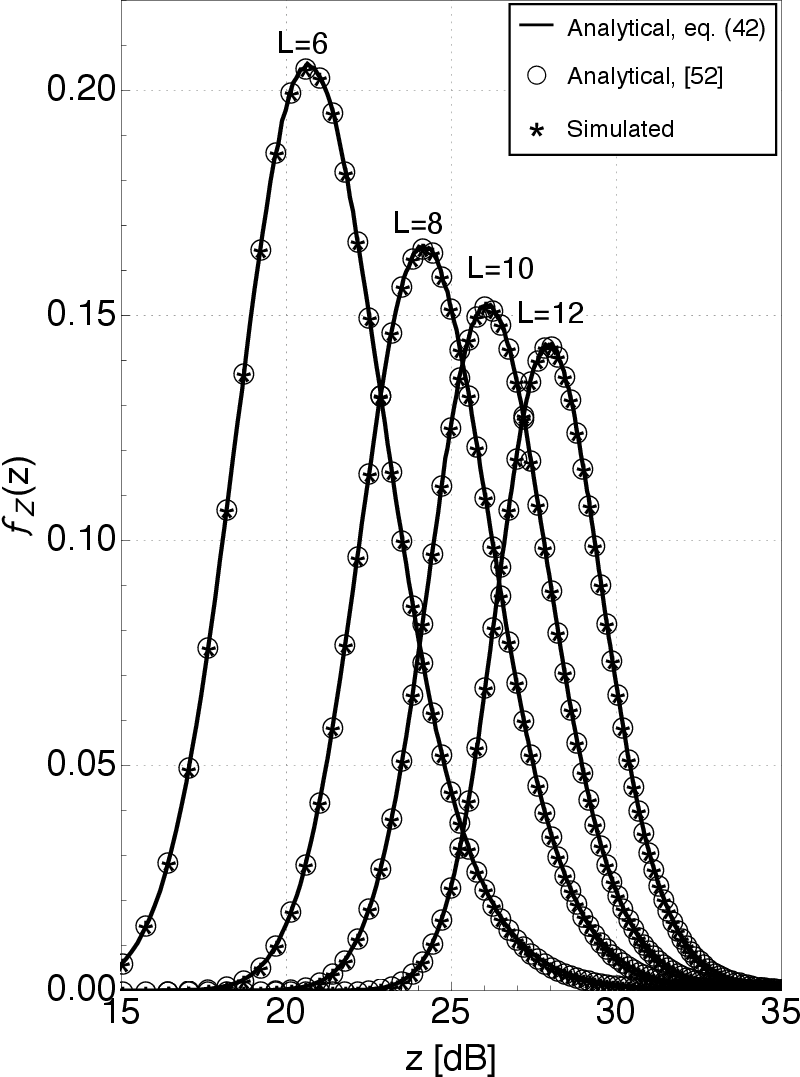}
      \\[-1mm]{\footnotesize (c) $\bar{\alpha}$-$\mathcal{F}$}
    \end{minipage}
    \hspace{-0.05cm}
    \begin{minipage}[c]{0.22\textwidth}\centering
      \includegraphics[trim={0cm 0cm 0cm 0cm},clip,scale=0.30]{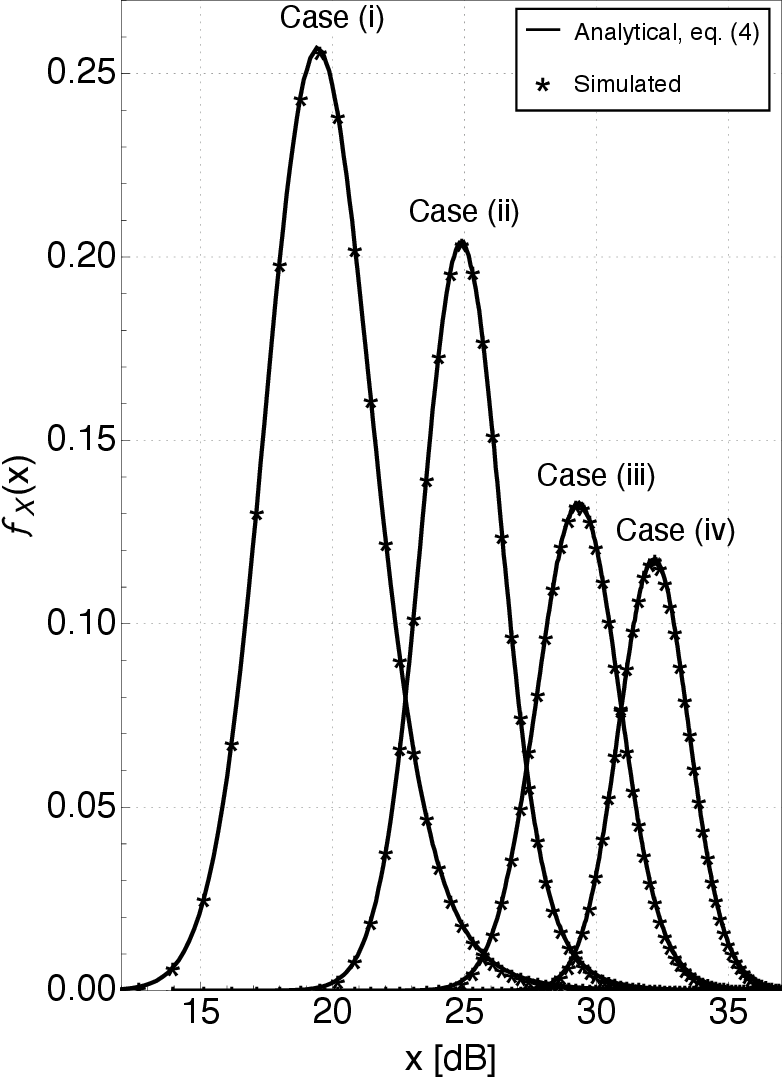}
      \\[-1mm]{\footnotesize (d) Mixed sum}
    \end{minipage}
  \end{tabular}
  \caption{Sum PDF of $L$ fading distributions.}
  \label{fig: Sum PDF}
\end{figure}
\begin{figure}[t!]
  \centering
  \begin{tabular}[c]{cc}
    \begin{minipage}[c]{0.22\textwidth}\centering
    \includegraphics[trim={0cm 0cm 0cm 0cm},clip,scale=0.30]{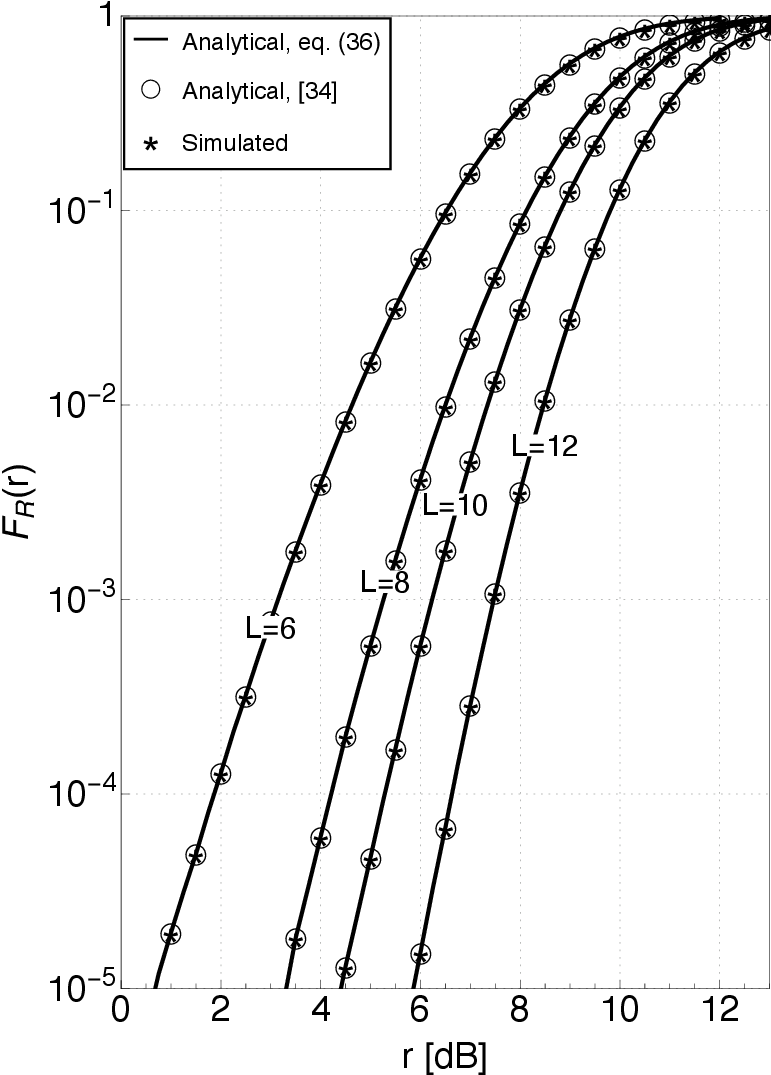}
      \\[-1mm]{\footnotesize (a) Squared FTR}
    \end{minipage}
    \hspace{-0.05cm}
    \begin{minipage}[c]{0.22\textwidth}\centering
    \includegraphics[trim={0cm 0cm 0cm 0cm},clip,scale=0.30]{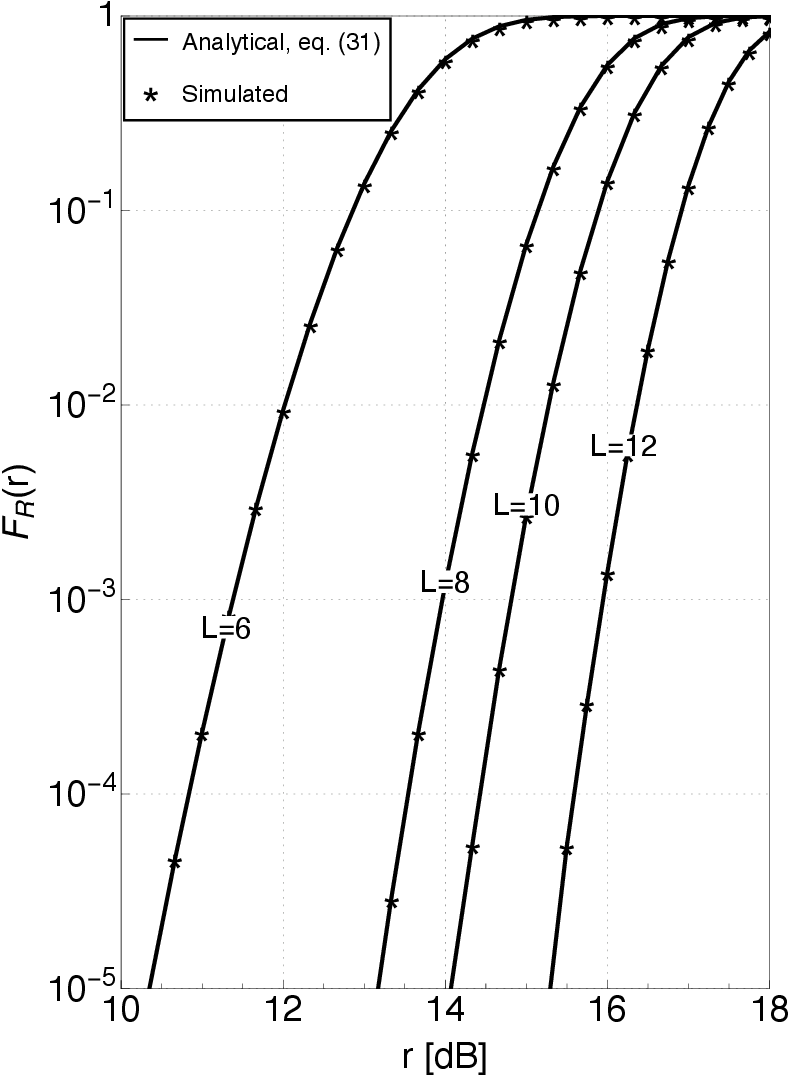}
      \\[-1mm]{\footnotesize (b) $\bar{\alpha}$-$\bar{\eta}$-$\kappa$-$\bar{\mu}$}
    \end{minipage}
    \vspace{0.3cm}
    \\
    \hspace{-0.4cm}
    \begin{minipage}[c]{0.22\textwidth}\centering
     \includegraphics[trim={0cm 0cm 0cm 0cm},clip,scale=0.30]{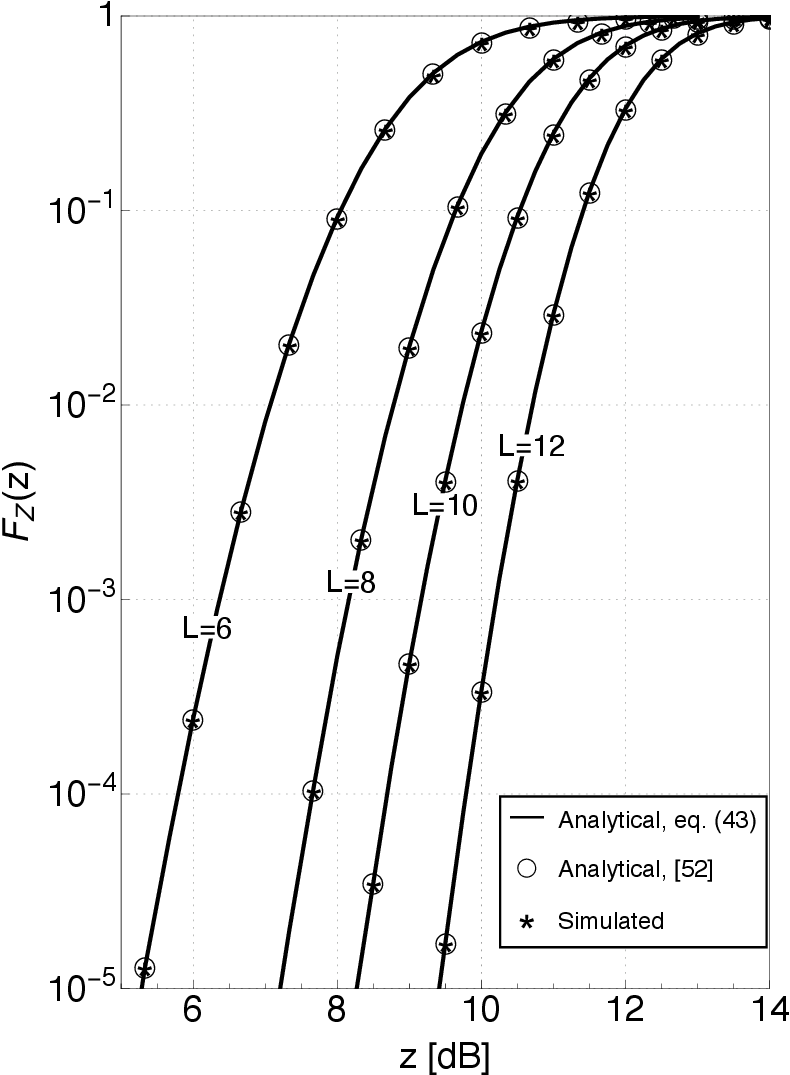}
      \\[-1mm]{\footnotesize (c) $\bar{\alpha}$-$\mathcal{F}$}
    \end{minipage}
    \hspace{-0.05cm}
    \begin{minipage}[c]{0.22\textwidth}\centering
      \includegraphics[trim={0cm 0cm 0cm 0cm},clip,scale=0.30]{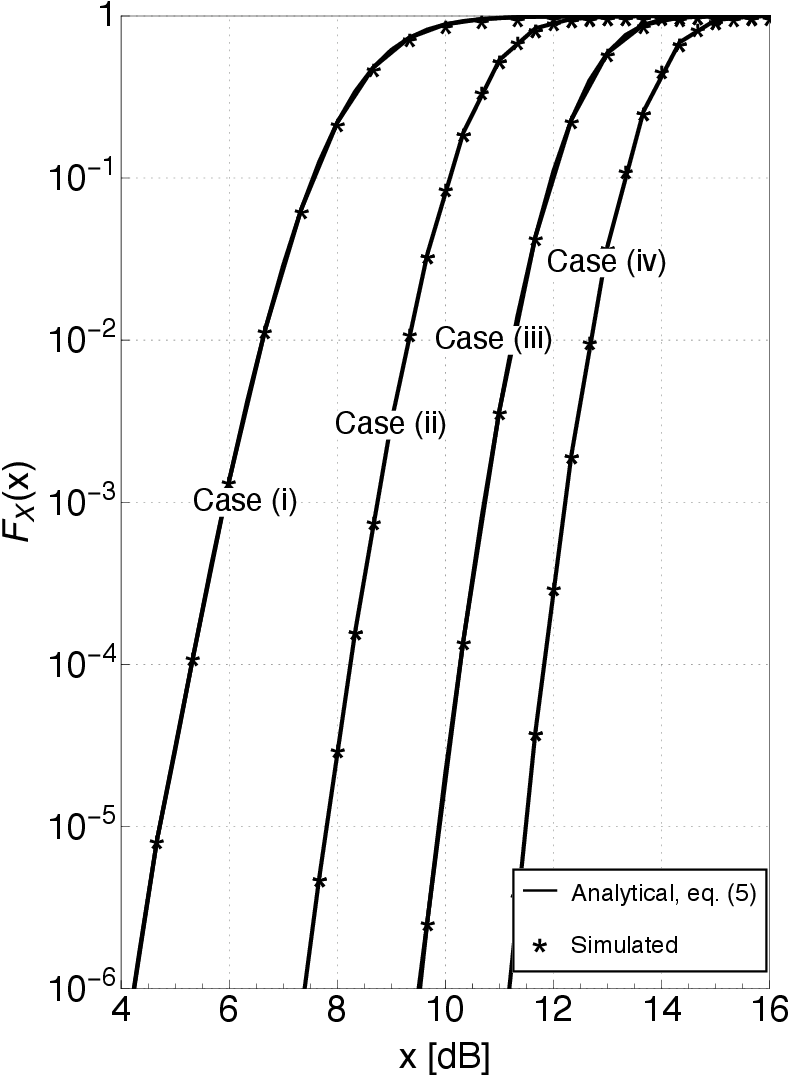}
      \\[-1mm]{\footnotesize (d) Mixed sum}
    \end{minipage}
  \end{tabular}
  \caption{Sum CDF of $L$ fading distributions.}
  \label{fig: Sum CDF}
\end{figure}
\begin{figure}[t!]
  \centering
  \begin{tabular}[c]{cc}
    \begin{minipage}[c]{0.22\textwidth}\centering
    \includegraphics[trim={0cm 0cm 0cm 0cm},clip,scale=0.30]{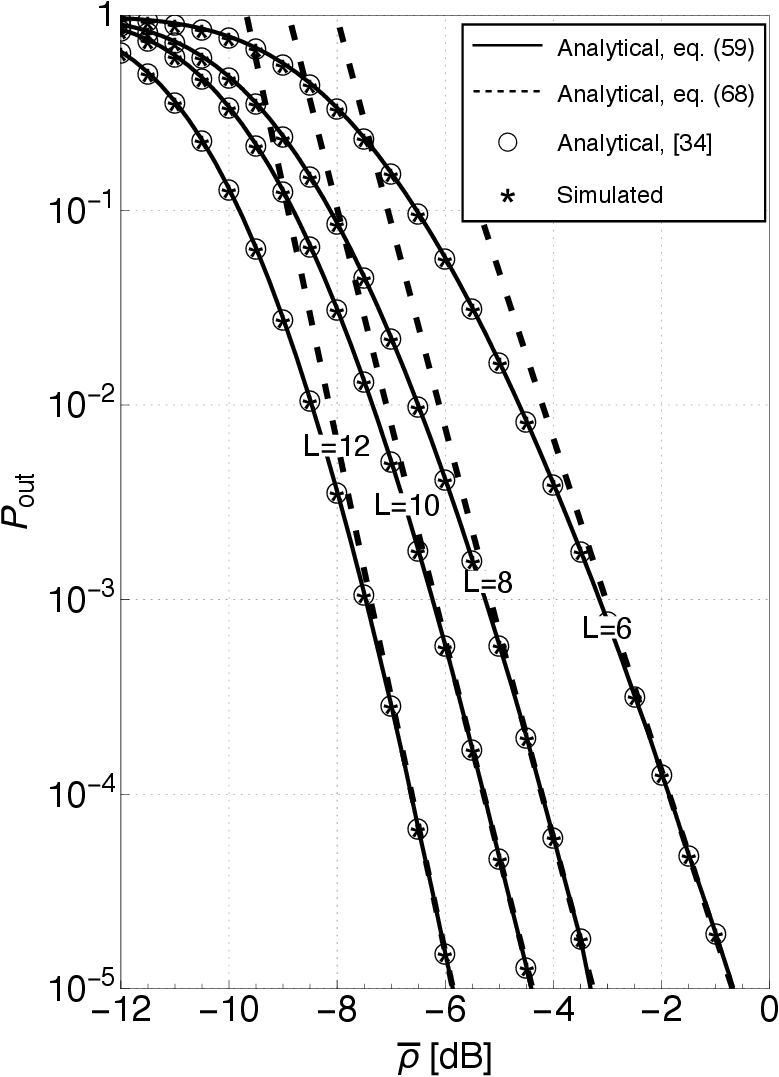}
      \\[-1mm]{\footnotesize (a) Squared FTR}
    \end{minipage}
    \hspace{-0.05cm}
    \begin{minipage}[c]{0.22\textwidth}\centering
        \includegraphics[trim={0cm 0cm 0cm 0cm},clip,scale=0.30]{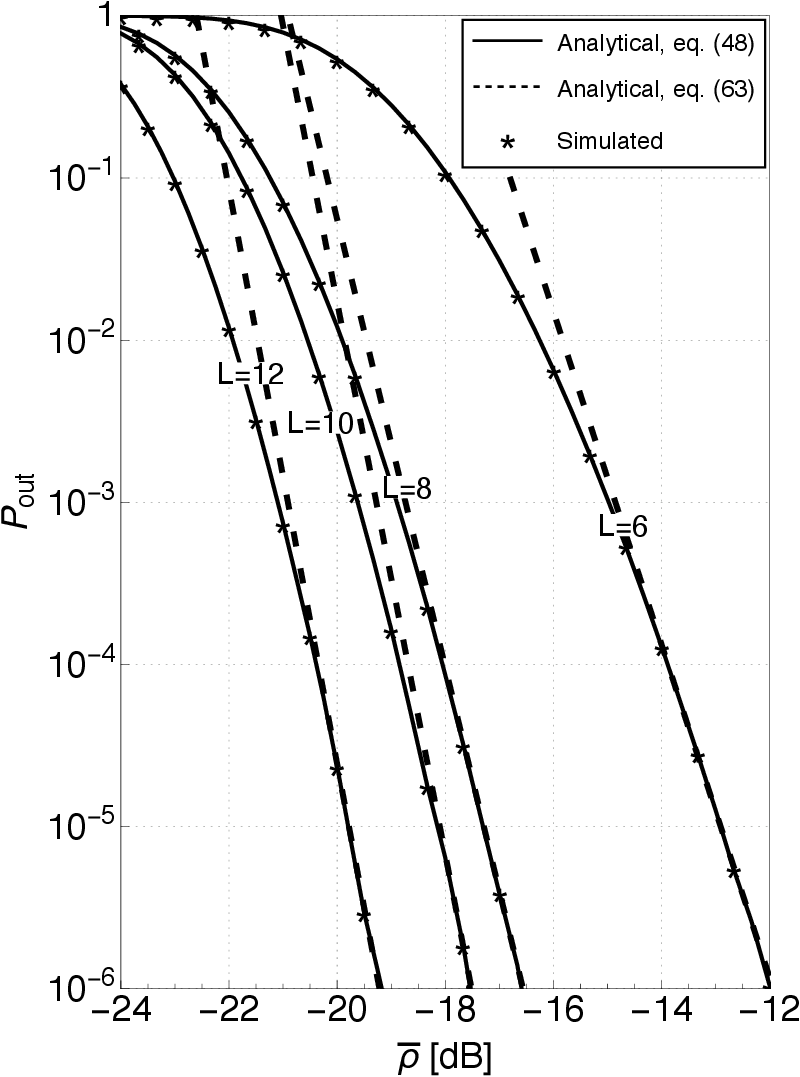}
      \\[-1mm]{\footnotesize (b) $\bar{\alpha}$-$\bar{\eta}$-$\kappa$-$\bar{\mu}$}
    \end{minipage}
    \vspace{0.3cm}
    \\
    \hspace{-0.4cm}
    \begin{minipage}[c]{0.22\textwidth}\centering
     \includegraphics[trim={0cm 0cm 0cm 0cm},clip,scale=0.30]{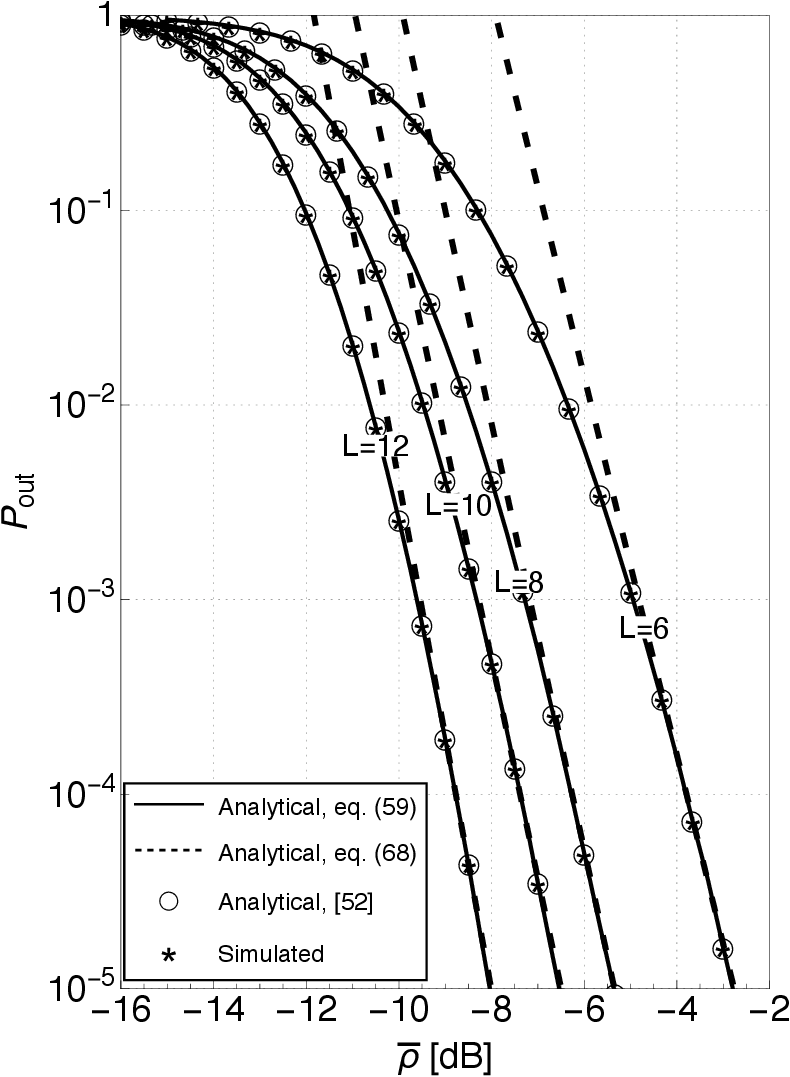}
      \\[-1mm]{\footnotesize (c) $\bar{\alpha}$-$\mathcal{F}$}
    \end{minipage}
    \hspace{-0.05cm}
    \begin{minipage}[c]{0.22\textwidth}\centering
      \includegraphics[trim={0cm 0cm 0cm 0cm},clip,scale=0.30]{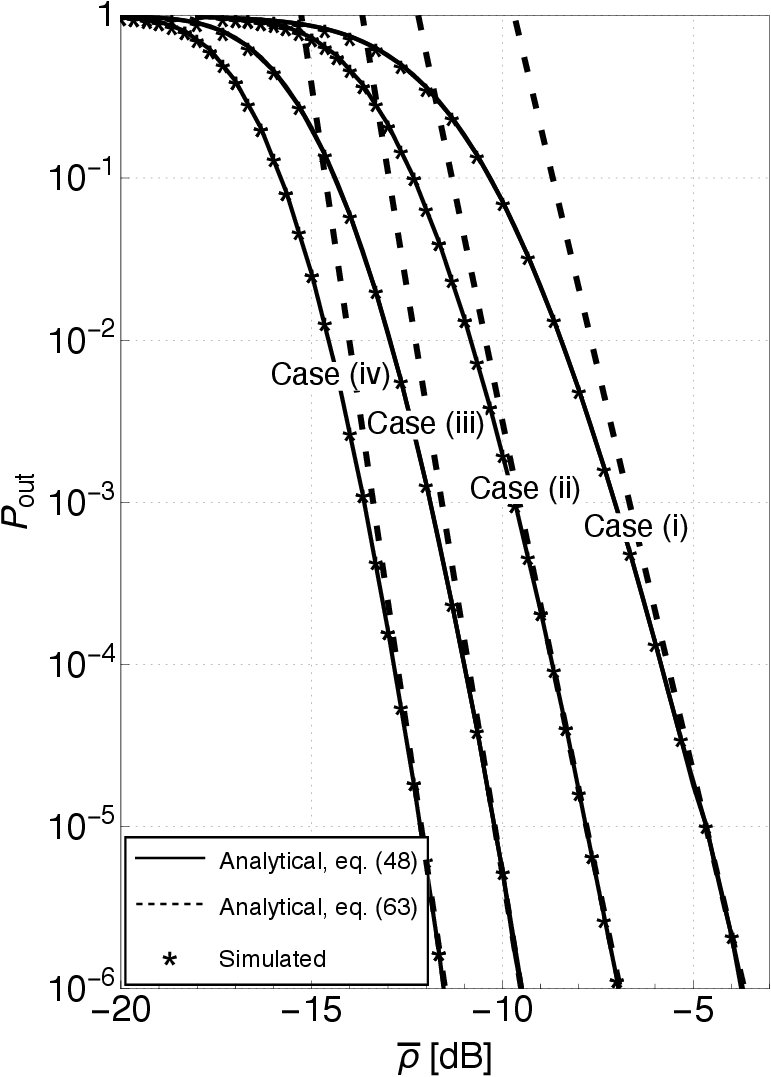}
      \\[-1mm]{\footnotesize (d) Mixed sum}
    \end{minipage}
  \end{tabular}
  \caption{OP as a function of the average SNR per symbol, $\bar{\rho}$, under different fading distributions.}
  \label{fig: OP}
\end{figure}
\begin{figure}[!t]
  \centering
  \begin{tabular}[c]{cc}
    \begin{subfigure}[c]{0.4\textwidth}
    \includegraphics[trim={0cm 0.03cm 0cm 0cm},clip,scale=0.5]{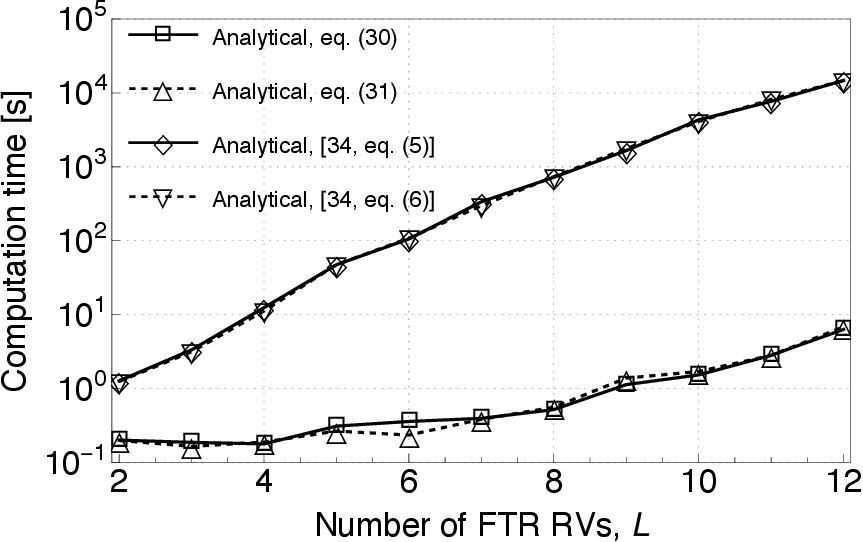}
      \caption{\centering
        Sum of i.non-i.d. squared FTR RVs.}
      \label{}
    \end{subfigure}
    \vspace{-0.3cm}
    \\
    \begin{subfigure}[c]{0.4\textwidth}
      \includegraphics[trim={0cm 0.03cm 0cm 0cm},clip,scale=0.5]{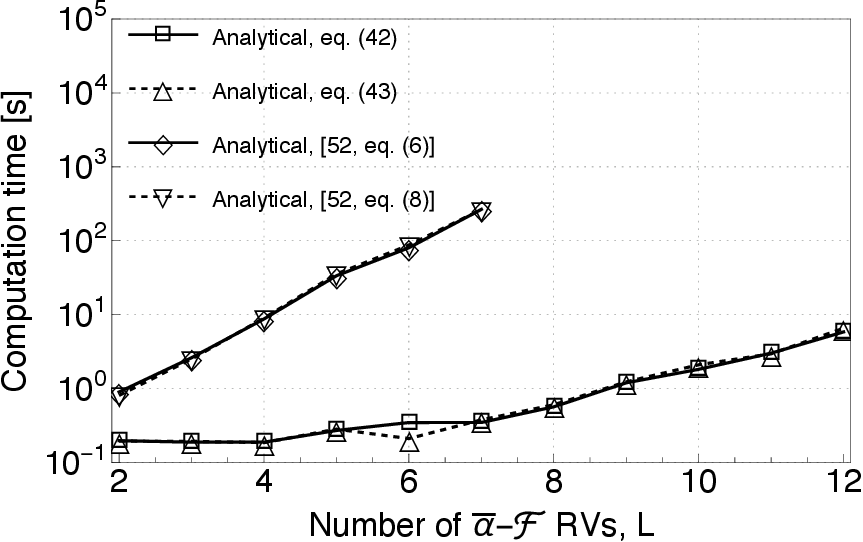}
      \caption{\centering
        Sum of i.n.i.d. $\bar{\alpha}$-$\mathcal{F}$ RVs.}
      \label{}
    \end{subfigure}
        \vspace{-0.3cm}
    \\
    \begin{subfigure}[c]{0.4\textwidth}
      \includegraphics[trim={0cm 0.03cm 0cm 0cm},clip,scale=0.5]{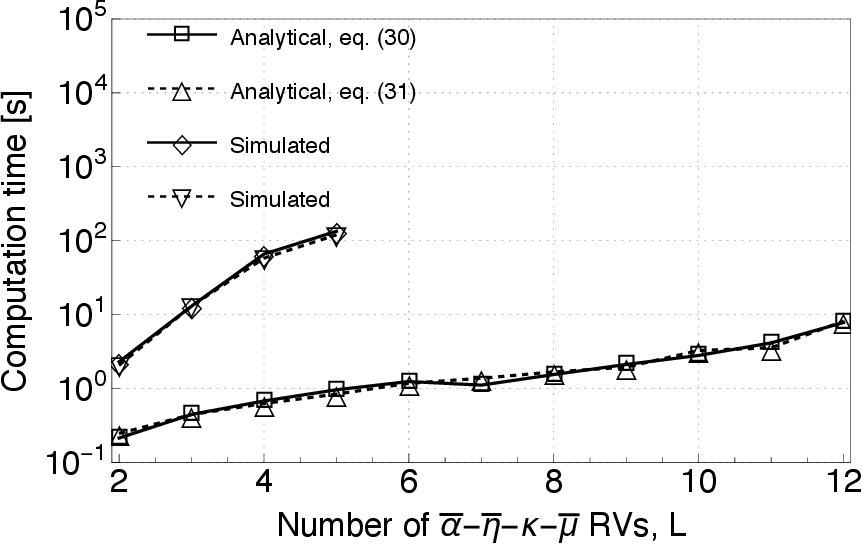}
      \caption{\centering
        Sum of i.n.i.d. $\bar{\alpha}$-$\bar{\eta}$-$\kappa$-$\bar{\mu}$ RVs.}
      \label{}
    \end{subfigure}
        \vspace{-0.3cm}
    \\
    \begin{subfigure}[c]{0.4\textwidth}
      \includegraphics[trim={0cm 0.03cm 0cm 0cm},clip,scale=0.5]{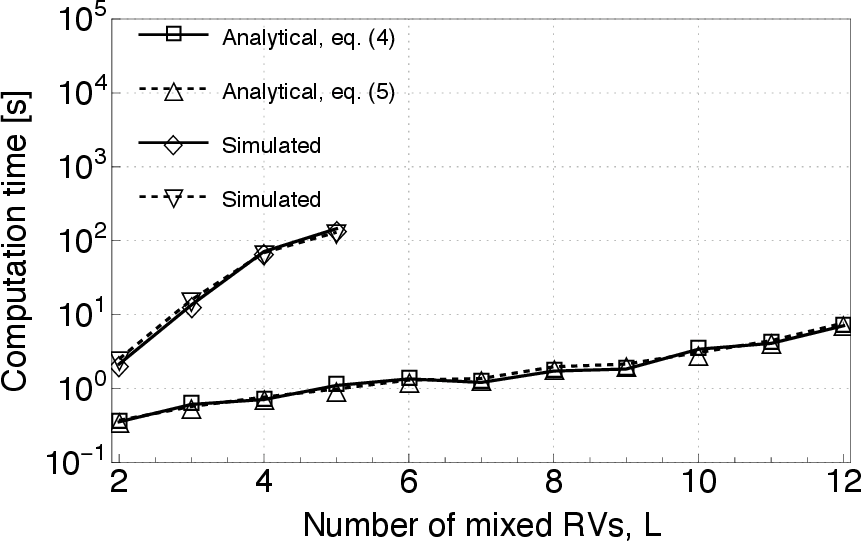}
      \caption{\centering
        Sum of mixed RVs.}
      \label{}
    \end{subfigure}
  \end{tabular}
  \caption{Computation time for the sum PDF and CDF of i.non-i.d., i.n.i.d., and mixed RVs at $x=10$ (an arbitrarily chosen evaluation point), with a relative error below $10^{-7}$.}
  \label{fig: computation time}
\end{figure}

In this section, we briefly validate our analytical findings through the numerical evaluation of the multifold Brennan's integral~\cite{Brennan59} or Monte Carlo simulations, wherever applicable---specifically, when Brennan's integral yields numerical inconsistencies or fails to produce a reliable numerical value, typically when $L>6$, although failure may occur for smaller $L$.
For benchmarking, our results are compared with those reported for the sum of squared FTR distributions in~\cite{Zheng19}, expressed in terms of the confluent multivariate hypergeometric function, and for the sum of $\bar{\alpha}$-$\mathcal{F}$ distributions in~\cite{Silva22}, expressed in terms of the multivariate Fox $H$-function. 
Notably, the sums of squared TWDP and Fisher-Snedecor~$\mathcal{F}$ distributions emerge as special cases of the FTR and $\bar{\alpha}$-$\mathcal{F}$ sums, respectively.

Figs.~\ref{fig: Sum PDF}(a)--\ref{fig: Sum PDF}(c), \ref{fig: Sum CDF}(a)--\ref{fig: Sum CDF}(c), and~\ref{fig: OP}(a)--\ref{fig: OP}(c), respectively, depict the simulated and theoretical PDFs, CDFs, and OPs of the sum for three general fading models: squared FTR, \mbox{$\bar{\alpha}$-$\bar{\eta}$-$\kappa$-$\bar{\mu}$}, and \mbox{$\bar{\alpha}$-$\mathcal{F}$}. 
The analysis
considers $L\in \left\{6,8,10,12\right\}$, with the twelve distinct parameter
configurations listed in Table~\ref{tab: PDF Parameters}.
Figs.~\ref{fig: Sum PDF}(d), \ref{fig: Sum CDF}(d), and
\ref{fig: OP}(d) present the corresponding simulated and analytical
results for mixed sums comprising FTR,
\mbox{$\bar{\alpha}$-$\bar{\eta}$-$\kappa$-$\bar{\mu}$}, and
\mbox{$\bar{\alpha}$-$\mathcal{F}$} fading RVs. Four mixed-sum
configurations are considered: (i) one RV from each distribution,
(ii) two RVs from each distribution, (iii) three RVs from each
distribution, and (iv) four RVs from each distribution. The
parameters are selected from Table~\ref{tab: PDF Parameters} as
follows: case~(i) uses $\ell=1$; case~(ii) uses
$\ell\in \left\{1,2\right\}$; case~(iii) uses $\ell\in \left\{1,2,3\right\}$; and
case~(iv) uses $\ell\in \left\{1,2,3,4\right\}$.
For the OP results, Fig.~\ref{fig: OP}(a) and \ref{fig: OP}(c) correspond to MRC
receivers, whereas Figs.~\ref{fig: OP}(b) and \ref{fig: OP}(d)
correspond to EGC receivers.
All figures show an excellent alignment between the analytical and simulated curves, confirming the analytical expressions. It is also worth noting that the PDFs, CDFs, and OPs corresponding to the sums of \mbox{$\bar{\alpha}$-$\bar{\eta}$-$\kappa$-$\bar{\mu}$} and mixed distributions were verified through Monte Carlo simulations or numerical integration, as no prior studies have addressed the sum of \mbox{$\bar{\alpha}$-$\bar{\eta}$-$\kappa$-$\bar{\mu}$} distributions---let alone the sum of mixed distributions.

The OP curves in Fig.~\ref{fig: OP} also provide useful
receiver-design insights. In all cases, the asymptotic expressions
closely approach the exact analytical results in the high-SNR
region, confirming the validity of the derived asymptotic
characterization. Increasing $L$ reduces the OP for a given average
SNR per symbol and produces a steeper high-SNR decay, because the
individual branch contributions to the diversity order are
additive. Moreover, fading configurations with the same aggregate
small-argument exponent may exhibit the same asymptotic slope while
remaining horizontally separated, which reflects identical
diversity orders but different coding gains. This distinction is
particularly relevant for mixed sums, in which Gaussian-class and
non-Gaussian-class branches contribute jointly to the diversity
order but affect the leading outage coefficient according to their
individual marginal parameters. Therefore, the proposed framework
allows the effects of the number of branches, fading severity, and
channel heterogeneity to be quantified separately in terms of
diversity order, coding gain, and the average SNR required to achieve
a prescribed outage level. Together with the unified EGC and MRC
formulations, it also provides a systematic basis for comparing
different diversity-receiver architectures under mixed fading
conditions.

Fig.~\ref{fig: computation time} compares the computation time
required to evaluate the proposed sum PDF and CDF expressions with those of available analytical benchmarks and, when no such benchmarks
exist, with the multifold Brennan integral. The number of summed RVs,
$L$, is varied along the horizontal axis, while the evaluation point
is fixed at $x=10$. For all methods, the computation time was recorded
once a relative error below $10^{-7}$ was achieved. The distribution
parameters were selected from Table~\ref{tab: PDF Parameters}.
Specifically, Fig.~\ref{fig: computation time}(a) considers sums of
i.non-i.d. squared FTR RVs,
Fig.~\ref{fig: computation time}(b) considers sums of i.n.i.d.
\mbox{$\bar{\alpha}$-$\mathcal{F}$} RVs,
Fig.~\ref{fig: computation time}(c) considers sums of i.n.i.d.
\mbox{$\bar{\alpha}$-$\bar{\eta}$-$\kappa$-$\bar{\mu}$} RVs, and
Fig.~\ref{fig: computation time}(d) considers mixed sums comprising
different fading distributions.

For the squared FTR case in Fig.~\ref{fig: computation time}(a), the
proposed expressions are compared with the analytical solutions
in~\cite{Zheng19}. Likewise, for the
\mbox{$\bar{\alpha}$-$\mathcal{F}$} case in
Fig.~\ref{fig: computation time}(b), they are compared with the
multivariate Fox $H$-function-based expressions in~\cite{Silva22}.
For the
\mbox{$\bar{\alpha}$-$\bar{\eta}$-$\kappa$-$\bar{\mu}$} and mixed-sum
cases in Figs.~\ref{fig: computation time}(c) and
\ref{fig: computation time}(d), respectively, no previously reported
analytical sum expressions are available; therefore, the multifold
Brennan integral is adopted as the computational benchmark.
The results demonstrate a substantial computational advantage of the
proposed framework. In all four scenarios, the PDF and CDF require similar evaluation times because both expressions rely on the same
coefficient recursion and differ only in the final outer-series
evaluation. More importantly, the computation time increases only
gradually with $L$ and remains on the order of a few seconds even for
$L=12$. This behavior agrees with the complexity analysis in
Section~\ref{sec: Complexity Analysis}, according to which the
dominant coefficient-construction cost for heterogeneous sums scales as
$\mathcal{O}(Lt_0^2)$.
By contrast, the competing analytical solutions exhibit a much steeper
increase in computation time as $L$ grows. For the squared FTR case,
the benchmark expressions require approximately $10^4$ seconds at
$L=12$, whereas the proposed expressions are evaluated in only a few
seconds. A similar trend is observed for the
\mbox{$\bar{\alpha}$-$\mathcal{F}$} case, for which the benchmark
becomes computationally prohibitive beyond a moderate number of RVs
and could be evaluated only up to $L=7$. This rapid growth is a direct
consequence of the multidimensional structure of the competing
confluent hypergeometric- and multivariate Fox $H$-function-based
formulations.
The Brennan-integral benchmarks in
Figs.~\ref{fig: computation time}(c) and
\ref{fig: computation time}(d) show a similarly pronounced growth and
remain computationally feasible only up to $L=5$. In contrast, the
proposed single-series representations remain practical throughout the
entire investigated range. Therefore, Fig.~\ref{fig: computation time}
confirms that the proposed framework combines exactness and analytical
tractability with substantial time savings and markedly improved
scalability relative to existing analytical and numerical alternatives.
These advantages are particularly relevant for mixed fading analysis,
where repeated evaluations over different numbers of branches,
distribution parameters, and operating conditions are often required.

\section{Conclusions}
\label{sec: Conclusions}

This paper addressed a longstanding and practically important problem: obtaining exact, unified, and tractable statistics for sums of an arbitrary number of mixed independent positive RVs. In wireless communications, these results establish a common analytical foundation for channel modeling and performance evaluation over heterogeneous fading environments.
Specifically, we derived novel expressions for the PDF and CDF of sums involving Gaussian-class and non-Gaussian-class fading distributions, covering i.i.d., i.n.i.d., and mixed configurations. We also introduced the $\alpha$-$\mu$ Mixture distribution as a unifying representation for Gaussian-class fading models. Compared with existing analytical approaches, the proposed solutions offer broader applicability, improved computational efficiency, and better scalability with the number of summed RVs.
Building on the derived sum statistics, we further developed a unified performance analysis for EGC and MRC receivers over mixed fading channels. Exact and high-SNR asymptotic OP expressions were obtained, explicitly revealing the corresponding diversity orders and coding gains. These results enable the separate assessment of the effects of the combining architecture, number of branches, fading severity, channel heterogeneity, and operating SNR. Consequently, the framework supports receiver comparison, reliability assessment, threshold selection, link-budget design, and the determination of the SNR required to meet a prescribed outage-probability target, without repeated model-specific derivations, costly numerical integrations, or extensive Monte Carlo simulations.

\begin{appendices}

\section{Proof of the Theorem}
\label{app: Theorem}

We prove the result directly in the $x$-domain.
For $L=1$, the result is immediate. 
Hence, we consider $L\geq2$.
Since $X_1,\ldots,X_L$ are independent nonnegative continuous RVs,
the PDF of $X$ is given by the $L$-fold convolution
\begin{align}
f_X(x)
&=
\int_{\mathcal S_{L-1}(x)}
\prod_{\ell=1}^{L}
f_{X_\ell}(t_\ell)
\,dt_1\cdots dt_{L-1},
\label{eq: convolution proof}
\end{align}
where
\begin{align}
\mathcal S_{L-1}(x)
&\triangleq
\left\{
(t_1,\ldots,t_{L-1})\in\mathbb R_+^{L-1}:
\sum_{\ell=1}^{L-1}t_\ell<x
\right\},
\label{eq: convolution simplex proof}
\end{align}
and $t_L \triangleq
x-\sum_{\ell=1}^{L-1}t_\ell$.
Thus, $t_\ell\geq0$ for all $\ell$ and
$\sum_{\ell=1}^{L}t_\ell=x$.

Substituting the marginal representation in
\eqref{eq: PDF marginal theorem} into \eqref{eq: convolution proof},
we obtain
\begin{align}
\nonumber  f_X(x)
&=
\left(\prod_{\ell=1}^{L}\Psi_\ell\right)
\int_{\mathcal S_{L-1}(x)}
\prod_{\ell=1}^{L} \\
& \times
\sum_{i_\ell=0}^{\infty}
\frac{
\eta_{i_\ell,\ell}
t_\ell^{\theta i_\ell+\beta_\ell-1}
}{
\Gamma(\theta i_\ell+\beta_\ell)
}
\,dt_1\cdots dt_{L-1}.
\label{eq: convolution substituted}
\end{align}

Appendix~\ref{app: Absolute Convergence X} establishes that, for every
$x>0$, the multiple series arising from the convolution expansion
in \eqref{eq: convolution substituted} is absolutely integrable. Therefore, Tonelli's theorem
applies to the corresponding nonnegative absolute series, and
Fubini's theorem justifies interchanging the infinite summations
with the convolution integral. Consequently,
\begin{align}
f_X(x)
&=
\left(\prod_{\ell=1}^{L}\Psi_\ell\right)
\sum_{i_1,\ldots,i_L=0}^{\infty}
\left(
\prod_{\ell=1}^{L}\eta_{i_\ell,\ell}
\right)
\nonumber\\
&\quad \times
\int_{\mathcal S_{L-1}(x)}
\prod_{\ell=1}^{L}
\frac{
t_\ell^{\theta i_\ell+\beta_\ell-1}
}{
\Gamma(\theta i_\ell+\beta_\ell)
}
\,dt_1\cdots dt_{L-1}.
\label{eq: termwise convolution}
\end{align}
By the Dirichlet integral identity~\cite{GuptaNagar2000}, we have
\begin{align}
&
\int_{\mathcal S_{L-1}(x)}
\prod_{\ell=1}^{L}
\frac{
t_\ell^{\theta i_\ell+\beta_\ell-1}
}{
\Gamma(\theta i_\ell+\beta_\ell)
}
\,dt_1\cdots dt_{L-1}
\nonumber\\
&\qquad =
\frac{
x^{\theta\sum_{\ell=1}^{L}i_\ell+B-1}
}{
\Gamma\!\left(\theta\sum_{\ell=1}^{L}i_\ell+B\right)
}.
\label{eq: Dirichlet identity proof}
\end{align}
Substituting \eqref{eq: Dirichlet identity proof} into
\eqref{eq: termwise convolution}, the resulting multiple series
depends on the multi-index $(i_1,\ldots,i_L)$ only through
$i_1+\cdots+i_L$ in the power of $x$ and in the Gamma function.
Therefore, collecting the finitely many multi-indices satisfying
$i_1+\cdots+i_L=i$, we obtain
\begin{align}
f_X(x)
&=
\left(\prod_{\ell=1}^{L}\Psi_\ell\right)
\sum_{i=0}^{\infty}
\frac{
\delta_i x^{\theta i+B-1}
}{
\Gamma(\theta i+B)
},
\qquad x>0,
\label{eq: PDF proof final}
\end{align}
where
\begin{align}
\delta_i
&=
\sum_{\substack{
i_1+\cdots+i_L=i\\
i_1,\ldots,i_L\in\mathbb N_0
}}
\prod_{\ell=1}^{L}\eta_{i_\ell,\ell}.
\label{eq: delta finite convolution proof}
\end{align}
For each fixed $i$, the sum in
\eqref{eq: delta finite convolution proof} is finite. Hence, the
coefficient $\delta_i$ is well defined, and
\eqref{eq: PDF proof final} is precisely the PDF representation
in \eqref{eq: PDF X Final}.

It remains to show that the recursive definition of $\delta_i$ in
\eqref{eq: delta coefficients} is equivalent to the
finite-convolution coefficient in
\eqref{eq: delta finite convolution proof}. Define the formal
power series
\begin{align}
\label{eq: g z}
g_\ell(z)
&\triangleq
\sum_{i=0}^{\infty}\eta_{i,\ell}z^i \\
D(z)
&\triangleq
\prod_{\ell=1}^{L}g_\ell(z)
=
\sum_{i=0}^{\infty}\delta_i z^i.
\label{eq: formal generating functions}
\end{align}
Then \eqref{eq: delta finite convolution proof} is exactly the
Cauchy-product coefficient of $D(z)$ in \eqref{eq: formal generating functions}. Since
$\eta_{0,\ell}>0$, each $g_\ell(z)$ has a nonzero constant term
and is therefore a unit in the ring of formal power series.
Hence, its reciprocal $1/g_\ell(z)$ exists as a formal power
series. The formal logarithmic derivative is consequently
well defined, and we can introduce the coefficients
$\phi_{h,\ell}$ through
\begin{align}
\frac{g_\ell'(z)}{g_\ell(z)}
&=
\sum_{h=0}^{\infty}\phi_{h,\ell}z^h,
\label{eq: formal logarithmic derivative g}
\end{align}
where the apostrophe symbol $(\,{}'\,)$ denotes formal
differentiation with respect to $z$.
Equating coefficients in
\begin{align}
g_\ell'(z)
&=
g_\ell(z)
\sum_{h=0}^{\infty}\phi_{h,\ell}z^h
\end{align}
gives the recurrence for the coefficients of the formal
logarithmic derivative. Indeed,
\begin{align}
g_\ell'(z)
&=
\sum_{h=0}^{\infty}(h+1)\eta_{h+1,\ell}z^h,
\end{align}
whereas the coefficient of $z^h$ in the Cauchy product
\begin{align}
g_\ell(z)
\sum_{r=0}^{\infty}\phi_{r,\ell}z^r
&=
\left(\sum_{t=0}^{\infty}\eta_{t,\ell}z^t\right)
\left(\sum_{r=0}^{\infty}\phi_{r,\ell}z^r\right)
\end{align}
is
\begin{align}
\sum_{t=0}^{h}\eta_{t,\ell}\phi_{h-t,\ell}
&=
\eta_{0,\ell}\phi_{h,\ell}
+
\sum_{t=1}^{h}\eta_{t,\ell}\phi_{h-t,\ell}.
\end{align}
Thus, for every $h\geq0$, we have
\begin{align}
(h+1)\eta_{h+1,\ell}
&=
\eta_{0,\ell}\phi_{h,\ell}
+
\sum_{t=1}^{h}\eta_{t,\ell}\phi_{h-t,\ell}.
\end{align}
Since $\eta_{0,\ell}>0$, the coefficients
$\phi_{h,\ell}$ are uniquely determined.  In particular, the case $h=0$ gives
\eqref{eq: phi 0}, while the cases $h\geq1$ give
\eqref{eq: phi i}.

Similarly, differentiating
$D(z)=\prod_{\ell=1}^{L}g_\ell(z)$, we obtain
\begin{align}
D'(z)
&=
D(z)
\sum_{\ell=1}^{L}
\frac{g_\ell'(z)}{g_\ell(z)}.
\label{eq: D log derivative}
\end{align}
Using $D'(z)=
\sum_{i=1}^{\infty}i\delta_i z^{i-1}$,
and equating the coefficients of $z^{i-1}$ in
\eqref{eq: D log derivative}, we obtain
\begin{align}
i\delta_i
&=
\sum_{h=1}^{i}
\delta_{i-h}
\sum_{\ell=1}^{L}\phi_{h-1,\ell},
\qquad i\geq 1.
\end{align}

Consequently, \eqref{eq: delta i} holds for every $i\geq1$.
The initial coefficient is obtained from the constant term of the
Cauchy product, yielding \eqref{eq: delta 0}.

Finally, the CDF is obtained by integrating
\eqref{eq: PDF proof final} over $[0,x]$. Term-by-term integration is justified by the absolute-convergence bound established in
Appendix~\ref{app: Absolute Convergence X}. Hence,
Fubini's theorem~\cite{fubini07} permits the interchange of the
summation and the integral, yielding
\begin{align}
F_X(x)
&=
\int_0^x f_X(u)\,du
\nonumber\\
&=
\left(\prod_{\ell=1}^{L}\Psi_\ell\right)
\sum_{i=0}^{\infty}
\frac{\delta_i}{\Gamma(\theta i+B)}
\int_0^x
u^{\theta i+B-1}
\,du.
\label{eq: CDF proof final}
\end{align}
After evaluating the integral in \eqref{eq: CDF proof final}, we obtain \eqref{eq: CDF X Final}, thereby completing the proof.

\section{Sufficient Condition for Absolute and Locally Uniform Convergence}
\label{app: Absolute Convergence X}

This appendix proves that the generalized coefficient-growth
condition in \eqref{eq: generalized coefficient growth} is
sufficient to guarantee the absolute and locally uniform
convergence, with respect to $x$, of the final PDF and CDF series
in \eqref{eq: PDF X Final} and \eqref{eq: CDF X Final},
respectively.

Assume that, for each $\ell\in\{1,\ldots,L\}$, there exist
constants $C_\ell,A_\ell,c_\ell>0$ and $0 \leq q_\ell<1$ such that
\begin{align}
|\eta_{i,\ell}|
&\leq
C_\ell A_\ell^i
\frac{
\Gamma(i\theta+\beta_\ell)\Gamma(q_\ell i+c_\ell)
}{
i!
},
\qquad i\in\mathbb N_0.
\label{eq: generalized eta bound appendix}
\end{align}
The constants are independent of $i$.

Applying the generalized coefficient-growth condition in
\eqref{eq: generalized eta bound appendix} and the triangle inequality to the series in
\eqref{eq: convolution substituted}, we obtain
\begin{align}
&
\sum_{i_1,\ldots,i_L=0}^{\infty}
\left(
\prod_{\ell=1}^{L}
|\eta_{i_\ell,\ell}|
\right)
\frac{
x^{\theta\sum_{\ell=1}^{L}i_\ell+B-1}
}{
\Gamma\!\left(\theta\sum_{\ell=1}^{L}i_\ell+B\right)
}
\nonumber\\
&\leq
\left(\prod_{\ell=1}^{L}C_\ell\right)
x^{B-1}
\sum_{i_1,\ldots,i_L=0}^{\infty}
\frac{
1
}{
\Gamma\!\left(\theta\sum_{\ell=1}^{L}i_\ell+B\right)
}
\nonumber\\
&\quad\times
\prod_{\ell=1}^{L}
\frac{
(A_\ell x^\theta)^{i_\ell}
\Gamma(\theta i_\ell+\beta_\ell)
\Gamma(q_\ell i_\ell+c_\ell)
}{
i_\ell!
}.
\label{eq: absolute bound step generalized}
\end{align}
Set $a_\ell
\triangleq
\theta i_\ell+\beta_\ell$.
Then, $a_\ell>0$ and $\sum_{\ell=1}^{L}a_\ell
=
\theta\sum_{\ell=1}^{L}i_\ell+B$.
By the multivariate beta-function identity~\cite{GuptaNagar2000}, we have
\begin{align}
\frac{
\prod_{\ell=1}^{L}\Gamma(\theta i_\ell+\beta_\ell)
}{
\Gamma\!\left(\theta\sum_{\ell=1}^{L}i_\ell+B\right)
}
&=
\mathrm B(a_1,\ldots,a_L).
\end{align}
Using the simplex representation of the multivariate beta
function~\cite{GuptaNagar2000}, we get
\begin{align}
\mathrm B(a_1,\ldots,a_L)
&=
\int_{\Delta_{L-1}}
\prod_{\ell=1}^{L}
u_\ell^{a_\ell-1}
\,du_1\cdots du_{L-1},
\end{align}
where
\begin{align}
\Delta_{L-1}
&\triangleq
\left\{
(u_1,\ldots,u_{L-1})\in\mathbb R_+^{L-1}:
\sum_{\ell=1}^{L-1}u_\ell<1
\right\},
\end{align}
and $u_L
\triangleq
1-\sum_{\ell=1}^{L-1}u_\ell$.
Thus, $0\leq u_\ell\leq1$ on $\Delta_{L-1}$. Since
$\beta_\ell , \theta i_\ell \geq 0$, it
follows that
\begin{align}
\prod_{\ell=1}^{L}u_\ell^{a_\ell-1}
&=
\left(
\prod_{\ell=1}^{L}u_\ell^{\beta_\ell-1}
\right)
\left(
\prod_{\ell=1}^{L}u_\ell^{\theta i_\ell}
\right)
\leq
\prod_{\ell=1}^{L}u_\ell^{\beta_\ell-1}
\end{align}
almost everywhere on $\Delta_{L-1}$ since
$u_\ell^{\theta i_\ell}\leq1$ for $0\leq u_\ell\leq1$.
Therefore, the multivariate beta
function can be bounded as
\begin{align}
\mathrm B(a_1,\ldots,a_L)
&\leq
\int_{\Delta_{L-1}}
\prod_{\ell=1}^{L}
u_\ell^{\beta_\ell-1}
\,du_1\cdots du_{L-1}
\nonumber\\
&=
\mathrm B(\beta_1,\ldots,\beta_L)
=
\frac{
\prod_{\ell=1}^{L}\Gamma(\beta_\ell)
}{
\Gamma(B)
}.
\end{align}
Consequently,
\begin{align}
\frac{
\prod_{\ell=1}^{L}\Gamma(\theta i_\ell+\beta_\ell)
}{
\Gamma\!\left(\theta\sum_{\ell=1}^{L}i_\ell+B\right)
}
&\leq
\frac{
\prod_{\ell=1}^{L}\Gamma(\beta_\ell)
}{
\Gamma(B)
}\triangleq
C_{\boldsymbol\beta}.
\label{eq: gamma ratio bound generalized}
\end{align}

Substituting \eqref{eq: gamma ratio bound generalized} into
\eqref{eq: absolute bound step generalized}, we obtain
\begin{align}
&
\sum_{i_1,\ldots,i_L=0}^{\infty}
\left(
\prod_{\ell=1}^{L}
|\eta_{i_\ell,\ell}|
\right)
\frac{
x^{\theta\sum_{\ell=1}^{L}i_\ell+B-1}
}{
\Gamma\!\left(\theta\sum_{\ell=1}^{L}i_\ell+B\right)
}
\nonumber\\
\nonumber &\leq
\left(\prod_{\ell=1}^{L}C_\ell\right)
\frac{\prod_{\ell=1}^{L}\Gamma(\beta_\ell)}{\Gamma(B)}
x^{B-1} \\
& \times
\sum_{i_1,\ldots,i_L=0}^{\infty}
\prod_{\ell=1}^{L}
\frac{
\Gamma(q_\ell i_\ell+c_\ell)
(A_\ell x^\theta)^{i_\ell}
}{
i_\ell!
}.
\label{eq: generalized product majorant intermediate}
\end{align}
Since all terms in the last multiple series are nonnegative,
Tonelli's theorem permits the factorization of the multiple series
into the product of its one-dimensional marginal series. Hence,
\begin{align}
&
\sum_{i_1,\ldots,i_L=0}^{\infty}
\left(
\prod_{\ell=1}^{L}
|\eta_{i_\ell,\ell}|
\right)
\frac{
x^{\theta\sum_{\ell=1}^{L}i_\ell+B-1}
}{
\Gamma\!\left(\theta\sum_{\ell=1}^{L}i_\ell+B\right)
}
\nonumber\\
\nonumber &\leq
\left(\prod_{\ell=1}^{L}C_\ell\right)
\frac{\prod_{\ell=1}^{L}\Gamma(\beta_\ell)}{\Gamma(B)}
x^{B-1} \\
& \times
\prod_{\ell=1}^{L}
\sum_{i_\ell=0}^{\infty}
\frac{
\Gamma(q_\ell i_\ell+c_\ell)
(A_\ell x^\theta)^{i_\ell}
}{
i_\ell!
}.
\label{eq: generalized product majorant}
\end{align}

Using the finite-convolution representation of $\delta_i$, together
with the triangle inequality applied to \eqref{eq: PDF X Final},
and the bound in \eqref{eq: generalized product majorant}, we obtain, for
every $x>0$,
\begin{align}
&
\sum_{i=0}^{\infty}
\frac{
|\delta_i|x^{\theta i+B-1}
}{
\Gamma(\theta i+B)
}
\nonumber\\
&\quad \leq
C_{\boldsymbol\beta}
\left(\prod_{\ell=1}^{L}C_\ell\right)
x^{B-1}
\prod_{\ell=1}^{L}
\sum_{i_\ell=0}^{\infty}
\frac{
\Gamma(q_\ell i_\ell+c_\ell)
(A_\ell x^\theta)^{i_\ell}
}{
i_\ell!
}.
\label{eq: PDF generalized majorant appendix}
\end{align}

Each one-dimensional series in
\eqref{eq: PDF generalized majorant appendix} converges for every
finite $x$.
To see this, define its $i$th nonnegative term as
\begin{align}
b_{i,\ell}(x)
&\triangleq
\frac{\Gamma(q_\ell i+c_\ell)(A_\ell x^\theta)^i}{i!},
\qquad i\in\mathbb N_0 .
\end{align}
Then,
\begin{align}
\frac{b_{i+1,\ell}(x)}{b_{i,\ell}(x)}
&=
\frac{A_\ell x^\theta}{i+1}
\frac{\Gamma(q_\ell(i+1)+c_\ell)}
{\Gamma(q_\ell i+c_\ell)}.
\end{align}
If $0<q_\ell<1$, the Gamma-ratio asymptotic relation gives
\begin{align}
\frac{\Gamma(q_\ell(i+1)+c_\ell)}
{\Gamma(q_\ell i+c_\ell)}
&\sim
(q_\ell i)^{q_\ell},
\qquad i\to\infty,
\end{align}
implying that
\begin{align}
\frac{b_{i+1,\ell}(x)}{b_{i,\ell}(x)}
&\sim
A_\ell x^\theta q_\ell^{q_\ell} i^{q_\ell-1}
\longrightarrow 0.
\end{align}
If $q_\ell=0$, then the Gamma ratio is equal to one, and hence
\[
\frac{b_{i+1,\ell}(x)}{b_{i,\ell}(x)}
=
\frac{A_\ell x^\theta}{i+1}
\longrightarrow 0.
\]
Thus, for every $0\leq q_\ell<1$, the ratio test~\cite{Kreyszig10} implies that
each one-dimensional series in
\eqref{eq: PDF generalized majorant appendix} converges for every
finite $x$. Consequently, the PDF series in
\eqref{eq: PDF X Final} converges absolutely for every $x>0$.

Applying
\[
\Gamma(1+\theta i+B)
=
(\theta i+B)\Gamma(\theta i+B)
\geq
B\Gamma(\theta i+B),
\]
we obtain, for every $x>0$,
\begin{align}
&
\sum_{i=0}^{\infty}
\frac{
|\delta_i|x^{\theta i+B}
}{
\Gamma(1+\theta i+B)
}
\nonumber\\
&\quad \leq
\frac{
C_{\boldsymbol\beta}
\left(\prod_{\ell=1}^{L}C_\ell\right)
x^{B}
}{B}
\prod_{\ell=1}^{L}
\sum_{i_\ell=0}^{\infty}
\frac{
\Gamma(q_\ell i_\ell+c_\ell)
(A_\ell x^\theta)^{i_\ell}
}{
i_\ell!
}.
\label{eq: CDF generalized majorant appendix}
\end{align}
The product on the right-hand side is finite by the same ratio-test
argument used for \eqref{eq: PDF generalized majorant appendix}.
Therefore, the CDF series in \eqref{eq: CDF X Final} converges
absolutely for every $x>0$. At $x=0$, each term is zero because
$\theta i+B>0$, and hence the CDF series converges absolutely on
$[0,\infty)$.

It remains to prove local uniform convergence. Let
$K_f\Subset(0,\infty)$ be compact. Then, there exist
$0<a_f\leq b_f<\infty$ such that $K_f\subset[a_f,b_f]$. Define
\begin{align}
M_f
&\triangleq
\max\{a_f^{B-1},b_f^{B-1}\}.
\end{align}
For every $x\in K_f$, the absolute majorant in
\eqref{eq: PDF generalized majorant appendix} satisfies
\begin{align}
&
C_{\boldsymbol\beta}
\left(\prod_{\ell=1}^{L}C_\ell\right)
x^{B-1}
\prod_{\ell=1}^{L}
\sum_{i_\ell=0}^{\infty}
\frac{
\Gamma(q_\ell i_\ell+c_\ell)
(A_\ell x^\theta)^{i_\ell}
}{
i_\ell!
}
\nonumber\\
&\quad \leq
C_{\boldsymbol\beta}
\left(\prod_{\ell=1}^{L}C_\ell\right)
M_f
\prod_{\ell=1}^{L}
\sum_{i_\ell=0}^{\infty}
\frac{
\Gamma(q_\ell i_\ell+c_\ell)
(A_\ell b_f^\theta)^{i_\ell}
}{
i_\ell!
}
<\infty .
\end{align}
Since the right-hand side is finite and independent of $x$, the
Weierstrass $M$-test~\cite{Kreyszig10} implies uniform absolute convergence of the
PDF series on $K_f$. Hence, the PDF series converges locally
uniformly on $(0,\infty)$.

Likewise, let $K_F\Subset[0,\infty)$ be compact. Then there exists
$b_F<\infty$ such that $K_F\subset[0,b_F]$. For every
$x\in K_F$, the absolute majorant in
\eqref{eq: CDF generalized majorant appendix} satisfies
\begin{align}
&
\frac{
C_{\boldsymbol\beta}
\left(\prod_{\ell=1}^{L}C_\ell\right)
x^{B}
}{B}
\prod_{\ell=1}^{L}
\sum_{i_\ell=0}^{\infty}
\frac{
\Gamma(q_\ell i_\ell+c_\ell)
(A_\ell x^\theta)^{i_\ell}
}{
i_\ell!
}
\nonumber\\
\nonumber &\qquad \leq
\frac{
C_{\boldsymbol\beta}
\left(\prod_{\ell=1}^{L}C_\ell\right)
b_F^{B}
}{B} \\
& \qquad \times
\prod_{\ell=1}^{L}
\sum_{i_\ell=0}^{\infty}
\frac{
\Gamma(q_\ell i_\ell+c_\ell)
(A_\ell b_F^\theta)^{i_\ell}
}{
i_\ell!
}
<\infty .
\end{align}
Therefore, the Weierstrass $M$-test implies uniform absolute
convergence of the CDF series on every compact set
$K_F\Subset[0,\infty)$. Hence, the CDF series converges locally
uniformly on $[0,\infty)$.

We conclude that the generalized coefficient-growth condition in
\eqref{eq: generalized eta bound appendix} is sufficient to
guarantee the absolute and locally uniform convergence of the
final PDF and CDF series in \eqref{eq: PDF X Final} and
\eqref{eq: CDF X Final}, respectively. This completes the proof.

\section{Proof of Corollary 1}
\label{app: IID case}

Upon substituting $\Psi _{\ell }= \Psi$,  $\beta_\ell = \beta$, and $\eta_{i,\ell}= \eta_i$ into \eqref{eq: PDF X Final}, \eqref{eq: CDF X Final}, and \eqref{eq: delta 0}, it immediately follows that $f_X(x)$, $F_X(x)$, and $\delta_0$ readily reduce to \eqref{eq: PDF X Final iid}, \eqref{eq: CDF X Final iid}, and \eqref{eq: delta 0 iid}, respectively. Moreover, after replacing $\phi_{i,\ell}=\phi_{i}$ into \eqref{eq: phi coefficients}, we obtain
\begin{subequations}
\label{eq: phi coefficients IID}
\begin{align}
    \label{eq: phi 0 IID}
    \phi_{0} =& \frac{\eta _{1}}{\eta _{0}} & \\ 
    \label{eq: phi i IID}
     \phi_{h} = & \frac{1}{\eta _{0 }} \left[(h+1) \, \eta _{h+1 }-\sum _{t=1}^h \eta _{t } \, \phi_{h-t }\right], & h \geq 1.
\end{align}
\end{subequations}
\normalsize

Plugging \eqref{eq: phi coefficients IID} into \eqref{eq: delta i} and followed by some algebraic manipulations, $\delta_i$ for $i \geq 1$ simplifies to \eqref{eq: delta i iid}.

\section{Truncation Error Bounds}
\label{app: Truncation error Bounds}

Applying the triangle
inequality, the truncation errors defined in \eqref{eq: Truncation Error PDF CDF} can be bounded as
\begin{align}
\left|
\Xi_{\vartheta}^{(t_0)}(x)
\right|
&\leq
\left(\prod_{\ell=1}^{L}\Psi_\ell\right)
x^{B+\varrho-1}
\sum_{i=t_0}^{\infty}
\frac{
|\delta_i|x^{i\theta}
}{
\Gamma(i\theta+B+\varrho)
}.
\label{eq: bound Xi step 1}
\end{align}

Assume that the marginal coefficients satisfy the generalized
coefficient-growth condition in
\eqref{eq: generalized coefficient growth}. Using the
finite-convolution representation of $\delta_i$ in
\eqref{eq: delta finite convolution proof}, together with the
triangle inequality, we obtain
\begin{align}
&
\sum_{i=t_0}^{\infty}
\frac{
|\delta_i|x^{i\theta}
}{
\Gamma(i\theta+B+\varrho)
}
\nonumber\\
&\leq
\left(\prod_{\ell=1}^{L}C_\ell\right)
\sum_{\substack{
i_1,\ldots,i_L\geq0\\
i_1+\cdots+i_L\geq t_0
}}
\frac{
x^{\theta\sum_{\ell=1}^{L}i_\ell}
}{
\Gamma\!\left(\theta\sum_{\ell=1}^{L}i_\ell+B+\varrho\right)
}
\nonumber\\
&\quad\times
\prod_{\ell=1}^{L}
\frac{
A_\ell^{i_\ell}
\Gamma(\theta i_\ell+\beta_\ell)
\Gamma(q_\ell i_\ell+c_\ell)
}{
i_\ell!
}.
\label{eq: truncation multiindex step}
\end{align}

Next, applying the Gamma-product bound in
\eqref{eq: gamma ratio bound generalized}, and using
the fact that $\sum_{\ell=1}^{L}i_\ell\geq t_0$, we have
\begin{align}
\frac{
\Gamma\!\left(\theta\sum_{\ell=1}^{L}i_\ell+B\right)
}{
\Gamma\!\left(\theta\sum_{\ell=1}^{L}i_\ell+B+\varrho\right)
}
&\leq
(t_0\theta+B)^{-\varrho}.
\label{eq: truncation gamma ratio generalized}
\end{align}
Therefore,
\begin{align}
&
\sum_{i=t_0}^{\infty}
\frac{
|\delta_i|x^{i\theta}
}{
\Gamma(i\theta+B+\varrho)
}
\nonumber\\
&\leq
C_{\boldsymbol\beta}
\left(\prod_{\ell=1}^{L}C_\ell\right)
(t_0\theta+B)^{-\varrho}
\mathcal T_{t_0}(x),
\label{eq: truncation tail bound generalized}
\end{align}
where
\begin{align}
\mathcal T_{t_0}(x)
&\triangleq
\sum_{\substack{
i_1,\ldots,i_L\geq0\\
i_1+\cdots+i_L\geq t_0
}}
\prod_{\ell=1}^{L}
\frac{
\Gamma(q_\ell i_\ell+c_\ell)
(A_\ell x^\theta)^{i_\ell}
}{
i_\ell!
}.
\label{eq: generalized multiindex tail}
\end{align}

Finally, replacing \eqref{eq: truncation tail bound generalized}
into \eqref{eq: bound Xi step 1}, we get
\begin{align}
\left|
\Xi_{\vartheta}^{(t_0)}(x)
\right|
&\leq
C_{\boldsymbol\beta}
\left(\prod_{\ell=1}^{L}C_\ell\Psi_\ell\right)
(t_0\theta+B)^{-\varrho}
x^{B+\varrho-1}
\mathcal T_{t_0}(x).
\label{eq: Truncation bound generalized tail}
\end{align}

We now derive a closed-form upper bound for
$\mathcal T_{t_0}(x)$. Define
\begin{align}
H_\ell(y)
&\triangleq
\sum_{i=0}^{\infty}
\frac{
\Gamma(q_\ell i+c_\ell)y^i
}{
i!
},
\qquad y\geq0 .
\label{eq: H ell definition}
\end{align}
Since $0\leq q_\ell<1$, the ratio test implies that
$H_\ell(y)$ is finite for every finite $y$. Moreover, for any
$\chi>1$,
\begin{align}
\mathcal T_{t_0}(x)
&\leq
\chi^{-t_0}
\prod_{\ell=1}^{L}
H_\ell\!\left(\chi A_\ell x^\theta\right).
\label{eq: generalized tail chi bound}
\end{align}
Indeed, whenever $i_1+\cdots+i_L\geq t_0$, one has
$\chi^{-t_0}\chi^{i_1+\cdots+i_L}\geq1$. Multiplying each term in
$\mathcal T_{t_0}(x)$ by this factor and then extending the sum to
all multi-indices in $\mathbb N_0^L$ gives
\eqref{eq: generalized tail chi bound}.

To bound $H_\ell(y)$ explicitly, use the Gamma integral
representation~\cite[Eq. (06.05.07.0001.01)]{Mathematica}
\begin{align}
\Gamma(q_\ell i+c_\ell)
&=
\int_{0}^{\infty}
t^{q_\ell i+c_\ell-1}e^{-t}\,dt,
\qquad q_\ell i+c_\ell>0 .
\end{align}
Since $c_\ell>0$ and $q_\ell\geq0$, this representation is valid
for every $i\in\mathbb N_0$. Hence, for $y\geq0$,
\begin{align}
H_\ell(y)
&=
\sum_{i=0}^{\infty}
\frac{y^i}{i!}
\int_{0}^{\infty}
t^{q_\ell i+c_\ell-1}e^{-t}\,dt
\nonumber\\
&=
\sum_{i=0}^{\infty}
\int_{0}^{\infty}
\frac{
y^i t^{q_\ell i+c_\ell-1}e^{-t}
}{
i!
}
\,dt .
\end{align}
The integrand is nonnegative for $y\geq0$. Therefore, Tonelli's
theorem permits the interchange of the summation and the integral,
which yields
\begin{align}
H_\ell(y)
&=
\int_{0}^{\infty}
t^{c_\ell-1}e^{-t}
\sum_{i=0}^{\infty}
\frac{
\left(y t^{q_\ell}\right)^i
}{
i!
}
\,dt
\nonumber\\
&=
\int_{0}^{\infty}
t^{c_\ell-1}e^{-t}
\exp\!\left(y t^{q_\ell}\right)
\,dt .
\label{eq: H ell integral}
\end{align}

For any $\varepsilon_\ell\in(0,1)$ and $0\leq q_\ell<1$, the
Young-type inequality~\cite{HardyLittlewoodPolya1952}
\begin{align}
y t^{q_\ell}
&\leq
\varepsilon_\ell t
+
\omega_\ell(\varepsilon_\ell,y),
\qquad t,y\geq0,
\label{eq: Young bound appendix}
\end{align}
holds, where
\begin{align}
\omega_\ell(\varepsilon_\ell,y)
&\triangleq
(1-q_\ell)
q_\ell^{\frac{q_\ell}{1-q_\ell}}
\varepsilon_\ell^{-\frac{q_\ell}{1-q_\ell}}
y^{\frac{1}{1-q_\ell}} .
\label{eq: omega definition appendix}
\end{align}
This bound follows by maximizing
$y t^{q_\ell}-\varepsilon_\ell t$ over $t\geq0$. When
$q_\ell=0$, \eqref{eq: Young bound appendix} is interpreted by
continuity. In this case,
$\omega_\ell(\varepsilon_\ell,y)=y$, and the inequality reduces to
$y\leq\varepsilon_\ell t+y$.

Substituting \eqref{eq: Young bound appendix} into
\eqref{eq: H ell integral}, we obtain
\begin{align}
H_\ell(y)
&\leq
\exp\!\left[\omega_\ell(\varepsilon_\ell,y)\right]
\int_{0}^{\infty}
t^{c_\ell-1}e^{-(1-\varepsilon_\ell)t}\,dt
\nonumber\\
&=
\Gamma(c_\ell)
(1-\varepsilon_\ell)^{-c_\ell}
\exp\!\left[\omega_\ell(\varepsilon_\ell,y)\right].
\label{eq: H ell closed bound}
\end{align}

Using \eqref{eq: H ell closed bound} into \eqref{eq: generalized tail chi bound}, we have
\begin{align}
\mathcal T_{t_0}(x)
&\leq
\chi^{-t_0}
\prod_{\ell=1}^{L}
\Gamma(c_\ell)(1-\varepsilon_\ell)^{-c_\ell}
\nonumber\\
&\quad\times
\exp\!\left[
\sum_{\ell=1}^{L}
\omega_\ell\!\left(
\varepsilon_\ell,
\chi A_\ell x^\theta
\right)
\right].
\label{eq: generalized tail closed bound}
\end{align}

Finally, combining \eqref{eq: Truncation bound generalized tail} with
\eqref{eq: generalized tail closed bound} yields the closed-form
truncation bound in \eqref{eq: closed truncation bound main}. This concludes the proof.

\section{Proof of Proposition~\ref{lemma: Marginal LT}}
\label{app: table theorem verification}

This appendix verifies directly that four of the fading models
listed in Table~\ref{tab: PDF-CDF parameters} satisfy the assumptions
of the Theorem, namely, the Extended
$\bar{\eta}$-$\bar{\mu}$,
$\bar{\alpha}$-$\kappa$-$\bar{\mu}$ Shadowed, MFTR, and
$\bar{\alpha}$-$\bar{\eta}$-$\kappa$-$\bar{\mu}$ fading models.
Throughout this appendix, all distribution parameters are assumed to
belong to their admissible ranges. In particular, all shape and scale
parameters are positive.
The remaining two models follow directly as special cases of these more
general distributions.
The verification is performed
through Proposition~\ref{lemma: Marginal LT}.

From Proposition~\ref{lemma: Marginal LT}, each marginal PDF can be
written in the canonical form required by the
Theorem. Hence, comparing
\eqref{eq: PDF marginal def} with \eqref{eq: PDF marginal theorem},
we identify $\eta_{n,\ell}
=
\lambda_{n,\ell}$, $\theta
=
\alpha$, and $\beta_\ell
=
\alpha\mu_\ell$.
Moreover, since $\varphi_{0,\ell}>0$ and $\alpha,\mu_\ell>0$, we
have
\begin{align}
\lambda_{0,\ell}
&=
\frac{\Gamma(\alpha\mu_\ell)}{\Gamma(\mu_\ell)}
\varphi_{0,\ell}
>0.
\label{eq: lambda zero positive appendix}
\end{align}
Hence, the leading-coefficient condition of the
Theorem is satisfied.

It remains to verify the generalized coefficient-growth condition
in \eqref{eq: generalized coefficient growth}. We first establish a
useful reduced bound for the coefficients $\varphi_{i,\ell}$.
Suppose that, for a given marginal model, there exist constants
$M_\ell,\rho_\ell>0$, independent of $i$, such that
\begin{align}
\frac{|\varphi_{i,\ell}|}{\Gamma(i+\mu_\ell)}
&\leq
M_\ell
\frac{\rho_\ell^i}{i!},
\qquad i\in\mathbb N_0 .
\label{eq: reduced coefficient bound appendix}
\end{align}
Then, using the definition of $\lambda_{n,\ell}$ in
\eqref{eq: lambda coefficient}, we obtain
\begin{align}
|\lambda_{n,\ell}|
&\leq
\frac{
\Gamma(\alpha n+\beta_\ell)
}{
\hat r_\ell^{\alpha n}
}
\sum_{j=0}^{n}
\frac{
|\varphi_{j,\ell}|
}{
(n-j)!\Gamma(j+\mu_\ell)
}
\nonumber\\
&\leq
\frac{
\Gamma(\alpha n+\beta_\ell)
}{
\hat r_\ell^{\alpha n}
}
M_\ell
\sum_{j=0}^{n}
\frac{\rho_\ell^j}{j!(n-j)!}
\nonumber\\
&=
M_\ell
\left(
\frac{1+\rho_\ell}{\hat r_\ell^\alpha}
\right)^n
\frac{\Gamma(\alpha n+\beta_\ell)}{n!}.
\label{eq: lambda growth appendix}
\end{align}
Therefore, \eqref{eq: generalized coefficient growth} holds as the
special case $C_\ell
=
M_\ell$, $A_\ell
=
\frac{1+\rho_\ell}{\hat r_\ell^\alpha}$, $q_\ell
=
0$, and $c_\ell
=
1$.
Indeed, with $q_\ell=0$ and $c_\ell=1$, we get 
$\Gamma(q_\ell n+c_\ell)=\Gamma(1)=1$.

We now verify that the coefficients associated with the considered fading models in
Table~\ref{tab: PDF-CDF parameters} satisfy
\eqref{eq: reduced coefficient bound appendix}.
We repeatedly use the following elementary consequence of the
gamma-ratio asymptotic formula: for any $a,b>0$ and any $\tau>1$,
there exists $K_{a,b,\tau}>0$ such that
\begin{align}
\frac{\Gamma(i+a)}{\Gamma(i+b)}
&\leq
K_{a,b,\tau}\tau^i,
\qquad i\in\mathbb N_0.
\label{eq: gamma ratio appendix}
\end{align}

\subsection{Extended $\bar{\eta}$-$\bar{\mu}$ Fading Model}

For the Extended $\bar{\eta}$-$\bar{\mu}$ model,
Table~\ref{tab: PDF-CDF parameters} gives
\begin{align}
\varphi_i
&=
\frac{1}{i!}
\left(
\frac{p}{\bar{\eta}}
\right)^{\frac{\bar{\mu}p}{p+1}}
\left(
\frac{\bar{\eta}-p}{\bar{\eta}}
\right)^i
\left(
\frac{\bar{\mu}p}{p+1}
\right)_i,
\qquad i\in\mathbb N_0.
\label{eq: extended eta mu coefficient appendix}
\end{align}
Define
\begin{align}
a
&\triangleq
\frac{\bar{\mu}p}{p+1}\\
d
&\triangleq
\frac{\bar{\eta}-p}{\bar{\eta}}.
\label{eq: extended eta mu constants appendix}
\end{align}
Under the admissible parameter conditions
$p,\bar{\eta},\bar{\mu}>0$, we have $a>0$. Moreover,
using
\begin{align}
(a)_i
&=
\frac{\Gamma(i+a)}{\Gamma(a)},
\end{align}
the coefficient in
\eqref{eq: extended eta mu coefficient appendix} can be rewritten as
\begin{align}
\varphi_i
&=
\left(
\frac{p}{\bar{\eta}}
\right)^a
\frac{\Gamma(i+a)}{\Gamma(a)}
\frac{d^i}{i!}.
\label{eq: extended eta mu coefficient gamma appendix}
\end{align}
In particular,
\begin{align}
\varphi_0
&=
\left(
\frac{p}{\bar{\eta}}
\right)^a
>0,
\end{align}
so the leading-coefficient condition is satisfied.

Taking absolute values in
\eqref{eq: extended eta mu coefficient gamma appendix} gives
\begin{align}
\frac{|\varphi_i|}{\Gamma(i+\bar{\mu})}
&=
\frac{1}{\Gamma(a)}
\left(
\frac{p}{\bar{\eta}}
\right)^a
\frac{\Gamma(i+a)}{\Gamma(i+\bar{\mu})}
\frac{|d|^i}{i!}.
\label{eq: extended eta mu reduced coefficient step appendix}
\end{align}
Applying the Gamma-ratio bound in
\eqref{eq: gamma ratio appendix} with
\begin{align}
a
&=
\frac{\bar{\mu}p}{p+1}\\
b
&=
\bar{\mu},
\end{align}
we conclude that, for any fixed $\tau>1$, there exists a constant
$K_{a,\bar{\mu},\tau}>0$, independent of $i$, such that
\begin{align}
\frac{\Gamma(i+a)}{\Gamma(i+\bar{\mu})}
&\leq
K_{a,\bar{\mu},\tau}\tau^i,
\qquad i\in\mathbb N_0.
\label{eq: extended eta mu gamma ratio appendix}
\end{align}
Substituting
\eqref{eq: extended eta mu gamma ratio appendix} into
\eqref{eq: extended eta mu reduced coefficient step appendix}, we
obtain
\begin{align}
\frac{|\varphi_i|}{\Gamma(i+\bar{\mu})}
&\leq
\frac{
K_{a,\bar{\mu},\tau}
}{
\Gamma(a)
}
\left(
\frac{p}{\bar{\eta}}
\right)^a
\frac{
\left(\tau |d|\right)^i
}{
i!
}.
\label{eq: extended eta mu coefficient preliminary bound appendix}
\end{align}
Therefore,
\begin{align}
\frac{|\varphi_i|}{\Gamma(i+\bar{\mu})}
&\leq
M_\ell
\frac{\rho_\ell^i}{i!},
\qquad i\in\mathbb N_0,
\label{eq: extended eta mu reduced coefficient bound appendix}
\end{align}
where
\begin{align}
M_\ell
&\triangleq
\frac{
K_{a,\bar{\mu},\tau}
}{
\Gamma(a)
}
\left(
\frac{p}{\bar{\eta}}
\right)^a,
\\
\rho_\ell
&\triangleq
\tau
\left(
1+
\left|
\frac{\bar{\eta}-p}{\bar{\eta}}
\right|
\right).
\label{eq: extended eta mu bound constants appendix}
\end{align}
The additional term $1$ in the definition of $\rho_\ell$ ensures that
$\rho_\ell>0$, including the boundary case $p=\bar{\eta}$. Since
\[
\left|
\frac{\bar{\eta}-p}{\bar{\eta}}
\right|^i
\leq
\left(
1+
\left|
\frac{\bar{\eta}-p}{\bar{\eta}}
\right|
\right)^i,
\]
the bound remains valid for all admissible parameter values.

Thus, the reduced coefficient bound in
\eqref{eq: reduced coefficient bound appendix} holds for the
Extended $\bar{\eta}$-$\bar{\mu}$ fading model.

\subsection{$\bar{\alpha}$-$\kappa$-$\bar{\mu}$ Shadowed Fading Model}

For the $\bar{\alpha}$-$\kappa$-$\bar{\mu}$ Shadowed model,
Table~\ref{tab: PDF-CDF parameters} gives
\begin{align}
\varphi_i
&=
\frac{m^m(m)_i}{i!(\kappa\bar{\mu}+m)^m}
\left(
\frac{\kappa\bar{\mu}}{\kappa\bar{\mu}+m}
\right)^i,
\qquad i\in\mathbb N_0 .
\end{align}
Since $(m)_i=\Gamma(i+m)/\Gamma(m)$, we have
\begin{align}
\frac{|\varphi_i|}{\Gamma(i+\bar{\mu})}
&=
\frac{m^m}{(\kappa\bar{\mu}+m)^m\Gamma(m)}
\frac{\Gamma(i+m)}{\Gamma(i+\bar{\mu})}
\frac{1}{i!}
\left(
\frac{\kappa\bar{\mu}}{\kappa\bar{\mu}+m}
\right)^i .
\end{align}
Using the Gamma-ratio bound in \eqref{eq: gamma ratio appendix}
with $a=m$ and $b=\bar{\mu}$, there exists a constant
$K_{m,\bar{\mu},\tau}>0$ such that, for any fixed $\tau>1$,
\begin{align}
\frac{\Gamma(i+m)}{\Gamma(i+\bar{\mu})}
&\leq
K_{m,\bar{\mu},\tau}\tau^i,
\qquad i\in\mathbb N_0 .
\end{align}
Therefore,
\begin{align}
\frac{|\varphi_i|}{\Gamma(i+\bar{\mu})}
&\leq
M
\frac{\rho^i}{i!},
\qquad i\in\mathbb N_0,
\end{align}
where
\begin{align}
M
&\triangleq
\frac{m^m K_{m,\bar{\mu},\tau}}
{(\kappa\bar{\mu}+m)^m\Gamma(m)},
&
\rho
&\triangleq
\tau
\frac{\kappa\bar{\mu}}{\kappa\bar{\mu}+m}.
\end{align}
Thus, the reduced coefficient bound in
\eqref{eq: reduced coefficient bound appendix} holds for suitable
constants $M_\ell,\rho_\ell>0$.

\subsection{MFTR Fading Model}

For the MFTR model, $K\geq0$ and $0\leq\Delta<1$. Define
\begin{align}
z_\Delta
&\triangleq
\frac{2\Delta K\bar{\mu}}
{(1-\Delta)K\bar{\mu}+m}
\geq0 .
\end{align}
The hypergeometric function appearing in
Table~\ref{tab: PDF-CDF parameters} has argument $-z_\Delta$.
Since $i+m>0$, $s+1/2>0$, and $s+1>s+1/2$, Euler's integral
representation gives~\cite{Olver10}
\begin{align}
0
\leq
{}_2F_1\!\left(
i+m,s+\frac{1}{2};s+1;-z_\Delta
\right)
\leq
1 .
\label{eq: MFTR hypergeometric bound appendix}
\end{align}
Moreover,
\begin{align}
\frac{\Gamma(s+\frac12)}{\Gamma(s+1)}
&\leq
\sqrt{\pi},
\qquad s\in\mathbb N_0 .
\end{align}
Therefore, the finite sum in the MFTR coefficient satisfies
\begin{align}
\nonumber &
\sum_{s=0}^{i}
\binom{i}{s}
\frac{\Gamma(s+\frac12)}{\Gamma(s+1)}
\left(
\frac{2\Delta}{1-\Delta}
\right)^s \\
& \qquad \times
\left|
{}_2F_1\!\left(
i+m,s+\frac12;s+1;-z_\Delta
\right)
\right|
\nonumber\\
&\qquad \leq
\sqrt{\pi}
\left(
1+\frac{2\Delta}{1-\Delta}
\right)^i .
\end{align}
Replacing this bound into the MFTR coefficient, we obtain
\begin{align}
\nonumber |\varphi_i|
&\leq
\frac{
m^m
}{
\Gamma(m)\left((1-\Delta)K\bar{\mu}+m\right)^m
} \\
& \times
\frac{\Gamma(i+m)}{i!}
\left(
\frac{K\bar{\mu}(1+\Delta)}
{(1-\Delta)K\bar{\mu}+m}
\right)^i .
\end{align}
Consequently,
\begin{align}
\nonumber \frac{|\varphi_i|}{\Gamma(i+\bar{\mu})}
&\leq
\frac{
m^m
}{
\Gamma(m)\left((1-\Delta)K\bar{\mu}+m\right)^m
} \\
& \times
\frac{\Gamma(i+m)}{\Gamma(i+\bar{\mu})}
\frac{1}{i!}
\left(
\frac{K\bar{\mu}(1+\Delta)}
{(1-\Delta)K\bar{\mu}+m}
\right)^i .
\end{align}
Using the Gamma-ratio bound in \eqref{eq: gamma ratio appendix}
with $a=m$ and $b=\bar{\mu}$, there exists
$K_{m,\bar{\mu},\tau}>0$ such that, for any fixed $\tau>1$,
\begin{align}
\frac{\Gamma(i+m)}{\Gamma(i+\bar{\mu})}
&\leq
K_{m,\bar{\mu},\tau}\tau^i,
\qquad i\in\mathbb N_0 .
\end{align}
Hence,
\begin{align}
\frac{|\varphi_i|}{\Gamma(i+\bar{\mu})}
&\leq
M
\frac{\rho^i}{i!},
\qquad i\in\mathbb N_0,
\end{align}
where
\begin{align}
M
&\triangleq
\frac{
m^m K_{m,\bar{\mu},\tau}
}{
\Gamma(m)\left((1-\Delta)K\bar{\mu}+m\right)^m
}\\
\rho
&\triangleq
\tau
\frac{K\bar{\mu}(1+\Delta)}
{(1-\Delta)K\bar{\mu}+m}
\end{align}
Thus, the reduced coefficient bound in
\eqref{eq: reduced coefficient bound appendix} holds for suitable
constants $M_\ell,\rho_\ell>0$.

\subsection{$\bar{\alpha}$-$\bar{\eta}$-$\kappa$-$\bar{\mu}$ Fading Model}

Finally, consider the
$\bar{\alpha}$-$\bar{\eta}$-$\kappa$-$\bar{\mu}$ model. From
Table~\ref{tab: PDF-CDF parameters}, its coefficient can be written
in the compact form
\begin{align}
\varphi_i
&=
C_0^\dagger p^{-i}
\sum_{k=0}^{i}
\frac{P_k E^{i-k}}{(i-k)!},
\label{eq: eta kappa mu compact appendix}
\end{align}
where
\begin{align}
E
&=
\frac{\kappa\bar{\mu}p^2q}{\delta},
\\
C_0^\dagger
&=
\exp\!\left[
-\frac{\kappa\bar{\mu}(pq+1)}{\delta}
\right]
\frac{
\bar{\eta}^{\bar{\mu}}
\left(\frac{p}{\bar{\eta}}\right)^{
\frac{\bar{\mu}p}{p+1}
}
}{
p^{\bar{\mu}}
},
\\
P_k
&=
(p-\bar{\eta})^k
L_k^{\frac{\bar{\mu}}{p+1}-1}
\!\left(
\frac{\bar{\eta}\kappa\bar{\mu}}
{\delta(\bar{\eta}-p)}
\right).
\label{eq: Pk definition appendix}
\end{align}

We first consider the case $p\neq\bar{\eta}$. In this case, the
order and argument of the generalized Laguerre polynomial in
\eqref{eq: Pk definition appendix} are fixed. Its generating function
is given by
\begin{align}
\sum_{k=0}^{\infty}L_k^{a}(x)z^k
&=
(1-z)^{-a-1}
\exp\!\left(
-\frac{xz}{1-z}
\right),
\qquad |z|<1.
\label{eq: Laguerre generating function appendix}
\end{align}
For any fixed $r\in(0,1)$, Cauchy's coefficient estimate applied to
\eqref{eq: Laguerre generating function appendix} yields
\begin{align}
\left|L_k^{a}(x)\right|
&\leq
K_r r^{-k},
\qquad k\in\mathbb N_0,
\end{align}
where $K_r>0$ is independent of $k$. After including the factor
$|p-\bar{\eta}|^k$, it follows that there exist constants
$K_\ell,\widetilde{\sigma}_\ell>0$, independent of $k$, such that
\begin{align}
|P_k|
&\leq
K_\ell\widetilde{\sigma}_\ell^k,
\qquad k\in\mathbb N_0.
\label{eq: Pk geometric appendix}
\end{align}

We next consider the boundary case $p=\bar{\eta}$. In this case,
\eqref{eq: Pk definition appendix} is interpreted through the limit
$p\to\bar{\eta}$. Define
\begin{align}
a_p
&\triangleq
\frac{\bar{\mu}}{p+1}-1\\
C^\dagger
&\triangleq
\frac{\bar{\eta}\kappa\bar{\mu}}{\delta}.
\end{align}
Using the finite expansion
\begin{align}
L_k^{a_p}(x)
&=
\sum_{j=0}^{k}
\binom{k+a_p}{k-j}
\frac{(-x)^j}{j!},
\end{align}
and noting that
\begin{align}
\frac{C^\dagger}{\bar{\eta}-p}
&=
-\frac{C^\dagger}{p-\bar{\eta}},
\end{align}
we obtain
\begin{align}
&(p-\bar{\eta})^k
L_k^{a_p}\!\left(
\frac{C^\dagger}{\bar{\eta}-p}
\right)
\nonumber\\
&\qquad=
\sum_{j=0}^{k}
\binom{k+a_p}{k-j}
\frac{(C^\dagger)^j}{j!}
(p-\bar{\eta})^{k-j}.
\label{eq: Laguerre boundary expansion appendix}
\end{align}
For each fixed $k$, all terms in
\eqref{eq: Laguerre boundary expansion appendix} with $j<k$
vanish as $p\to\bar{\eta}$, whereas the term with $j=k$ remains.
Consequently,
\begin{align}
&\lim_{p\to\bar{\eta}}
(p-\bar{\eta})^k
L_k^{\frac{\bar{\mu}}{p+1}-1}
\!\left(
\frac{\bar{\eta}\kappa\bar{\mu}}
{\delta(\bar{\eta}-p)}
\right)
\nonumber\\
&\qquad=
\frac{1}{k!}
\left(
\frac{\bar{\eta}\kappa\bar{\mu}}{\delta}
\right)^k.
\label{eq: Laguerre boundary limit appendix}
\end{align}
Thus, when $p=\bar{\eta}$,
\begin{align}
|P_k|
&=
\frac{(C^\dagger)^k}{k!}
\leq
\widetilde{\sigma}_\ell^k,
\qquad k\in\mathbb N_0,
\end{align}
where
\begin{align}
\widetilde{\sigma}_\ell
&\triangleq
\max\{1,C^\dagger\}.
\end{align}
Therefore, after increasing $K_\ell$ and
$\widetilde{\sigma}_\ell$ if necessary, the geometric estimate in
\eqref{eq: Pk geometric appendix} holds for both
$p\neq\bar{\eta}$ and $p=\bar{\eta}$.

Using \eqref{eq: Pk geometric appendix} in
\eqref{eq: eta kappa mu compact appendix}, we obtain
\begin{align}
|\varphi_i|
&\leq
|C_0^\dagger|K_\ell p^{-i}
\sum_{k=0}^{i}
\frac{
\widetilde{\sigma}_\ell^k |E|^{i-k}
}{
(i-k)!
}
\nonumber\\
&\leq
|C_0^\dagger|K_\ell p^{-i}
\sum_{k=0}^{i}
\binom{i}{k}
\widetilde{\sigma}_\ell^k |E|^{i-k}
\nonumber\\
&=
|C_0^\dagger|K_\ell
\left(
\frac{\widetilde{\sigma}_\ell+|E|}{p}
\right)^i.
\label{eq: eta kappa mu coefficient bound appendix}
\end{align}
In the second inequality, we used
\begin{align}
\frac{1}{(i-k)!}
&\leq
\frac{i!}{k!(i-k)!}
=
\binom{i}{k},
\qquad 0\leq k\leq i.
\end{align}
Consequently,
\begin{align}
\frac{|\varphi_i|}{\Gamma(i+\bar{\mu})}
&\leq
|C_0^\dagger|K_\ell
\frac{\Gamma(i+1)}{\Gamma(i+\bar{\mu})}
\left(
\frac{\widetilde{\sigma}_\ell+|E|}{p}
\right)^i
\frac{1}{i!}.
\end{align}
Applying the Gamma-ratio bound in
\eqref{eq: gamma ratio appendix} with $a=1$ and
$b=\bar{\mu}$, for any fixed $\tau>1$ there exists a constant
$K_{1,\bar{\mu},\tau}>0$ such that
\begin{align}
\frac{\Gamma(i+1)}{\Gamma(i+\bar{\mu})}
&\leq
K_{1,\bar{\mu},\tau}\tau^i,
\qquad i\in\mathbb N_0.
\end{align}
Therefore,
\begin{align}
\frac{|\varphi_i|}{\Gamma(i+\bar{\mu})}
&\leq
M_\ell
\frac{\rho_\ell^i}{i!},
\qquad i\in\mathbb N_0,
\end{align}
where
\begin{align}
M_\ell
&\triangleq
|C_0^\dagger|K_\ell K_{1,\bar{\mu},\tau}\\
\rho_\ell
&\triangleq
\tau
\frac{\widetilde{\sigma}_\ell+|E|}{p}.
\end{align}
Thus, the reduced coefficient bound in
\eqref{eq: reduced coefficient bound appendix} holds for the
$\bar{\alpha}$-$\bar{\eta}$-$\kappa$-$\bar{\mu}$ model.

\subsection{Conclusion of the Proof}

Consequently, the coefficient families of the Extended
$\bar{\eta}$-$\bar{\mu}$,
$\bar{\alpha}$-$\kappa$-$\bar{\mu}$ Shadowed, MFTR, and
$\bar{\alpha}$-$\bar{\eta}$-$\kappa$-$\bar{\mu}$ fading models
satisfy the reduced bound in
\eqref{eq: reduced coefficient bound appendix}, with the
remaining two models covered as special cases of the distributions
verified above.
Hence, by \eqref{eq: lambda growth appendix}, the induced
coefficients $\lambda_{n,\ell}$ for each of these fading families
satisfy
\begin{align}
|\lambda_{n,\ell}|
&\leq
C_\ell A_\ell^n
\frac{
\Gamma(\alpha n+\beta_\ell)\Gamma(q_\ell n+c_\ell)
}{
n!
},
\qquad n\in\mathbb N_0,
\label{eq: final lambda theorem bound appendix}
\end{align}
with the special choice $q_\ell
=
0$ and $c_\ell
=
1$.
Therefore, the generalized coefficient-growth condition of the
Theorem, given in
\eqref{eq: generalized coefficient growth}, is satisfied.
Thus, for any collection of independent marginals from the
verified fading families, the Theorem applies directly
to $R
=
\sum_{\ell=1}^{L}R_\ell$.
Equivalently, the PDF and CDF of $R$ are precisely those stated in
Proposition~\ref{lemma: Marginal LT}. The coefficients $\delta_n$
are computed using the recursion in
Proposition~\ref{lemma: Marginal LT}. Moreover, the absolute and
locally uniform convergence of the resulting PDF and CDF series
follows from Appendix~\ref{app: Absolute Convergence X}. This
completes the proof.

\section{Proof of Proposition~\ref{prop: sum ratio}}
\label{app: sum ratio}

This appendix shows that Proposition~\ref{prop: sum ratio} follows
as a direct application of the Theorem. In
particular, we verify that the marginal ratio PDFs satisfy the
canonical PDF representation and the generalized coefficient-growth
condition in \eqref{eq: generalized coefficient growth}.

Let
\begin{align}
\tilde{q}
&\triangleq
\frac{\alpha_M}{\alpha_Q}.
\label{eq: q ratio appendix}
\end{align}
Since Proposition~\ref{prop: sum ratio} assumes
$\alpha_M<\alpha_Q$, it follows that $0<\tilde{q}<1$.
For each $\ell\in\{1,\ldots,L\}$, define
\begin{align}
\beta_\ell
&\triangleq
\alpha_M\mu_{M,\ell}\\ 
\Psi_\ell
&\triangleq
\frac{\alpha_M}
{\Gamma(\mu_{M,\ell})\Gamma(\mu_{Q,\ell})},
\label{eq: ratio canonical parameters}
\end{align}
Additionally, define
\begin{align}
u_{i,\ell}
&\triangleq
\frac{(-1)^i
\Gamma\!\left(\alpha_M(i+\mu_{M,\ell})\right)}
{i!}
\widetilde{\varrho}_{\ell}^{\alpha_M(i+\mu_{M,\ell})}
\nonumber\\
&\quad\times
\Gamma\!\left(
\tilde{q}(i+\mu_{M,\ell})+\mu_{Q,\ell}
\right),
\qquad i\in\mathbb N_0 .
\label{eq: u ratio appendix}
\end{align}
Then, the marginal PDF in \eqref{eq: pdf_ratio} can be rewritten as
\begin{align}
f_{Z_\ell}(z)
&=
\Psi_\ell
\sum_{i=0}^{\infty}
u_{i,\ell}
\frac{
z^{\alpha_M i+\beta_\ell-1}
}{
\Gamma(\alpha_M i+\beta_\ell)
},
\qquad z>0.
\label{eq: canonical ratio marginal PDF}
\end{align}
Thus, \eqref{eq: canonical ratio marginal PDF} has exactly the
canonical form required by the Theorem, with 
$\theta
=
\alpha_M$, $\eta_{i,\ell}
=
u_{i,\ell}$, and $\beta_\ell
=
\alpha_M\mu_{M,\ell}$.
Furthermore,
\begin{align}
u_{0,\ell}
&=
\Gamma(\alpha_M\mu_{M,\ell})
\widetilde{\varrho}_{\ell}^{\alpha_M\mu_{M,\ell}}
\Gamma(\tilde{q}\mu_{M,\ell}+\mu_{Q,\ell})
>0,
\label{eq: u zero positive appendix}
\end{align}
because all factors are positive. Hence, the leading-coefficient
condition of the Theorem is satisfied.

It remains to verify \eqref{eq: generalized coefficient growth}.
From \eqref{eq: u ratio appendix}, we obtain 
\begin{align}
|u_{i,\ell}|
&=
\widetilde{\varrho}_{\ell}^{\alpha_M\mu_{M,\ell}}
\left(\widetilde{\varrho}_{\ell}^{\alpha_M}\right)^i
\frac{
\Gamma(\alpha_M i+\beta_\ell)
\Gamma(\tilde{q} i+c_\ell)
}{
i!
},
\label{eq: u growth exact appendix}
\end{align}
where
\begin{align}
c_\ell
&\triangleq
\tilde{q} \mu_{M,\ell}+\mu_{Q,\ell}
>0 .
\label{eq: c ratio appendix}
\end{align}
Therefore, the generalized coefficient-growth condition holds with
\begin{align}
C_\ell
&=
\widetilde{\varrho}_{\ell}^{\alpha_M\mu_{M,\ell}}\\
A_\ell
&=
\widetilde{\varrho}_{\ell}^{\alpha_M}\\
q_\ell
&=
\tilde{q}
=
\frac{\alpha_M}{\alpha_Q}\\
c_\ell
&=
\frac{\alpha_M}{\alpha_Q}\mu_{M,\ell}
+\mu_{Q,\ell}.
\label{eq: ratio theorem constants appendix}
\end{align}
Since $0<q_\ell<1$, all assumptions of the
Theorem are fulfilled.
Consequently, the Theorem applies directly to $Z
=
\sum_{\ell=1}^{L}Z_\ell$.
Since
\begin{align}
B
&\triangleq
\sum_{\ell=1}^{L}\beta_\ell
=
\alpha_M\sum_{\ell=1}^{L}\mu_{M,\ell},
\end{align}
the PDF and CDF of $Z$ are
\begin{align}
f_Z(z)
&=
\left(
\prod_{\ell=1}^{L}\Psi_\ell
\right)
\sum_{i=0}^{\infty}
\frac{
\delta_i z^{\alpha_M i+B-1}
}{
\Gamma(\alpha_M i+B)
},
\qquad z>0,
\label{eq: sum PDF ratio theorem proof}
\\
F_Z(z)
&=
\left(
\prod_{\ell=1}^{L}\Psi_\ell
\right)
\sum_{i=0}^{\infty}
\frac{
\delta_i z^{\alpha_M i+B}
}{
\Gamma(1+\alpha_M i+B)
},
\qquad z\geq0.
\label{eq: sum CDF ratio theorem proof}
\end{align}
Substituting the definition of $\Psi_\ell$ in
\eqref{eq: ratio canonical parameters} into
\eqref{eq: sum PDF ratio theorem proof} and
\eqref{eq: sum CDF ratio theorem proof}, we obtain
\eqref{eq: sum PDF z} and \eqref{eq: sum CDF z}, respectively.
Finally, the coefficients $\delta_i$ are the coefficients of the
Theorem with $\eta_{i,\ell}=u_{i,\ell}$.
The absolute and locally uniform convergence of the PDF and CDF
series follows from Appendix~\ref{app: Absolute Convergence X}.
This completes the proof.

\end{appendices}

\bibliographystyle{IEEEtran}
\bibliography{Bibliography}

\end{document}